\documentclass[10pt,a4paper]{article}
\usepackage[margin=1in]{geometry}
\usepackage{cancel}
\usepackage[utf8]{inputenc}
\usepackage{amsmath}
\usepackage{dblfnote}
\usepackage{amsfonts}
\usepackage{amssymb}
\usepackage{graphicx}
\usepackage{mathtools}
\usepackage{hyperref}
\usepackage{cleveref}
\usepackage[outercaption]{sidecap}   

\usepackage{bm}
\usepackage{feynmp-auto}
\numberwithin{equation}{section}
\usepackage{authblk}
\usepackage[backend=biber,style=nature,url=false,eprint=false,isbn=false]{biblatex} 

\AtEveryBibitem{
	\clearfield{note}
}
\AtEveryBibitem{\clearfield{month}}
\AtEveryBibitem{\clearfield{day}}

\makeatletter
\newcommand{\vast}{\bBigg@{4}}
\newcommand{\Vast}{\bBigg@{5}}
\makeatother

\title{\textbf{ \Large Accurate dynamics from self-consistent memory in \\stochastic chemical reactions with small copy numbers} }
\author[1]{\textbf{Moshir Harsh} \footnote{
		email: moshir.harsh@uni-goettingen.de}}
	
\author[1,2]{\textbf{Peter Sollich} \footnote{email: peter.sollich@uni-goettingen.de}}
\affil[1]{Instit{\"u}t f{\"u}r Theoretische Physik\protect\\ Georg-August-Universit{\"a}t G{\"o}ttingen, Germany \protect\vspace{10pt}}
\affil[2]{Department of Mathematics\protect\\ King’s College London, WC2R 2LS, UK}

\begin{document}
	\maketitle
	
	\begin{abstract}		
		We present a method that captures the fluctuations beyond mean field in chemical reactions in the regime of small copy numbers and hence large fluctuations, using self-consistently determined \textit{memory}: by integrating information from the past we can systematically improve our approximation for the dynamics of chemical reactions. This memory emerges from a perturbative treatment of the effective action of the Doi-Peliti field theory for chemical reactions. By dressing only the response functions and by the self-consistent replacement of bare responses by the dressed ones, we show how a very small class of diagrams contributes to this expansion, with clear physical interpretations. From these diagrams, a large sub-class can be further resummed to infinite order, resulting in a method that is stable even for large values of the expansion parameter or equivalently large reaction rates. We demonstrate this method and its accuracy on single and multi-species binary reactions across a range of reaction constant values.		
	\end{abstract}

\section{Introduction}
	
An important problem with wide ranging applications, especially in biology and synthetic chemistry, is the one of treating strong stochasticity in chemical reaction networks. This is most pronounced when the copy number of participating molecules are small, typically of the order of a few molecules~\cite{cardelli_stochastic_2016,gillespie_perspective_2013,kurtz_strong_1978,lecca_stochastic_2013}. Indeed, thinking of copy number fluctuations as Poissonian the mean $\bar{n}$ and the standard deviation $\sqrt{\bar{n}}$ are of the same order when $\bar{n}=O(1)$, so that fluctuations can never be neglected. Equivalently, the time evolution of lower order moments of the copy number distributions are hierarchically coupled to higher order moments, and this hierarchy cannot be truncated by neglecting relative copy number fluctuations, or by treating them as small enough to be Gaussian.

Recent experiments in living cells, in \textit{gene} and \textit{protein} regulation networks~\cite{ozbudak_regulation_2002,bar-even_noise_2006, elowitz_stochastic_2002,golding_real-time_2005,blake_noise_2003} have shown the importance of intrinsic stochasticity originating from the operation of these biochemical networks in the limit of small copy number of participating molecules. Indeed, this limit is the natural regime of operation for many of these networks. For example there are only a few copies of each gene coding for a protein in each cell, which transcribe a few copies of mRNA that are later translated into proteins. Sometimes even a few copies of a signalling molecule or transcription factor are enough to trigger signalling pathways in a cell~\cite{ladbury_noise_2012}. This regime of small copy numbers is dominated by large fluctuations in the time courses of the molecule numbers, where mean field or mass action kinetics, which is the standard description of chemical reaction dynamics, becomes inaccurate~\cite{mcquarrie_stochastic_1967, renyi_treating_1953}, and very few methods~\cite{schnoerr_approximation_2017} exist that are able to accurately calculate the time-dependent moments of the stochastic process. Experiments have even shown that that random fluctuations in gene expression can lead to very different behaviours of otherwise identical cells~\cite{elowitz_stochastic_2002, ozbudak_regulation_2002}. Understanding the dynamics in the regime of strong fluctuations is thus essential if we want to be able to infer the relevant biology from experimental measurements of such systems. 
	
The challenges arising from strong fluctuations become even more acute for spatially resolved dynamics, i.e.\ reaction-diffusion systems. A common approach~\cite{schnoerr_cox_2016} to theoretically treat and make inferences from such systems is to divide space into small compartments, and model diffusion as molecules being destroyed in one compartment and created in a neighbouring one. In dilute mixtures, or generally for small enough compartments the number of molecules in each compartment is  then always small enough and fluctuations again large. In the context of population dynamics the intrinsic stochasticity from such small populations can lead to features such as noise-induced Turing patterns that are absent in the deterministic limit~\cite{karig_stochastic_2018}. Large fluctuations in the dynamics can also lead to extinction or fixation of stochastic populations~\cite{assaf_extinction_2010,assaf_wkb_2017}. Similar rare events also play a role in epidemic models, affecting e.g.\ the distribution of outbreak sizes~\cite{hindes_outbreak_2022}.

To track large fluctuations, the Chemical Master Equation (CME) is the widely accepted theoretical description for stochastic chemical reactions~\cite{gillespie_rigorous_1992}. It gives the time evolution of the probability of the system to be in a certain state specified by the copy numbers of all species, starting from some initial distribution across states. However, analytical solutions to the CME are available only for very specific cases. Instead one usually has to rely on stochastic simulations such as the rejection-free Gillespie algorithm~\cite{gillespie_exact_1977}, which can exactly simulate and sample the underlying distribution. In the regime of large fluctuations, a very large number of such simulations have to be carried out to get accurate and reliable statistics, which becomes computationally extremely expensive, especially in the case of multiple reacting species where high-dimensional distributions need to be sampled. In addition, stochastic simulations do not allow one to extract a likelihood function for the probability of a given time course of copy numbers, given a set of reaction constants. Hence they cannot be used to infer such dynamical parameters, which is often an important step towards understanding e.g.\ the biological function of a reaction network.
	
Even though analytical solutions of the CME are rare, a few landmark results exist. In particular Renyi~\cite{renyi_treating_1953} gave the time dependent solution of the $A+B \rightarrow C$ binary reaction starting from deterministic initial conditions, and showed that the mass action kinetics description is only approximately valid and breaks down for small copy numbers. McQuarrie et al.~\cite{mcquarrie_kinetics_1964} fully solved some other simple binary reactions starting from deterministic initial conditions using the technique of generating functions; we refer the reader to the review by McQuarrie~\cite{mcquarrie_stochastic_1967} for a detailed historical overview of the development of chemical reaction stochastics. The stochastic solution of the Michaelis-Menten enzyme dynamics with a single enzyme was given by Arányi and Tóth~\cite{aranyi_full_1977}. As regards the simpler question of steady state distributions, a straightforwardly solvable case is that of a single species where only one molecule can be created or destroyed in a reaction~\cite{gardiner_handbook_1985}. Steady state solutions are also available for a number of multi-species systems with binary reactions such as gene regulation or multi-enzyme Michaelis-Menten reactions; we refer the reader to~\cite{schnoerr_approximation_2017} for a full list.

Jahnke and Huisinga~\cite{jahnke_solving_2007} solved the CME for a reaction network with an arbitrary number of species with time dependent rates but undergoing only birth, death or conversion reactions with only one reactant and one product molecule. These results were based on guessing the form of the solution, using carefully crafted transformations or exploiting special properties of unary reaction networks, which preserve the Poisson character of copy number distributions. These results are thus not generalizable to include other reactions.
	
Given the scarcity of exact solutions, approximate approaches to the CME have been widely explored~\cite{schnoerr_approximation_2017}. One popular approximation scheme consists of approximating the CME, a continuous-time Markov jump process on a discrete state space of copy numbers, by a diffusion process for concentrations that can take any non-negative real value. Such a description is obtained by a second order expansion of the CME, originally developed by Kramers~\cite{kramers_brownian_1940} and Moyal~\cite{moyal_stochastic_1949}, and yields the so-called Chemical Fokker-Planck equation, with an associated Chemical Langevin equation. The simulation of these equations can be easier than the CME but many challenges remain in the low copy number regime, including the lack of a natural boundary condition at zero copy number and the consequent appearance of imaginary noise terms in the Chemical Langevin Equation. Schnoerr et al~\cite{schnoerr_complex_2014} showed that some of these problems can be circumvented by formally extending the state space of the Chemical Langevin Equation to complex-valued concentrations.

Another popular approximation method called the system size expansion is due to van Kampen~\cite{van_kampen_expansion_1976} and based on a perturbative expansion of the CME in inverse system volume. This leads to mean concentrations that are given by the solutions of the macroscopic rate equations, with fluctuations scaling as the inverse square root of the system volume. To the leading order in this small parameter the fluctuations are Gaussian, and only retaining these yields the so-called Linear Noise Approximation (LNA)~\cite{elf_fast_2003,paulsson_models_2005}; in the limit of small copy numbers this can have severe deviations from the solution of the underlying CME~\cite{thomas_how_2013}. Keeping the next order in the expansion of the CME results in Effective Mesoscopic Rate Equations (EMRE)~\cite{grima_effective_2010} that, unlike the LNA, contain corrections to the mean copy numbers. Higher order corrections in inverse system size can be obtained but are quite cumbersome and computationally expensive to calculate; diagrammatic perturbation theory can be used to make the expansion easier to use~\cite{thomas_system_2014}. A related technique is the WKB approximation~\cite{assaf_wkb_2017}, which again studies large systems but instead of the typical Gaussian fluctuations concentrates on large deviations that are exponentially rare in system size.
	
Moment closure approximations, widely deployed for various stochastic systems, remain one of the most commonly used techniques to deal with stochastic reaction systems~\cite{schnoerr_approximation_2017}. The starting point for these is the hierarchically coupled system of equations for the time evolution of the moments (mean concentrations, mean square concentrations etc.) that can be derived from the CME. In these equations the time evolution of lower order moments depends on higher order moments so they have to be \textit{closed} by hand at some order, e.g.\ by assuming some form for the $n^{\text{th}}$ moments or by setting cumulants beyond some order to zero. The most common approach is the normal moment closure~\cite{gomez-uribe_mass_2007}, which assumes higher than second order cumulants to be zero, thus effectively imposing a Gaussian form of the distribution of the concentrations, for all times. These approximations, however, have well-documented problems~\cite{schnoerr_approximation_2017} such as concentrations that diverge in time or become negative, negative variances etc.
	
Vastola~\cite{vastola_solving_2021} has recently used the Doi-Peliti path integral approach to re-derive the results of Jahnke and Huisinga~\cite{jahnke_solving_2007}, including also arbitrary unary 
 reactions in his analysis. We will also use the Doi-Peliti path integral technique in this work, but will then deploy the tools of statistical field theory and show that under certain approximation, we can obtain very accurate and general results that allows us to treat generic reaction networks made up of binary 
 reactions in addition to arbitrary unary reactions. 
 
In this paper we develop novel approximation methods for chemical reaction networks in the challenging regime of small copy numbers or, equivalently, large fluctuations. The starting point is the Doi-Peliti~\cite{doi_second_1976,peliti_path_1985} path integral, which exploits the correspondence between classical statistical systems and quantum systems by using ladder operators to represent the creation and annihilation of molecules. We then apply diagrammatic perturbation theory around the Gaussian part of the path integral, which can capture all unary reactions and corresponds exactly to Poissonian copy number distributions. The perturbation expansion is formally set up using the rates of binary and higher order~\cite{laidler_glossary_1996} reactions as small parameters. Nonetheless the applicability of our approach is not limited to this parameter regime because we capture many non perturbative effects by resummation and self-consistency; we will justify this theoretically and also demonstrate it numerically.

Our method relies on identifying a series of key diagrams in the perturbation expansion that can be efficiently resummed. In addition, we self-consistently replace the ``bare'' response functions that appear by their perturbatively corrected or ``dressed'' versions. The resulting approximation, which we call self-consistent bubble resummation (SBR), allows us to capture many non-perturbative effects, giving the method a broad scope of applicability. We focus throughout on the first and the second copy number moments, namely the means and the two-time correlation and response functions of the process and show how these can be very accurately calculated at a computational cost that is small relative to that of stochastic simulations. We benchmark all of our results against numerically exact solutions of the underlying chemical master equation.

This manuscript is organized as follows: in \cref{section_path_integral} we start from the chemical master equation and construct the Doi-Peliti path integral for a generic chemical reaction network. We demonstrate also some properties of the path integral that will be relevant later in our analysis. We then move on to identify a \textit{baseline} set of reactions around which we will set up the perturbation theory. In \cref{effective_action_section} we introduce the effective action and vertex functions corresponding to the path integral description, and derive a general equation for calculating the time evolution of the mean copy numbers. In \cref{section_AAA} we consider as a paradigmatic example a binary single species reaction $A+A\rightarrow A$. We demonstrate the derivation of our approximation method for this case, and show how it performs in numerical tests against the exact CME benchmark. We follow this by explaining how our approach extends to the case where other single species reactions are included. In \cref{section_ABC} we extend the method to the multi-species binary reaction $A+B \rightarrow C$ and again demonstrate its numerical performance. We conclude with a summary and outlook in \cref{section_discussions}.

\section{Coherent state path integral for reaction networks}{\label{section_path_integral}}
We consider a system of $N$ molecular species $X_i$ indexed by $i = 1,2,\dots,N$ with $n_i$ denoting the number of molecules of species $i$ in the system. The state of the system is given by specifying the number of molecules of all species, $ \bm{n} = (n_1,n_2,\dots n_N)$. We allow a general system of reactions where reaction $\beta$ converts $r_1^\beta$ copies of species $X_1$ together with $r_2^\beta $ copies of $X_2$ etc.\ into $s_1^\beta$ copies of  $X_1$, $s_2^\beta$ copies of $X_2$ etc., or in shorthand
\begin{equation}{\label{general_reaction_form}}
	\sum_{i=1}^{N} r_i^\beta X_i \xrightarrow{k_\beta} \sum_{i=1}^{N} s_i^\beta X_i
\end{equation} 
Here $k_\beta$ is the rate for the reaction, with units of inverse time. The order of the $\beta^{\text{th}}$ reaction~\cite{laidler_glossary_1996} is defined by to be $\sum_i r_i^\beta$. We collect the $r_i^\beta$ and $s_i^\beta$ into 	
vectors $\bm{r}^\beta$ and $\bm{s}^\beta$ for the reactant and product stochiometry, respectively. The probability that this reaction will take place in a time interval $(t,t+dt)$ is given by the microscopic \textit{propensity function}, which depends on the state of the system $\bm{n}$ as
	\begin{equation}{\label{propensity_1}}
		f_\beta(\bm{n}) = k_\beta \prod_i \frac{n_i !}{(n_i - r_i^\beta)! }
	\end{equation}
The ratio of factorials takes care of the appropriate combinatorics, whereby e.g.\ for the reaction $2X_1\to X_2$ the reaction probability is proportional to the number $n_1(n_1-1)/2=n_1!/(n_1-2)!/2$ of pairs of $X_1$ molecules that can react. The factor 1/2 that appears here compared to \cref{propensity_1}, which in the general case would be $1/\prod_i r_i^\beta!$, has been included in the definition of $k_\beta$ to make the following expressions shorter.

To illustrate the notation above, a single chemical reaction where a molecule of A reacts with a B molecule to form a C molecule would be represented by
\begin{equation}{\label{ABC_reaction_example}}
	A + B \xrightarrow{k} C
\end{equation} 
where $r_A = 1, r_B = 1, s_C =1 \text{ and } r_C = s_A = s_B = 0$, with reaction constant $k_\beta = k$ and propensity function $f_\beta(n_A,n_B,n_C) = k n_A n_B$.
	
\subsection{Chemical Master Equation}
Given a set of reactions $\beta$ defined as above, the 
probability of the state of the system $P(\bm{n},\tau)$ evolves according to the \textit{Chemical Master Equation} (CME)~\cite{gillespie_rigorous_1992} given by
	\begin{equation}{\label{Master_equation}}
		\frac{\partial P(\bm{n},\tau)}{\partial \tau} = \sum_\beta f_\beta(\bm{n} - \bm{s}^\beta + \bm{r}^\beta) P(\bm{n} - \bm{s}^\beta + \bm{r}^\beta,\tau) - \sum_\beta f_\beta(\bm{n})P(\bm{n},\tau)  
	\end{equation}
	Following the seminal work by Doi and Peliti~\cite{doi_second_1976,peliti_path_1985}, one can -- as the first step towards a path integral formulation -- cast the master equation in a quantum mechanical ``second quantized'' form~\cite{tauber_applications_2005} by introducing annihilation and creation operators for species $i$, $\hat{a}_i$ and $\hat{a}_i^\dagger$, respectively, which obey the following commutation relations
	\begin{eqnarray}
		\left[\hat{a}_i^\dagger,\hat{a}_j^\dagger \right] = \left[ \hat{a}_i,\hat{a}_j \right] =0 \quad & \left[ \hat{a}_i,\hat{a}_j^\dagger \right] = \delta_{ij}
	\end{eqnarray}
	We introduce a ket $| \bm{n} \rangle = |n_1,n_2,\dots, n_i,\dots,n_N\rangle$ that defines the state of the system. The creation and annihilation operators act on the state ket $| \bm{n} \rangle$ as (notice that the normalization differs from the standard choice used in quantum mechanics)
	\begin{eqnarray}
		\hat{a}_i |\bm{n} \rangle &= \ n_i & \! \! \!\! |n_1,n_2,\dots, n_i-1,\dots,n_N\rangle \\
		\hat{a}_i^\dagger |\bm{n}\rangle &\!= \ \ \ {} 
		& \!\!\!\!  |n_1,n_2,\dots, n_i+1,\dots,n_N\rangle 
	\end{eqnarray}
	The state $ |\bm{n}\rangle$ can therefore be obtained by acting on the zero or ``vacuum'' state $| \bm{0} \rangle$ with the appropriate product of creation operators $\hat{a}_i^\dagger$,
	\begin{equation}
		|\bm{n}\rangle = \prod_i (\hat{a}_i^\dagger)^{n_i}|\bm{0}\rangle
\label{n_from_0}	
	\end{equation}
	The operator $\hat{n}_i$ that counts the number of molecules of species $i$ is given by
	\begin{equation}
		\hat{n}_i = \hat{a}_i^\dagger \hat{a}_i
	\end{equation}
	To obtain the quantum mechanical form of the CME, one identifies the probability distribution $P(\bm{n},\tau)$ across states with the vector $|P(\tau)\rangle = \sum_{\bm{n}} P(\bm{n},\tau)|\bm{n}\rangle$. As the CME is a linear equation for the $P(\bm{n},\tau)$, it can then equivalently be written in the form of 
an imaginary time Schr\"odinger equation,
	 \begin{equation}
	 	\partial_\tau | P(\tau) \rangle = \hat{H} | P(\tau) \rangle
	 \end{equation}
with an effective Hamilton operator $\hat{H}$. Comparing with the original CME, one can read off $\hat{H}$ (see  \cite{tauber_applications_2005}) as
	\begin{equation}{\label{Hamiltonian_operator}}
		\hat{H} = \sum_\beta k_\beta \left[ \prod_i (\hat{a}_i^\dagger)^{s_i^\beta} (\hat{a}_i)^{r_i^\beta} - \prod_i (\hat{a}_i^\dagger)^{r_i^\beta} (\hat{a}_i)^{r_i^\beta} \right]
	\end{equation}	
which e.g.\ for the system defined in \cref{ABC_reaction_example} would reduce to
	\begin{equation}
		\hat{H} = k \left[ \hat{a}^\dagger_C - \hat{a}^\dagger_A  \hat{a}^\dagger_B \right] \hat{a}_A \hat{a}_B
	\end{equation}
	The formal solution for the time evolution of the system to time $t$ is then generally given by
	\begin{equation}
		| P(t) \rangle = e^{\hat{H}t} | P(0) \rangle 		
	\end{equation}
	We will consider throughout an initial state in which the copy number of each molecular species $i$ has independent Poisson fluctuations around a mean $\bar{n}_{0i}$. Such states can be written in the simple form	
	\begin{equation}{\label{Poisson_initial_state}}
		| P(0) \rangle = \sum_{\bm{n}} \prod_i \left(
		e^{-\bar{n}_{0i}}\frac{\bar{n}_{0i}^{n_i}}{n_i!}
		\right)	|\bm{n}\rangle = e^{\sum_i \bar{n}_{0i} (\hat{a}_i^\dagger - 1)} | \bm{0} \rangle 
	\end{equation} 
where the second version follows from \cref{n_from_0}.
	 
	 \subsection{The path integral and the action}
	To construct the path integral, one can split the quantum mechanical time evolution by $e^{\hat{H}t}$ into small discrete time intervals $\Delta t$, and insert at each time step resolutions of the identity operator, expressed as integrals over an (overcomplete) set of coherent states~\cite{doi_second_1976,peliti_path_1985,tauber_applications_2005}. We generalize this construction using two sets of generating fields ${\theta}$ and ${\tilde{\theta}}$ defined for each species at each time step, namely $\theta_i(\tau), \tilde{\theta}_i(\tau)$. Inserting then factors of 	
		\begin{equation}			
e^{ \sum_i \theta_i(\tau)\Delta t\,\hat{a}_i} \, \mathbb{I} \, e^{ \sum_i \tilde\theta_i(\tau)\Delta t\,(\hat{a}^\dagger_i -1 )}
		\end{equation}
	at each discretized time $\tau$ allows one to generate averages such as means, correlation functions etc.\ by taking derivatives w.r.t\ the generating fields. Defining here the $\tilde{\theta}$ factor with $(\hat{a}^\dagger - 1)$ rather than $\hat{a}^\dagger$ will simplify a number of formulas below. The steps of the method are detailed in  \cref{app_path_integral} and lead to the following path integral for the \textit{generating function} or \textit{partition function},
	\begin{equation}{\label{path_integral_main_text}}			
			\mathcal{Z}( \tilde{\theta}  , \theta ) = \lim_{\Delta t \rightarrow 0} \mathcal{N}^{-1} \int \prod_{i,\tau} d \phi_i^*(\tau) d\phi_{i}(\tau)e^{S\left(  \phi^* ,  \phi  \right)} 
	\end{equation}
The normalization constant $\mathcal{N}$ ensures that $\mathcal{Z}(0,0)=1$.	The integrations are over the (time-discretized) paths of the complex fields $\phi_i(\tau)$ defining the coherent states. The action $S$ depends on these fields and their complex conjugates $\phi_i^*(\tau)$ and reads
	\begin{equation}{\label{discrete_action}}
	\begin{split}
		S\left(  \phi^* ,  \phi  \right) &= \Delta t \sum_{\tau=\Delta t}^{t} H(\bm{\phi}^*(\tau), \bm{\phi}(\tau_-)) + \sum_i \Big\lbrace \phi_i(t) + \bar{n}_{0i} (\phi_i^*(0)-1) - \phi_i(0)\phi_i^*(0) \\ 
		&- \Delta t \sum_{\tau=\Delta t}^{t} \phi_i^*(\tau) \Delta_\tau \phi_i(\tau)
		+ \Delta t \sum_{\tau=0}^{t} \left[ \tilde{\theta}_i(\tau) \left( \phi_i^*(\tau)-1 \right)  + \theta_i(\tau) \phi_i(\tau)  \right] \Big\rbrace
	\end{split}	
\end{equation}
where $\tau_-\equiv \tau-\Delta t$, $t$ is the total time of the dynamics considered, and we use $\Delta_\tau \phi_i(\tau) = \frac{1}{\Delta t} (\phi_i(\tau) - \phi(\tau_-))$ as a shorthand for the discrete time derivative. $ H(\bm{\phi}^*(\tau), \bm{\phi}(\tau_-))$ is obtained from the Hamiltonian $\hat{H}$ in \cref{Hamiltonian_operator} by replacing $\hat{a}_i^\dagger$ by $\phi^*_i(\tau)$ and $\hat{a}$ by $\phi_i(\tau_-)$.

One can now apply a \textit{Doi-shift}~\cite{tauber_applications_2005,falkovich_john_2008}  by replacing $\phi_i^*(\tau) = 1 + \tilde{\phi}_i(\tau)$. This turns out to make the average of $\bm{\tilde{\phi}} = 0$ and can be justified by appropriate rearrangements in the partition function before the coherent states are introduced~\cite{tauber_applications_2005}. If we continue to use the same label for the function $H$ evaluated at $\bm{\phi}^* = 1 + \bm{\tilde{\phi}}$, then in terms of the new variables the action reads
	\begin{equation}{\label{Action_discrete}}
		\begin{split}
		S(  \tilde{\phi},  \phi) &= \Delta t \sum_{\tau=\Delta t}^{t} H(\bm{\tilde{\phi}}(\tau), \bm{\phi}(\tau_-)) + \sum_i \Big\lbrace  \bar{n}_{0i} \tilde{\phi}_i(0) - \phi_i(0) \tilde{\phi}_i(0)  \\ 
			&- \Delta t \sum_{\tau=\Delta t}^{t} \tilde{\phi}_i(\tau) \Delta_\tau \phi_i(\tau)
			+ \Delta t \sum_{\tau=0}^{t} \left[ \tilde{\theta}_i(\tau) \tilde{\phi}_i(\tau)  + \theta_i(\tau) \phi_i(\tau) \right] \Big\rbrace
		\end{split}	
	\end{equation}
	We will work in discrete time and only take the continuous time limit in the final equations of motion, but one can also take this limit in the expression for the action to obtain 
	\begin{equation}{\label{action_continuous}}
		\begin{split}
				S( \tilde{\phi} ,  \phi  ) =  \int_0^{t} d\tau\, H(\bm{\tilde{\phi}}(\tau), \bm{\phi}(\tau_-))  + \sum_i \Big( & \bar{n}_{0i} \tilde{\phi}_i(0) - \phi_i(0) \tilde{\phi}_i(0) \\
				&+ \int_0^{t} d\tau \left[ -\tilde{\phi}_i(\tau) \partial_\tau \phi_i(\tau) + \tilde{\theta}_i(\tau) \tilde{\phi}_i(\tau)  + \theta_i(\tau) \phi_i(\tau)  \right] \Big)		
			\end{split}	
\end{equation}


	\subsection{Response and correlation functions}{\label{response_correlation_relations_section}}
	Once the path integral is defined over the fields $\bm{\phi}$ and $\bm{\tilde{\phi}}$, we need to show how the average values of observables such as mean copy numbers of species $i$, its variance or other two time quantities are related to the statistics of these fields. We start by taking derivatives w.r.t.\ $\tilde\theta_i(\tau_+)$ (with $\tau_+\equiv \tau+\Delta t$) and $\theta_i(\tau)$ in \cref{path_integral1} such that the operators are \textit{normal ordered}, 
	\begin{equation}
		\frac{1}{(\Delta t)^2} \left( \frac{\partial }{\partial \tilde\theta_i(\tau_+)} + 1 \right) \frac{\partial }{\partial \theta_i(\tau)} \mathcal{Z} \Bigg|_{\theta,\tilde{\theta}=0} = 
		\langle \bm{1}|\hat{a}_i^\dagger e^{\hat{H}\Delta t}\hat{a}_i e^{\hat{H}\tau} |P(0)\rangle = \langle n_i(\tau) \rangle
	\end{equation}
	 where $n_i(\tau)$ is the copy numbers of species $i$ at time $\tau$. The last equality applies in the limit $\Delta t\to 0$, where the short-time propagator $e^{\hat{H}\Delta t}$ can be ignored. In the path integral one can take the same derivatives, which gives	
	\begin{equation}
	\frac{1}{(\Delta t)^2} \left( \frac{\partial }{\partial \tilde\theta_i(\tau_+)} + 1 \right) \frac{\partial }{\partial \theta_i(\tau)} \mathcal{Z} \Bigg|_{\theta,\tilde{\theta}=0} = 	\langle [\tilde\phi_i(\tau_+)+1]\phi_i(\tau)\rangle
\end{equation}
This average simplifies to  $\langle \phi_i(\tau)\rangle$ because of a {\em causality property} of the path integral: any average of a product of $\phi$ or $\tilde\phi$ fields vanishes when the last factor -- the one associated with the latest of all times that occur -- is a $\tilde\phi$.

To see this causality property, consider a reaction where a molecule of species $i$ is created spontaneously with rate $k_{1i}$. From \cref{Hamiltonian_operator}, this corresponds to a term $k_{1i}(a_i^\dagger-1)$ in $\hat{H}$, and hence, generalizing to a time-dependent creation rate, to a contribution $\Delta t \sum_\tau k_{1i}(\tau)\tilde\phi_i(\tau)$ in the action. Comparing this with the generating term 
$\Delta t \sum_\tau \tilde\theta_i(\tau) \tilde\phi_i(\tau)$ 
from the $\tilde\theta_i(\tau)$-field shows that, whenever we take derivatives of the partition function, we have the identity
\begin{equation}
\frac{1}{\Delta t}\frac{\partial}{\partial k_{1i}(\tau)} = \frac{1}{\Delta t}\frac{\partial}{\partial \tilde\theta_{i}(\tau)}
\end{equation}
Now any product $f_{<\tau}$ of fields evaluated at times before $\tau$ can be generated by a sequence of appropriate derivatives of $\mathcal{Z}$. Taking then an additional derivative as in our last identity and setting the generating fields to zero afterwards shows that
\begin{equation}
\frac{1}{\Delta t}\frac{\partial}{\partial k_{1i}(\tau)} \langle f_{<\tau}\rangle = \langle \tilde\phi_i(\tau)f_{<\tau}\rangle
\label{response_field_relation}
\end{equation}
However, the l.h.s.\ vanishes by causality -- the average of $f_{<\tau}$ cannot depend on a creation rate at a later time -- and therefore so does the r.h.s., as claimed.

Summarizing so far, then, and introducing the symbol $\mu$ for the means of our fields, we have
	\begin{equation}
		\begin{split}
				 \mu_i(\tau) &= \langle \phi_i(\tau)\rangle = \langle n_i(\tau) \rangle \\
			\tilde{\mu}_i(\tau) &= \langle \tilde{\phi}_i(\tau)\rangle = 0
		\end{split}
	\end{equation}
The second line follows again from the causality property. In fact, applying the property inductively, one sees that the average of any product of $\tilde\phi$-factors vanishes, e.g.\ $\langle \tilde{\phi}(\tau) \tilde{\phi}(\tau') \rangle = 0 \text{ } \forall \text{ } \tau,\tau'$.

Similarly calculating the fourth order derivatives in terms of the number operator and from the path integral we have for $\tau'<\tau_-$, 
	 \begin{equation}{\label{fourth_order_derivatives}}
	 	\begin{split}
	\frac{1}{(\Delta t)^4}	\left(\frac{\partial}{\partial \tilde{\theta}_i(\tau_+)} + 1 \right) \frac{\partial}{\partial \theta_i(\tau)} \left( \frac{\partial}{\partial \tilde{\theta_i}(\tau'_+)} + 1 \right) \frac{\partial}{\partial \theta_i(\tau')} \mathcal{Z} \Bigg|_{\theta,\tilde{\theta}=0} 
	&= \langle n_i(\tau) n_i(\tau') \rangle \\ &= \langle [\tilde{\phi}_i(\tau_+)+1]\phi_i(\tau) [\tilde{\phi}(\tau'_+)+1] \phi_i(\tau') \rangle 
\end{split}
\end{equation}
while the analogue for $\tau'=\tau$ is
\begin{equation}{\label{fourth_order_derivatives_equal_time}}
	\frac{1}{(\Delta t)^4}	\left(\frac{\partial}{\partial \tilde{\theta}_i(\tau_+)} + 1 \right)\frac{\partial}{\partial \theta_i(\tau)} \left( \frac{\partial}{\partial \tilde{\theta_i}(\tau_+)} +1 \right) \frac{\partial}{\partial \theta_i(\tau)} \mathcal{Z} \Bigg|_{\theta,\tilde{\theta}=0} 
	= \langle n_i^2(\tau) \rangle - \langle n_i(\tau) \rangle = \langle [\tilde{\phi}_i(\tau_+)+1]^2 \phi^2_i(\tau) \rangle 
\end{equation}
These can now be linked to the connected response and correlation functions for each species $i$, 
	\begin{equation}{\label{mean_resp_corr}}
		\begin{split}
			R_i(\tau,\tau') &= \langle \phi_i(\tau) \tilde{\phi}_i(\tau')  \rangle - \langle \phi_i(\tau) \rangle \langle \tilde{\phi}_i(\tau') \rangle = \langle \delta \phi_i(\tau) \delta \tilde{\phi}_i(\tau') \rangle\\
			C_i(\tau,\tau') &= \langle \phi_i(\tau) \phi_i(\tau') \rangle - \langle \phi_i(\tau) \rangle \langle \phi_i(\tau') \rangle = \langle \delta \phi_i(\tau) \delta \phi_i(\tau') \rangle
		\end{split}
	\end{equation}
The corresponding disconnected functions do not have the subtraction of the product of the averages; note though that since $\langle \tilde{\phi}\rangle = 0$, the connected and the disconnected response functions are identical.  On the r.h.s.\ of \cref{fourth_order_derivatives_equal_time} we have, by causality again, the disconnected correlator $\langle \phi_i^2(\tau)\rangle = C_i(\tau,\tau)+\mu_i(\tau)^2$. Subtracting the squared mean shows 
	\begin{equation}
		\begin{split}{\label{num_correlator_relation_1}}
		\text{Var}(n_i(\tau)) - \langle n_i(\tau)\rangle &= C_i(\tau,\tau)
	\end{split}
	\end{equation}
The l.h.s.\ vanishes for Poissonian copy number fluctuations, so a non-zero equal-time correlator of the $\phi$-field indicates deviations from Poisson statistics.	

For the $\tau'<\tau_-$ case in \cref{fourth_order_derivatives} we have from causality
\begin{equation}
\langle n_i(\tau) n_i(\tau') \rangle  = \langle \phi_i(\tau) \phi_i(\tau') \rangle + \langle \phi_i(\tau) \tilde{\phi}(\tau'_+) \phi_i(\tau') \rangle
\end{equation}
or
\begin{equation}
		\begin{split}
		\langle \delta n_i(\tau) \delta n_i(\tau') \rangle =\langle n_i (\tau)n_i(\tau')\rangle - \langle n_i (\tau) \rangle \langle n_i(\tau')\rangle &= 
C_i(\tau,\tau') + 
 \langle \phi_i(\tau) \tilde{\phi}(\tau'_+) \phi_i(\tau') \rangle		
		\end{split}
	\end{equation}
where the copy number fluctuations are defined as $\delta n(\tau) = n(\tau) - \mu(\tau)$. 
The final three-point correlator cannot be simplified in general, but for Gaussian field statistics one can use Wick's theorem and causality to express it as
$\langle \phi_i(\tau) \tilde{\phi}(\tau'_+) \phi_i(\tau') \rangle = \langle \phi_i(\tau) \tilde{\phi}(\tau'_+)\rangle \langle \phi_i(\tau') \rangle $ so that 
		\begin{equation}
		\begin{split}{\label{num_correlator_relation_2}}
		\langle \delta n_i(\tau) \delta n_i(\tau') \rangle 
		 &= R_i(\tau,\tau')\mu_i(\tau') + C_i(\tau,\tau') \quad \text{ for } \tau'<\tau\\
		\end{split}
	\end{equation}

	\subsection{Baseline action and Gaussian path integral}{\label{baseline_action_section}}
	Equipped with the path integral for describing chemical reactions, we now look at a concrete set of chemical reactions that we call the \textit{baseline} because the Hamiltonian of \cref{Hamiltonian_operator} associated with these reactions will be quadratic in creation and annihilation operators. We will treat other reactions as perturbations of this quadratic baseline Hamiltonian. If we consider the set of only creation and (unary) destruction reactions for each of the species with the following rate constants
	 \begin{eqnarray}
	 	\emptyset \xrightarrow{k_{1i}} X_i & & X_i \xrightarrow{k_{2i}} \emptyset
	 \end{eqnarray}
	then from \cref{Hamiltonian_operator} with the operators replaced by the fields, the Doi-shifted Hamiltonian in the action is 
	\begin{equation}
		H_0(\bm{\tilde{\phi}}, \bm{\phi} ) = \sum_i 
\left( k_{1i}\tilde{\phi}_i - k_{2i} \tilde{\phi}_i \phi_i \right)
	\end{equation}
It is clearly decoupled across the different species and only has linear and quadratic terms in $\phi,\tilde{\phi}$. We will later use it a baseline for perturbation theory (with the linear terms omitted), hence the notation $H_0$. The corresponding baseline action $S_0$ from \cref{Action_discrete} becomes
	\begin{eqnarray}
		S_0(  \tilde{\phi},  \phi)  &=
		&\sum_i \Big\lbrace \bar{n}_{0i} \tilde{\phi}_i(0) - \phi_i(0) \tilde{\phi}_i(0) +   \Delta t \sum_{\tau=\Delta t}^{t}  \left[-\tilde{\phi}_i(\tau) \Delta_\tau \phi_i(\tau) + k_{1i}\tilde{\phi}_i(\tau) - k_{2i} \tilde{\phi}_i(\tau) \phi_i(\tau_-) \right] 
\nonumber\\ 		
&&{}+ \Delta t \sum_{\tau=0}^{t} \left[ \tilde{\theta}_i(\tau) \tilde{\phi}_i(\tau)  + \theta_i(\tau) \phi_i(\tau)  \right] \Big\rbrace
\label{baseline_action}
	\end{eqnarray}
Because the path integral is Gaussian, we can calculate the means of the fields by extremizing the action
	 \begin{align}
	 		\frac{\partial S_0}{\partial \tilde{\phi}_i(\tau)}\Bigg|_{\substack{\phi_i(\tau) = \mu_i(\tau) \\ \tilde{\phi}_i(\tau) = \tilde{\mu}_i(\tau) }} = 0 \quad & \text{and} \quad  \frac{\partial S_0}{\partial \phi_i(\tau) } \Bigg|_{\substack{\phi_i(\tau) = \mu_i(\tau) \\ \tilde{\phi}_i(\tau) = \tilde{\mu}_i(\tau) }} = 0
	 \end{align}
The derivatives with respect to $\tilde\phi_i(\tau)$ for $0 < \tau \leq t$  give the equations of motion for the means as 
	 \begin{equation}
	 	\mu_i(\tau) - \mu_i(\tau-\Delta t) = \Delta t \left( k_{1i} - k_{2i} \mu_i(\tau-\Delta t) + \tilde{\theta}_i(\tau) \right) 
	 \end{equation}
while for $\tau=0$ one obtains the initial condition
\begin{equation}
		\mu_i(0) = \bar{n}_{0i} + \tilde{\theta}_i(0)\Delta t
\end{equation}
Conversely, the $\phi_i(\tau)$ derivatives yield for $0\leq \tau<t$ 	 
 	\begin{equation}
 		\tilde{\mu}_i(\tau) - \tilde{\mu}_i(\tau+\Delta t) = \Delta t \left( -k_{2i} \tilde{\mu}_i(\tau+\Delta t) + \theta_i(\tau)  \right)
 	\end{equation}	
	and for $\tau = t$ one obtains a final condition
	\begin{equation}
		\tilde{\mu}_i(t) = \theta_i(t) \Delta t
	\end{equation}
The equations for the physical means thus have to be solved forwards in time as expected, while the ones for the conjugate means are solved backwards starting from the final condition.

The physical problem corresponds to zero value of generating fields, from which we can recover the initial condition that species $i$ has initial mean $\bar{n}_{0i}$ and the fact that $\tilde{\mu}_i(\tau) = 0 \text{ } \forall \text{ } \tau $ as expected. In the continuous time limit one also obtains for the physical means the expected equation of motion
	\begin{equation}{\label{mu_eom_baseline}}
		\partial_\tau \mu_i(\tau) = k_{1i} - k_{2i} \mu_i(\tau)
	\end{equation}
with two terms on the r.h.s.\ reflecting creation and destruction of particles of species $i$. This equation is also valid if we generalize to time-dependent creation rates $k_{1i}(\tau)$
	\begin{equation}
		\partial_\tau \mu_i(\tau) = k_{1i}(\tau) - k_{2i} \mu_i(\tau)
	\end{equation}
and gives the solution
\begin{equation}
\mu_i(\tau) = \bar{n}_{0i} e^{-k_{2i}\tau} + \int_0^\tau d\tau' e^{-k_{2i}(\tau-\tau')} k_{1i}(\tau') 
\label{mu0_solution}
\end{equation}

We can calculate the response function using the relation demonstrated in \cref{response_field_relation}, which incidentally has a direct analogue for the Martin-Siggia-Rose-Jansen-de Dominicis (MSRJD) path integral~\cite{martin_statistical_1973,janssen_lagrangean_1976,dominicis_techniques_1976,hertz_path_2017}:
\begin{equation}
R_{0i}(\tau,\tau') = \langle \phi(\tau)\tilde\phi(\tau')\rangle_0 = 
\frac{1}{\Delta t} \frac{\partial \langle \phi_i(\tau) \rangle_0 }{\partial k_{1i}{(\tau')}}
=  \frac{1}{\Delta t} \frac{\partial \mu_i(\tau) }{\partial k_{1i}(\tau')}
\end{equation}
where all averages are evaluated at zero generating fields, $\tilde\theta=\theta=0$, and the zero subscripts indicate evaluation within the baseline path integral. In the continuous time limit the derivative becomes a functional derivative w.r.t.\ $k_{1i}(\tau')$ and one finds explicitly from \cref{mu0_solution}
\begin{equation}{\label{R_0}}
	R_{0i}(\tau,\tau') = e^{-k_{2i}(\tau-\tau')} \Theta(\tau-\tau')
\end{equation}
with $\Theta(\cdot)$ the Heaviside step function. This response function is causal as expected,
	\begin{equation}{\label{response_causality}}
		R_{0i} (\tau,\tau') = 0 \text{ if } \tau < \tau' 
	\end{equation} 
i.e.\ the real field $\phi$ needs to be ahead in time of the conjugate field $\tilde\phi$, and its equal-time limit is unity, $\lim_{\tau' \rightarrow \tau^-} R_0(\tau,\tau') = 1$. We note for later also the operator inverse of the response function
	\begin{equation}{\label{R_inv_defn}}
		R_{0i}^{-1}(\tau,\tau') = (\partial_{\tau} + k_{2i}) \delta(\tau-\tau')
	\end{equation}
For a direct calculation of the response function in discrete time by inversion of the precision matrix of the Gaussian path integral, see \cref{matrix_action}. 

We consider finally the correlation functions calculated within our baseline path integral. These are zero because the baseline action only has quadratic terms of the form $\tilde{\phi} \phi$ (see \cref{matrix_action}), i.e.\ the \textit{bare correlation functions} are 
\begin{equation}{\label{bare_correlation_eqn}}
C_{0i}(\tau,\tau') = \langle \phi_i(\tau) \phi_i(\tau') \rangle_0 - \langle \phi_i(\tau)\rangle_0 \langle \phi_i(\tau') \rangle_0 = 0
\end{equation}
In our subsequent analysis we will consider only the bare correlation function, that is we will approximate $C_i $ as generally defined in \cref{mean_resp_corr} to be always equal to $C_{0i}$. From \cref{num_correlator_relation_1} this implies that $\text{Var}(n_i(\tau)) = \langle n_i(\tau) \rangle$, so we are then effectively approximating the variance of every copy number by its value for a Poisson distribution. 
The full distribution can in fact also be shown to be Poissonian for our baseline path integral (see \cref{poiss_dist_proof} for a proof). In the general case $C_i(\tau,\tau)$ quantifies the deviation from this behaviour and our approximation is valid if $C_i(\tau,\tau) \ll \langle n_i(\tau) \rangle$; 
see \cref{numerics_AAA} and \cref{fig10} for a numerical test of this approximation.
 		
\section{Interactions and effective action}{\label{effective_action_section}}

We now consider higher order reactions. The associated ``interacting'' Hamiltonian $H_{\text{int}}$ will no longer be quadratic and we will treat it as a perturbation to our baseline $H_0$, introducing a parameter $\alpha$ to set up a perturbation theory:
 	\begin{equation}
 		H_\alpha = H_0 +\alpha H_{\text{int}} 
 	\end{equation}
We will call the associated action $S_\alpha$. 
 To simplify the perturbation theory it is useful to have a baseline with zero means. We therefore define $H_0$ and $S_0$ from here on to contain only the {\em quadratic} terms from the baseline Hamiltonian and action, respectively. Any linear terms such as $\int d\tau\, k_{1i} \tilde{\phi}_i(\tau)$ 
 from particle creation will be included in $H_{\text{int}}$ and treated perturbatively.

We now need a formalism to calculate means and response functions within the interacting, non-Gaussian path integral. A useful starting point is the generating function of the connected $n$-point functions $\mathcal{W}$, also called the free energy and given by
		\begin{equation}
			\mathcal{W}( \tilde{\theta}  , \theta )  = \ln \mathcal{Z}( \tilde{\theta}  , \theta ) 
		\end{equation}
This is a function of the set of generating fields $\tilde{\theta}$ and $\theta$ for all species $i$  and all times $\tau$. The one- and two-point functions are just the by now familiar field means and response and correlation functions
   \begin{equation}
		\langle \phi_i(\tau)\rangle = \frac{\delta \mathcal{W}( \tilde{\theta} ,  \theta )
		}{\delta \theta_i(\tau)}, \quad  \langle \tilde{\phi}_i(\tau)\rangle = \frac{\delta \mathcal{W}( \tilde{\theta} ,  \theta )
		}{\delta \tilde{\theta}_i(\tau)}
		\label{1point}
\end{equation}	
\begin{equation}
		R_{ij}(\tau,\tau')  = \langle\delta\phi_i(\tau)\delta\tilde{\phi}_j(\tau')\rangle = \frac{\delta^2 \mathcal{W}( \tilde{\theta} ,  \theta )}{\delta \theta_i(\tau) \delta \tilde{\theta}_j(\tau')},\quad C_{ij}(\tau,\tau')  =
		\langle\delta\phi_i(\tau)\delta\phi_j(\tau')\rangle = \frac{\delta^2 \mathcal{W}( \tilde{\theta} ,  \theta )}{\delta \theta_i(\tau) \delta \theta_j(\tau')}
		\label{2point}
	\end{equation}
and generally depend on $\tilde{\theta}$, $\theta$. We use continuous time notation in this section for brevity.

From the free energy one can define the \textit{effective action} $\Gamma$ as the Legendre transform of the free energy with respect to generic field means (also known as  ``background fields'' in field theory) that we collectively denote $\tilde{\mu}$ and $\mu$, to represent the set of means for all species and all times,  
\begin{equation}{\label{LT}}
		\Gamma ( \tilde{\mu} ,  \mu )  = \substack{ \text{\normalsize extr} \\  \tilde{\theta}  ,  \theta } \left\lbrace \ln \int d\tilde{\phi}\, d\phi \,\exp \left[ S_\alpha(  \tilde{\phi} , \phi ) - \sum_i \left( \int d\tau\, \tilde{\mu}_i(\tau) \tilde{\theta}_i(\tau) +  \int d\tau\, \mu_i(\tau) \theta_i(\tau)  \right) \right] \right\rbrace 
\end{equation}
The extremization conditions yield 
	\begin{equation}
		\mu_i(\tau) = \frac{\delta \mathcal{W}( \tilde{\theta} ,  \theta )}{\delta \theta_i(\tau)} \hspace{20pt} \tilde{\mu}_i(\tau) = \frac{\delta \mathcal{W}( \tilde{\theta} ,  \theta )}{\delta \tilde{\theta}_i(\tau)} 
	\end{equation}	
which by comparison with \cref{1point} shows that $\mu$ and $\tilde{\mu}$ are indeed the means of the field variables. The generating fields $ \tilde{\theta},  \theta $ can be viewed as Lagrange multipliers whose value is determined by extremizing the free energy with the constraint that the fields $\tilde{\phi}_i (\tau)$ and $\phi_i(\tau)$  have mean values $\tilde{\mu}_i(\tau)$ and $\mu_i(\tau)$ $\forall \text{ } i,\tau$. From the Legendre transform property, the values of the generating fields are given by
		\begin{equation}{\label{fields}}
		\theta_i(\tau) = -\frac{\delta \Gamma ( \tilde{\mu} ,  \mu )}{\delta \mu_i(\tau) } \hspace{20pt}  \tilde{\theta}_i(\tau) = -\frac{\delta \Gamma ( \tilde{\mu} ,  \mu )}{\delta \tilde{\mu}_i(\tau) } 
	\end{equation} 
From \cref{fields} one then obtains the \textit{variational equations} for the means,
	\begin{equation}{\label{variational_eqns}}
		\frac{\delta \Gamma ( \tilde{\mu} ,  \mu )}{\delta \mu_i(\tau) } = 0\hspace{20pt}  \frac{\delta \Gamma ( \tilde{\mu} ,  \mu )}{\delta \tilde{\mu}_i(\tau) } = 0 
	\end{equation} 
by asserting that the physical problem has no generating fields, i.e.\ $\tilde{\theta}_i(\tau) = \theta_i(\tau) =0$ $\forall \text{ } i,\tau$.
These equations tell us that the physical means $\tilde{\mu}$ and $\mu$ are those that make the value of $\Gamma(\tilde{\mu},\mu)$ an extremum, motivating the name effective action for this quantity. The variational equations define the equations of motion of $\mu_i(\tau)$ and $\tilde{\mu}_i(\tau)$. For the conjugate means we know that the solution will be $\tilde{\mu}_i(\tau) = 0 \text{ } \forall \text{ } i,\tau$ from the arguments in \cref{response_correlation_relations_section}.

By considering more general derivatives of the effective action, one obtains the so-called vertex functions. We write the definition for the case of a single species but this can be easily extended to the multiple species case with extra indices in $\Gamma^{l,m}$: 
	\begin{equation}
	\Gamma^{l,m}(\tau_1,\dots,\tau_l,\tau'_1,\dots,\tau'_m) = \frac{\delta^{(l+m)} \Gamma}{\delta \tilde{\mu}(\tau_1) \dots \delta \tilde{\mu}(\tau_l) \delta {\mu}(\tau'_{1}) \dots \delta {\mu}(\tau'_{m})} \Bigg|_{\mu(\tau)=\tilde{\mu}(\tau) = 0 \text{ } \forall \text{ } \tau}
	\end{equation}
Unlike in typical quantum field theory settings, the physical means of the $\phi$-fields will be nonzero in our case. Writing these as $\mu^*$, a second set of vertex functions can be defined via derivatives evaluated not at zero but at the physical means:
	\begin{equation}
		\Gamma^{l,m*}(\tau_1,\dots,\tau_l,\tau'_{1},\dots,\tau'_{m}) = \frac{\delta^{(l+m)} \Gamma}{\delta \tilde{\mu}(\tau_1) \dots \delta \tilde{\mu}(\tau_l) \delta {\mu}(\tau'_{1}) \dots \delta {\mu}(\tau'_{m})} \Bigg|_{\mu(\tau) = \mu^*(\tau), \tilde{\mu}(\tau) = 0 \text{ } \forall \text{ } \tau}
	\end{equation}
	
We will now describe how to calculate the physical means $\mu^*(\tau)$ using the vertex functions. The effective action $\Gamma$ can be reconstructed from the set of vertex functions by a Taylor expansion around $\mu=\tilde{\mu} = 0$:
	\begin{equation}{\label{taylor_exp_gamma}}
		\Gamma(\tilde{\mu},\mu) = \sum_{j,k} \frac{1}{k! j!} \Gamma^{k,j}: \tilde{\mu}^k \mu^j
	\end{equation}
where $\Gamma^{k,j}$ is the vertex function at zero means as before. The ``:'' notation indicates contraction across all the temporal ``indices'' of the vertex function and is a shorthand for
	\begin{equation}
		\Gamma^{k,j}:\tilde{\mu}^k\mu^j \equiv 
		\int d\tau_1\cdots d\tau_k d\tau'_1 \cdots d\tau'_j  \, 
		\Gamma^{k,j}(\tau_1,\dots,\tau_k,\tau'_1,\dots,\tau'_j) 
		\tilde{\mu}(\tau_1) \cdots  \tilde{\mu}(\tau_k) \mu(\tau'_1) \cdots \mu(\tau'_j) 
		\end{equation}
	Differentiating \cref{taylor_exp_gamma} w.r.t $\tilde{\mu}(\tau)$ and setting $\tilde{\mu} = \tilde{\mu}^*$ and $\mu = \mu^*$ we have for the first order vertex function at the physical means
	\begin{equation}
		\Gamma^{1,0*}(\tau) = \sum_{j,k} \frac{1}{(k-1)! j!} \Gamma^{k,j}(\tau,\dots): (\tilde{\mu}^*)^{k-1} (\mu^*)^j
	\end{equation}
Since the physical means of the conjugate fields vanish, $\tilde{\mu}^* = 0$, we only need to keep the $k=1$ term and obtain
	\begin{equation}
		\Gamma^{1,0*}(\tau) = \sum_{j} \frac{1}{j!} \Gamma^{1,j}(\tau,\dots): (\mu^*)^j
	\end{equation}
Now from the variational equation of motion~(\ref{variational_eqns}), $\Gamma^{1,0*}(\tau) = 0 \text{ } \forall \text{ } \tau$. Writing the $j=0$ and $j=1$ terms on the right explicitly then gives as the equation for the physical means $\mu^*$,
	\begin{equation}
		0 = \Gamma^{1,0}(\tau) +\Gamma^{1,1}(\tau,\cdot): \mu^*  + \sum_{j \geq 2} \frac{1}{j!} \Gamma^{1,j}(\tau,\dots) : (\mu^*)^j
	\end{equation}
The $\Gamma^{1,1}$-term can now be simplified using the generic Legendre transform relation that the second derivatives of $\mathcal{W}$ and $\Gamma$ -- which give the connected two-point correlation functions and the two-point vertex functions, respectively -- are, up to a minus sign, inverses of each other~\cite{kleinert_path_2009,bellac_quantum_1991}. Writing this result in $2\times2$-block form for the $\tilde{\theta}$ and $\theta$ components, respectively, we have for the physical ($\tilde{\theta}=\theta=0$) solution with response and correlation functions $R^*$ and $C^*$:
\begin{equation}
\begin{pmatrix}
0 & (R^*)^{\rm T} \\
R^* & C^*
\end{pmatrix}
=-\begin{pmatrix}
\Gamma^{2,0*} & \Gamma^{1,1*} \\
(\Gamma^{1,1*})^{\rm T} & \Gamma^{0,2*}
\end{pmatrix}^{-1}
\end{equation}
This implies $\Gamma^{1,1*}=-(R^*)^{-1}$ and $\Gamma^{0,2*}=0$. It turns out that the same relations also hold at zero means, i.e.\ $\Gamma^{1,1}=-R^{-1}$ and $\Gamma^{0,2}=0$. We thus have 
	\begin{equation}
		R^{-1}(\tau,\cdot): \mu^* = \Gamma^{1,0}(\tau) +  \sum_{j \geq 2} \frac{1}{j!} \Gamma^{1,j}(\tau,\dots) : (\mu^*)^j
\label{Rinv_mustar}
	\end{equation}
We can simplify further on the l.h.s.\ by using the Feynman-Dyson equation discussed below (see \cref{feynmann_dyson}), i.e.\ $R^{-1} = R_0^{-1} - \Sigma$ with the response self-energy $\Sigma$, like $R$, evaluated at $\mu=0$ and $R_0$ the bare propagator, 
\begin{equation}
	R^{-1}_0(\tau,\cdot):	\mu^* = \Gamma^{1,0}(\tau) + \Sigma(\tau,\cdot): \mu^* +  \sum_{j \geq 2} \frac{1}{j!} \Gamma^{1,j}(\tau,\ldots) : (\mu^*)^j
\end{equation}
We also write an equivalent form obtained by multiplying by the bare response function, 
which will have a direct diagrammatic analogue:
	\begin{equation}{\label{diagrammatic_analogue_eqn}}
		\mu^*(\tau) = R_0(\tau,\cdot):\Gamma^{1,0}(\cdot) + R_0(\tau,\cdot):\Sigma(\cdot,\cdot): \mu^* +  \sum_{j \geq 2} \frac{1}{j!} R_0(\tau,\cdot):\Gamma^{1,j}(\cdot,\ldots) : (\mu^*)^j
	\end{equation}
	If we define a new function of a single time argument, $\Omega^{1,j}(\tau)$, as the contraction of $\Gamma^{1,j}$ with $j$ factors of $\mu^*$, 
	\begin{equation}{\label{omega_defn}}
		\Omega^{1,j}(\tau) = \frac{1}{j!}\Gamma^{1,j}(\tau,\dots):(\mu^*)^j
	\end{equation}
then the equation of motion for the means can be written more explicitly as 
	\begin{equation}
		\int_0^\tau d\tau' \, R_0^{-1}(\tau,\tau') \mu^*(\tau') = \Gamma^{1,0}(\tau)  + \int_0^\tau d\tau'\, \Sigma(\tau,\tau') \mu^*(\tau') + \sum_{j\geq 2} \Omega^{1,j}(\tau) 
		\label{mu_eq_with_r0_inverse}
	\end{equation}
	Using the form of $R_0^{-1}$ from \cref{R_inv_defn} and dropping the asterisks on the physical means again for notational simplicity, we have finally
	\begin{equation}{\label{mu_star_equation}}
		(\partial_\tau + k_2) \mu(\tau) = \Gamma^{1,0}(\tau)  + \int_0^\tau d\tau'\, \Sigma(\tau,\tau') \mu(\tau') + \sum_{j\geq 2} \Omega^{1,j}(\tau) 
	\end{equation}
	We will deploy the above equation to determine the mean copy numbers of the dynamics, $\mu(\tau)$, by constructing the vertex functions $\Gamma^{1,m}$ (hence $\Omega^{1,m}$) and the self-energy $\Sigma$ using diagrammatic perturbation theory. Independently of specific calculations, what is notable is that \cref{mu_star_equation} contains {\em memory corrections} in the last two terms on the r.h.s., which depend on the entire history of the copy numbers up to time $\tau$. Such memory terms thus arise naturally when applying the effective action formalism to Doi-Peliti field theory.

	\section{Single species interactions: $A+A \rightarrow A$}{\label{section_AAA}}
We will now consider a system with particles of only a single species, $A$, with the baseline creation and destruction reactions $\varnothing \xrightleftharpoons[k_2]{k_1} A$ and in addition the binary coagulation reaction 
	\begin{equation}
		A+A \xrightarrow{k_3} A
	\end{equation}
with rate $k_3$. As throughout we will assume that the initial distribution of copy numbers is Poissonian. With the binary reaction present, this system is not exactly solvable in our formalism because the path integral is no longer Gaussian. Instead it must be dealt with perturbatively, as we will illustrate in this section.

	\subsection{Internal vertices and Feynman rules}
	\label{sec:internal_feynman}
	
	\begin{fmffile}{diagrams/2AA_vertices}
		
		\fmfcmd{
			style_def end_arrow expr p =
			cdraw p;			
			cfill (harrow (p, 0.5));
			enddef;}
		
		\fmfcmd{
			style_def end_arrow_two expr p =
			cdraw p;			
			cfill (harrow (p, 0.6));
			enddef;}
		
		\fmfcmd{
			style_def end_arrow_dash expr p =
			draw_dashes p;
			cfill (harrow (p, 0.4));
			enddef;}
		
		\fmfcmd{
			style_def end_arrow_dash_two expr p =
			draw_dashes p;
			cfill (harrow (p, 0.5));
			enddef;}
		
		\fmfcmd{
			style_def double_arrow expr p =
			draw_double p;
			cfill (arrow p);
			enddef;
		}
		
		\fmfcmd{
			style_def mixed_response_GB expr p =
			save oldpen;
			pen oldpen;
			oldpen := currentpen;
			pickup oldpen scaled 3;
			draw (subpath( (0 ,0.5)*length(p)) of p) withcolor green;
			draw (subpath( (0.5,1)*length(p)) of p) withcolor blue;
			pickup oldpen;
			cullit ; undraw p; cullit;
			cfill (marrow (p, 0.5));
			enddef;}
		
		\fmfcmd{
			style_def mixed_response_BG expr p =
			save oldpen;
			pen oldpen;
			oldpen := currentpen;
			pickup oldpen scaled 3;
			draw (subpath( (0 ,0.5)*length(p)) of p) withcolor blue;
			draw (subpath( (0.5,1)*length(p)) of p) withcolor green;
			pickup oldpen;
			cullit ; undraw p; cullit;
			cfill (marrow (p, 0.5));
			enddef;}
		
As explained above, we include the term $k_1 \tilde{\phi}$ from the creation reaction in the interaction Hamiltonian  $H_{\text{int}}$ so that $H_0$ is purely quadratic. 
The same applies to the initial condition term $\bar{n}_0\tilde\phi(0)$. For the coagulation reaction the Hamiltonian operator \cref{Hamiltonian_operator} is $k_3 [\hat{a}^\dagger \hat{a}^2 - (\hat{a}^\dagger)^2 \hat{a}^2]$, which after replacing operators by fields and Doi-shift becomes $-k_3(\tilde\phi+\tilde\phi^2)\phi^2$, giving for the interaction part of the action 
			\begin{equation}
				S_{\text{int}} = \Delta t \sum_{\tau=\Delta t}^{t} \left[ k_1  \tilde{\phi}(\tau) - k_3\tilde{\phi}(\tau) \phi(\tau_-) \phi(\tau_-) -k_3 \tilde{\phi}(\tau) \tilde{\phi}(\tau) \phi(\tau_-) \phi(\tau_-) \right] + \bar{n}_0\tilde\phi(0)
			\end{equation}
		 Diagrammatically this is represented by the following three internal vertices:
				\begin{equation}{\label{AAA_vertices}}
			S_{\text{int}} = \Delta t\sum_\tau \left( \rule{0cm}{1cm}   \quad
				\parbox{15mm}{
					\begin{fmfgraph*}(20,20)  
						\fmfleft{e1}
						\fmfright{v1}               
						\fmfv{label=$k_1,, \tau$, label.angle=105,label.dist=6}{v1}
						\fmfv{label=$\tilde{\phi}$ ,label.dist=0.5}{e1} 
						\fmf{fermion,tension=1}{v1,e1}    
						\fmfv{decor.shape=circle,decor.filled=full, decor.size=2thick}{v1 }		
				\end{fmfgraph*}}	+	\quad	\parbox{20mm}{\begin{fmfgraph*}(30,20)  
						\fmfleft{e1}              
						\fmfright{e2,e3} 
						\fmfv{label=$-k_3,, \tau$, label.angle=0,label.dist=10}{v1}
						\fmfv{label=$ \tilde{\phi} $,label.dist=1}{e1} 
						\fmfv{label=$ \phi $,label.dist=0.5}{e2} 
						\fmfv{label=$ \phi $,label.dist=0.5}{e3}   
						\fmf{fermion,tension=3}{v1,e1}    
						\fmf{plain,tension=2}{e2,v1}
						\fmf{plain,tension=2}{e3,v1}
						\fmfv{decor.shape=circle,decor.filled=full, decor.size=2thick}{v1 }		
				\end{fmfgraph*}}  \quad	\quad +  \parbox{20mm}{\begin{fmfgraph*}(35,20)  
						\fmfleft{e1,e2}              
						\fmfright{e3,e4} 
						\fmfv{label=$-k_3,, \tau$, label.angle=0,label.dist=10}{v1}
						\fmfv{label=$ \tilde{\phi} $,label.dist=0.5}{e1} 
						\fmfv{label=$ \tilde{\phi} $,label.dist=0.5}{e2} 
						\fmfv{label=$ \phi $,label.dist=0.5}{e3} 
						\fmfv{label=$ \phi $,label.dist=0.5}{e4}   
						\fmf{fermion,tension=2}{v1,e1}    
						\fmf{fermion,tension=2}{v1,e2} 
						\fmf{plain,tension=2}{e3,v1}
						\fmf{plain,tension=2}{e4,v1}
						\fmfv{decor.shape=circle,decor.filled=full, decor.size=2thick}{v1 }		
				\end{fmfgraph*}} \quad \quad \right)
		\end{equation}

Here we use outgoing arrows to indicate $\tilde{\phi}$ legs; legs without arrows are $\phi$ legs. The $\phi$ legs are always one time step behind the $\tilde{\phi}$ legs at the same vertex. 
We have three types of vertices: 

		\begin{enumerate}
			\item The one-legged vertex with just a $\tilde{\phi}$ leg is a source term and indicates the creation of $A$ with the rate $k_1(\tau)$ at time $\tau$. 
			We have included the initial condition term here by defining a time-dependent creation rate $k_1(\tau) = k_1 + \delta_{\tau,0}\bar{n}_0/\Delta t$.
			\item The three-legged vertex with one $\tilde{\phi}$ leg and two $\phi$ legs reflects the coagulation reaction we are actually considering, with two incoming A molecules at time $\tau_-$ that react at time $\tau$ with rate $k_3$ to form one A molecule.
			\item In analogy with the MSRJD path integral and the associated Langevin equation with multiplicative noise~\cite{falkovich_john_2008,itakura_two_2010}, the four-legged vertex can be interpreted as describing the noise in the system whereby two $A$ molecules meet at rate $k_3$ but do not react, resulting in two outgoing $A$ molecules. 
		\end{enumerate}
	
The perturbative expansion is now set up by expressing averages in the full non-Gaussian path integral as $\langle \ldots \rangle = \langle \ldots e^{\alpha S_{\text{int}}}\rangle_0$ where the second average is taken across the Gaussian baseline defined by the quadratic action $S_0$. One then expands in powers of $\alpha$ or equivalently $S_{\text{int}}$ and evaluates the resulting Gaussian averages using Wick's theorem. Each factor in a Wick pairing corresponds to a bare second order correlation function or ``propagator'', in our case specifically a bare response function.

We then have the following \textbf{rules for constructing diagrams} in the perturbative expansion:
		\begin{enumerate}
			\item Feynman diagrams can be constructed using the internal vertices of the interacting action; a summation over internal time indices of the vertex functions is always implied.	
			\item Given the causal nature of the bare response function, we can only join a $\tilde{\phi}$ leg at an earlier time, marked by an outgoing arrow, with a $\phi$ leg (without arrow) at the same or later time. This forms a response function, $R_0$, which connects two vertices.
			\item Two $\tilde{\phi}$ or two $\phi$ legs cannot be joined to each other because they result in $\langle \tilde{\phi} \tilde{\phi} \rangle_0$ or $\langle \phi \phi \rangle_0$ correlations, both of which are zero because the baseline action only has $\phi \tilde{\phi}$ terms.
			\item A $\tilde{\phi}$ leg cannot connect to $\phi$ leg at the same vertex because the $\phi$ leg is a time step behind, resulting in a vanishing response function.
			\item Again because of the causality of the response function, there must not be any closed time loops in the diagrams. This implies there must be a consistent flow of time along the response functions; we will draw this from right (early times) to left (later times) below.
			\item The vertex functions $\Gamma^{l,m}$ are drawn with $l$ amputated $\tilde{\phi}$ legs (with outgoing arrows) and $m$ amputated $\phi$ legs (without arrows) where propagators can connect, to form the appropriate diagrams that contribute to connected $n$-point correlation functions. When connecting propagators, the direction of the arrows must be respected. 
			 \item For the vertex functions $\Gamma^{l,m}$, it is known from field theory~\cite{bellac_quantum_1991} that these are constructed using only one particle irreducible (1PI) diagrams, that is diagrams that cannot be split into two separate components when any propagator (i.e.\ response) line is cut. 
				In particular such diagrams cannot contain tadpoles~\cite{kuester_tadpole_1997}, i.e.\ sub-diagrams with only internal vertices that are connected to the rest of the diagram by a single propagator line. For $\Gamma^{l,m*}$, on the other hand, tadpoles are included, i.e.\ here all diagrams contribute that are 1PI in the broader sense that they cannot be separated by cutting a single response line, into two components that each contain at least one external connection via an amputated leg.
		\end{enumerate}
	
			\subsection{Vertex functions and equation of motion for the mean}
			
			We begin by noting that the self-energy $\Sigma$ that appears in \cref{diagrammatic_analogue_eqn} vanishes for the reaction $A+A\rightarrow A$ because there is no vertex with only one outgoing and only one incoming leg; we will see this more explicitly in \cref{sigma_section}. The diagrammatic analogue of \cref{diagrammatic_analogue_eqn} is then 
			\begin{equation}{\label{diagrammtic_analogue}}
				\begin{split}
					\parbox{15mm}{ \begin{fmfgraph*}(35,30)
							\fmfleft{e1}
							\fmfright{e2}
							\fmf{plain,tension=2}{e1,e2}
							\fmfv{label=$\tau$,label.angle = 45,label.dist=0, decor.shape=circle, decor.filled=empty, decor.size= 8thick}{e1} 
					\end{fmfgraph*}} = 
					\parbox{20mm}{ \begin{fmfgraph*}(38,30)
							\fmfleft{eg1}
							\fmfright{eg2}
							\fmf{fermion,tension=2}{eg2,eg1}
							\fmfv{decor.shape=circle, decor.filled=shaded, decor.size= 8thick}{eg2} 
					\end{fmfgraph*}} +
					\parbox{30mm}{\begin{fmfgraph*}(75,30)
							\fmfleft{e1}
							\fmfright{e2,e3}
							\fmf{fermion,tension=2}{v1,e1}
							\fmf{dashes}{e2,v1}
							\fmf{dashes}{e3,v1}
							\fmfv{ decor.shape=circle, decor.filled=shaded, decor.size= 8thick}{v1}
							\fmfv{decor.shape=circle, decor.filled=empty, decor.size= 8thick}{e2}
							\fmfv{decor.shape=circle, decor.filled=empty, decor.size= 8thick}{e3}
					\end{fmfgraph*}} +
					\parbox{30mm}{\begin{fmfgraph*}(65,50)
							\fmfleft{e1}
							\fmfright{e2,e3,e4}
							\fmf{fermion,tension=2}{v1,e1}
							\fmf{dashes}{e2,v1}
							\fmf{dashes}{e3,v1}
							\fmf{dashes}{e4,v1}
							\fmfv{ decor.shape=circle, decor.filled=shaded, decor.size= 8thick}{v1}
							\fmfv{decor.shape=circle, decor.filled=empty, decor.size= 8thick}{e2}
							\fmfv{decor.shape=circle, decor.filled=empty, decor.size= 8thick}{e3}
							\fmfv{decor.shape=circle, decor.filled=empty, decor.size= 8thick}{e4} 
					\end{fmfgraph*}} +
					\parbox{30mm}{\begin{fmfgraph*}(65,60)
							\fmfleft{e1}
							\fmfright{e2,e3,e4,e5}
							\fmf{fermion,tension=2}{v1,e1}
							\fmf{dashes}{e2,v1}
							\fmf{dashes}{e3,v1}
							\fmf{dashes}{e4,v1}
							\fmf{dashes}{e5,v1}
							\fmfv{ decor.shape=circle, decor.filled=shaded, decor.size= 8thick}{v1}
							\fmfv{decor.shape=circle, decor.filled=empty, decor.size= 8thick}{e2}
							\fmfv{decor.shape=circle, decor.filled=empty, decor.size= 8thick}{e3}
							\fmfv{decor.shape=circle, decor.filled=empty, decor.size= 8thick}{e4} 
							\fmfv{decor.shape=circle, decor.filled=empty, decor.size= 8thick}{e5} 
					\end{fmfgraph*}} 
					+\dots
				\end{split}
			\end{equation}
			The l.h.s.\ diagrammatically represents the mean or one point function at time $\tau$, i.e\ $\mu(\tau)$. In the diagrams on the r.h.s.\ the shaded circles represent the vertex functions $\Gamma^{1,m}$ connected to the left external vertex $\phi(\tau)$, by a line in the middle which is the response function with the arrow specifying the direction of time, from right to left. The sum over the internal time index where the response function connects to the vertex function is implied. The empty circles represent $\mu$ as on the left, and we use dashed lines to connect them to the vertex functions, to indicate that there is no propagator there. These together make up the $\Omega^{1,m}$ as in \cref{omega_defn}, which are the vertex functions with the $m$ amputated $\phi(\tau_1),\dots,\phi(\tau_m)$ legs replaced 
			by $m$ factors of $\mu(\tau_{1}),\dots,\mu(\tau_{m})$ and summed over the internal times $\tau_1,\dots \tau_m$.
			
	The 1PI	one-point vertex function with one $\tilde{\phi}$ leg, i.e\ $\Gamma^{1,0}$, for our example system is represented diagrammatically as 
	\begin{eqnarray}
		\Gamma^{1,0} \quad = \quad \parbox{20mm}{\begin{fmfgraph*}(38,30)
				\fmfleft{eg1}
				\fmfright{eg2}
				\fmf{end_arrow_dash}{eg2,eg1}
				\fmfv{decor.shape=circle, decor.filled=shaded, decor.size= 7 thick}{eg2} 
		\end{fmfgraph*}} \quad = \quad \parbox{20mm}{ \begin{fmfgraph*}(38,30)  
				\fmfleft{e1}
				\fmfright{v1}               
				\fmfv{label=$k_1$, label.angle=105,label.dist=5}{v1}				
				\fmf{end_arrow_dash,tension=2}{v1,e1}    
				\fmfv{decor.shape=circle,decor.filled=full, decor.size=4thick}{v1 }		
		\end{fmfgraph*}}
	\end{eqnarray}
	The first diagram gives our generic diagrammatic notation for such a vertex function; the second equality asserts that for our system there is only a single 1PI diagram contributing to $\Gamma^{1,0}$ with the value $\alpha k_1(\tau)$, which is of $O(\alpha)$ as it contains one internal vertex; we do not write these powers of $\alpha$ explicitly in the diagrams. 
	The dashed lines indicate the amputated $\tilde\phi$ leg where a propagator can connect. The arrow also indicates the direction of the flow of time in the vertex function. 
	
Taking only the first term on the r.h.s\ of \cref{diagrammtic_analogue}, the mean is given by
	\begin{equation}
		\begin{split}
			\mu(\tau) = \Delta t \sum_{\tau'} R_0(\tau,\tau') \alpha k_1(\tau')
		\end{split}
	\end{equation}
	Multiplying by $R_0^{-1}$ we get 
	\begin{equation}
		\begin{split}			
			\Delta t \sum_\tau R_0^{-1}(\tau,\tau')\mu(\tau') =  \alpha k_1(\tau)
		\end{split}
	\end{equation}
	In continuous time this gives the equation of motion for the mean,
	\begin{equation}
		(\partial_\tau + k_2) \mu(\tau) = \alpha k_1(\tau)
	\end{equation}
	which is the same result as \cref{mu_eom_baseline} obtained in the baseline theory. 
	In continuous time $k_1(\tau)=k_1+\bar{n}_0\delta(\tau)$ and the $\delta(\tau)$ term just fixes the initial condition $\mu(\tau=0^+)=\bar{n}_0$ as expected. We leave this initial condition term implicit throughout the rest of the paper and write $k_1$ again instead of $k_1(\tau)$.
	
	To account for the effect of the interactions, we include the other terms on the r.h.s\ of \cref{diagrammtic_analogue}. For this, we consider the vertex functions $\Gamma^{1,m}$, 
	i.e.\ with one amputated $\tilde{\phi}$ leg and any number of amputated $\phi$ legs, at zero mean. The simplest case is $m=2$, which up to $O(\alpha^5)$ has the expansion $\Gamma^{1,2}$, given by	
	\begin{equation}{\label{Gamma_12_1}}
		\setlength{\jot}{20pt}
		\begin{split}
			\Gamma^{1,2} =  \parbox{15mm}{
				\begin{fmfgraph*}(55,30)
					\fmfleft{e1}
					\fmfright{e2,e3}
					\fmf{end_arrow_dash_two,tension=2}{v1,e1}
					\fmf{dashes,tension=2}{e2,v1}
					\fmf{dashes,tension=2}{e3,v1}
					\fmfv{decor.shape=circle, decor.filled=shaded, decor.size= 7thick}{v1}
				\end{fmfgraph*}
			}  
			&=  \parbox{20mm}{ \begin{fmfgraph*}(60,30)  
					\fmfleft{e1}              
					\fmfright{e2,e3} 
					\fmfv{label.angle=105,label.dist=6}{v1}
					\fmf{end_arrow_dash,tension=2}{v1,e1}    
					\fmf{dashes,tension=2}{e2,v1}
					\fmf{dashes,tension=2}{e3,v1}
					\fmfv{decor.shape=circle,decor.filled=full, decor.size=4thick}{v1 }		
			\end{fmfgraph*}} + 
			\parbox{30mm}{ 	\begin{fmfgraph*}(100,30)
					\fmfleft{e1}
					\fmfright{e2,e3}
					\fmf{end_arrow_dash,tension=1.5}{v1,e1}
					\fmf{dashes}{e2,v2}
					\fmf{dashes}{e3,v2}
					\fmf{fermion,left}{v2,v1}
					\fmf{fermion,right}{v2,v1}
					\fmfv{label.angle=125,decor.shape=circle, decor.filled=full, decor.size= 4thick}{v1}
					\fmfv{label.angle=180,decor.shape=circle, decor.filled=full, decor.size= 4thick}{v2}
			\end{fmfgraph*}}
			+ 
			\parbox{40mm}{
				\begin{fmfgraph*}(135,30)
					\fmfleft{e1}
					\fmfright{e2,e3}
					\fmf{end_arrow_dash,tension=1.5}{v1,e1}
					\fmf{dashes}{e2,v2}
					\fmf{dashes}{e3,v2}
					\fmf{fermion,left}{v3,v1}
					\fmf{fermion,right}{v3,v1}
					\fmf{fermion,left}{v2,v3}
					\fmf{fermion,right}{v2,v3}
					\fmfv{label.angle=125,decor.shape=circle, decor.filled=full, decor.size= 4thick}{v1}
					\fmfv{label.angle=180,decor.shape=circle, decor.filled=full, decor.size= 4thick}{v2}
					\fmfv{label.angle=180,decor.shape=circle, decor.filled=full, decor.size= 4thick}{v3}
				\end{fmfgraph*}
			} \\
			&+ \parbox{50mm}{
				\begin{fmfgraph*}(165,30)
					\fmfleft{e1}
					\fmfright{e2,e3}
					\fmf{end_arrow_dash,tension=1.5}{v1,e1}
					\fmf{dashes}{e2,v2}
					\fmf{dashes}{e3,v2}
					\fmf{fermion,left}{v3,v1}
					\fmf{fermion,right}{v3,v1}
					\fmf{fermion,left}{v4,v3}
					\fmf{fermion,right}{v4,v3}
					\fmf{fermion,left}{v2,v4}
					\fmf{fermion,right}{v2,v4}
					\fmfv{label.angle=125,decor.shape=circle, decor.filled=full, decor.size= 4thick}{v1}
					\fmfv{label.angle=180,decor.shape=circle, decor.filled=full, decor.size= 4thick}{v2}
					\fmfv{label.angle=180,decor.shape=circle, decor.filled=full, decor.size= 4thick}{v3}
					\fmfv{label.angle=180,decor.shape=circle, decor.filled=full, decor.size= 4thick}{v4}
				\end{fmfgraph*}
			} +  \quad
			\parbox{50mm}{
				\begin{fmfgraph*}(190,30)
					\fmfleft{e1}
					\fmfright{e2,e3}
					\fmf{end_arrow_dash,tension=1.5}{v1,e1}
					\fmf{dashes}{e2,v2}
					\fmf{dashes}{e3,v2}
					\fmf{fermion,left}{v3,v1}
					\fmf{fermion,right}{v3,v1}
					\fmf{fermion,left}{v4,v3}
					\fmf{fermion,right}{v4,v3}
					\fmf{fermion,left}{v5,v4}
					\fmf{fermion,right}{v5,v4}
					\fmf{fermion,left}{v2,v5}
					\fmf{fermion,right}{v2,v5}
					\fmfv{label.angle=125,decor.shape=circle, decor.filled=full, decor.size= 4thick}{v1}
					\fmfv{label.angle=180,decor.shape=circle, decor.filled=full, decor.size= 4thick}{v2}
					\fmfv{label.angle=180,decor.shape=circle, decor.filled=full, decor.size= 4thick}{v3}
					\fmfv{label.angle=180,decor.shape=circle, decor.filled=full, decor.size= 4thick}{v4}
					\fmfv{label.angle=180,decor.shape=circle, decor.filled=full, decor.size= 4thick}{v5}
				\end{fmfgraph*}
			} 
		\end{split}
	\end{equation}
	The first diagram, for example, which with one internal vertex is $O(\alpha)$, represents the contribution 
	\begin{equation}
		\Gamma^{1,2}(\tau,\tau',\tau'')|_{\alpha} = -2\alpha k_3 \frac{\delta_{\tau_-,\tau'}}{\Delta t} \frac{\delta_{\tau',\tau''}}{\Delta t}
		\label{Gamma12_Oalpha}
	\end{equation}	
	where the factor 2 arises from the two ways of associating the two amputated $\phi$ legs with the internal vertex. To leading order in $\alpha$ we then have $\Omega^{1,2}$ from \cref{Gamma12_Oalpha} as 
\begin{equation}
	\Omega^{1,2}(\tau)|_{\alpha} = \frac{1}{2!}(-2\alpha k_3) (\Delta t)^2 \sum_{\tau',\tau''} \frac{\delta_{\tau_-,\tau'}}{\Delta t} \frac{\delta_{\tau',\tau''}}{\Delta t} \mu(\tau') \mu(\tau'') = -\alpha k_3 \mu^2(\tau_-)
\end{equation}
Combining with the $O(\alpha)$ term from $\Gamma^{1,0}$ gives 
\begin{equation}
	\mu(\tau) =  \alpha k_1 \Delta t \sum_{\tau'} R_0(\tau,\tau') - \alpha k_3 \Delta t \sum_{\tau'} R_0(\tau,\tau') \mu^2(\tau'_-)
\end{equation}
Multiplying by $R_0^{-1}$ on both sides and going to continuous time gives the {\em mean field equation}
\begin{equation}{\label{AAA_MAK}}
	\partial_\tau  \mu(\tau) =  \alpha k_1 -  k_2 \mu(\tau) - \alpha k_3  \mu^2(\tau)
\end{equation}
This description, which only accounts for tree diagrams, i.e.\ diagrams without loops, corresponds directly to mass action kinetics (MAK) for particle creation at rate $\alpha k_1$, decay at rate $k_2$ and coagulation at rate $\alpha k_3$. 

We pause briefly here to note that the diagrams above have a clear physical interpretation that parallels the structure of the reaction we are considering. Time is propagated from right to left in the diagrams by response functions, with a consistent flow of time and no closed time loops. The $\Omega^{1,m}$ diagrams give all possible ways in which $m$ molecules of $A$  can react to form one molecule of $A$, using the two internal vertices.

So far we have only considered $O(\alpha)$ diagrams, which therefore have a single internal vertex. Higher order diagrams involve loops built using the four-legged internal vertex, such as the diagrams in \cref{Gamma_12_1}. The $O(\alpha^2)$ contribution there is 
\begin{equation}{\label{bare_O_alpha2}}
	\Gamma^{1,2}(\tau,\tau',\tau'')|_{\alpha^2} = 
	\parbox{30mm}{ 	\begin{fmfgraph*}(100,30)
			\fmfleft{e1}
			\fmfright{e2,e3}
			\fmf{end_arrow_dash,tension=1.5}{v1,e1}
			\fmf{dashes}{e2,v2}
			\fmf{dashes}{e3,v2}
			\fmf{fermion,left}{v2,v1}
			\fmf{fermion,right}{v2,v1}
			\fmfv{label.angle=125,decor.shape=circle, decor.filled=full, decor.size= 4thick}{v1}
			\fmfv{label.angle=180,decor.shape=circle, decor.filled=full, decor.size= 4thick}{v2}
	\end{fmfgraph*}} \quad
		=  2\times 2 (-\alpha k_3)^2 R_0^2(\tau_-,\tau'_+) \frac{\delta_{\tau',\tau''}}{\Delta t}
\end{equation}
where the symmetry factor of 2 comes from the two ways to connect the internal vertices via response functions. Multiplying by $\mu(\tau')\mu(\tau'')$ and summing gives $\Omega^{1,2}$ at this order as 
\begin{equation}
	\Omega^{1,2}(\tau)|_{\alpha^2} = \frac{1}{2!}2\times 2(-\alpha k_3)^2 \Delta t \sum_{\tau'} R_0^2(\tau_-,\tau'_+) \mu^2(\tau')
	= 2(-\alpha k_3)^2 \Delta t \sum_{\tau'} R_0^2(\tau_-,\tau') \mu^2(\tau'_-)
\end{equation}
By including this term we get the equation for the mean in continuous time to $O(\alpha^2)$ as
\begin{equation}{\label{AAA_alpha2}}
	\partial_\tau  \mu(\tau) =  \alpha k_1 -  k_2 \mu(\tau) - \alpha k_3  \mu^2(\tau) + 2(-\alpha k_3)^2 \int_0^{\tau} d\tau' R_0^2(\tau,\tau') \mu^2(\tau')
\end{equation}
At $O(\alpha^3)$ we have a further contribution to $\Omega^{1,2}$ from the two 
loop diagram in \cref{Gamma_12_1}, 
\begin{eqnarray}
	\Omega^{1,2}(\tau)|_{\alpha^3} = 4(-\alpha k_3)^3 (\Delta t)^2 \sum_{\tau',\tau''} R_0^2(\tau_-,\tau')R_0^2(\tau'_-,\tau'') \mu^2(\tau''_-)
\end{eqnarray}
The correction to the equation of motion from this diagram is 
\begin{equation}
	\partial_\tau  \mu(\tau) = \dots + 4(-\alpha k_3)^3 \int_{0}^{\tau} d\tau' \int_{0}^{\tau'} d\tau'' R_0^2(\tau,\tau') R_0^2(\tau',\tau'') \mu^2(\tau'')
\end{equation}
From $O(\alpha^3)$, however, also higher order vertices $\Gamma^{1,m}$, $m \geq 3$ will contribute, giving terms in $\partial_\tau \mu(\tau)$ of third and higher order in $\mu$. The simplest such contribution, $\Gamma^{1,3}$, to $O(\alpha^3)$ is
		\begin{equation}{\label{Gamma_13_1}}
			\begin{split}				
				\parbox{50mm}{
					\begin{fmfgraph*}(145,60)
						\fmfleft{e1}
						\fmfright{e2,ex3,ex4}
						\fmf{end_arrow_dash,tension=1.}{v1,e1}
						\fmf{dashes,tension=1}{ex3,v2}
						\fmf{dashes,tension=1}{ex4,v2}
						\fmf{fermion,right=0.4,tension=0.5}{v2,v1}
						\fmf{fermion,left=0.4,tension=0.2}{v2,v3,v1}
						\fmfforce{(0.5w,0.1h)}{v3}
						\fmfforce{(0.75w,0.5h)}{v2}
						\fmfforce{(0.25w,0.5h)}{v1}
						\fmffreeze
						\fmf{dashes,tension=5}{e2,v3}
						\fmfv{label.angle=130,decor.shape=circle, decor.filled=full, decor.size= 4thick}{v1}
						\fmfv{label.angle=-45,decor.shape=circle, decor.filled=full, decor.size= 4thick}{v2}
						\fmfv{label.angle=60,decor.shape=circle, decor.filled=full, decor.size= 4thick}{v3}
					\end{fmfgraph*}
				} 
			\end{split}
		\end{equation}
		In the next section we will see how to include some of these higher order corrections without the need to explicitly calculate them, by \textit{dressing} the response functions.
		
\subsection{Self-energy and equation of motion for the response function}{\label{sigma_section}}
		
As explained in~\cref{effective_action_section}, one in general needs to use the dressed response function, $R$, in the equation of motion of the mean, see e.g.~\cref{Rinv_mustar}. $R$ is obtained from the bare response function $R_0$ and the {\em self-energy} $\Sigma$ using the {\em Feynman-Dyson} equation~\cite{bellac_quantum_1991, hertz_path_2017}. $\Sigma$ here refers just to the response block of the self-energy, which consists of all 1PI diagrams with one amputated $\tilde{\phi}$ and one amputated $\phi$ leg where two propagators can connect. Using $\Sigma$, the full response (double line on the l.h.s.) can be expressed diagrammatically as
		\begin{equation}{\label{feynmann_dyson}}
			\begin{split}
				\parbox{18mm}{
					\begin{fmfgraph*}(40,30)
						\fmfleft{e1}
						\fmfright{e2}
						\fmf{dbl_plain_arrow,tension=2}{e2,e1}
					\end{fmfgraph*} 
				}  = 
				\parbox{25mm}{
					\begin{fmfgraph*}(60,30)
						\fmfleft{e1}
						\fmfright{e2}
						\fmf{fermion,tension=2}{v1,e1}
						\fmf{fermion,tension=2}{e2,v1}
						\fmfv{label=$\Sigma$,label.dist=0,decor.shape=circle, decor.filled=empty, decor.size= 10thick}{v1}
					\end{fmfgraph*} 
				} + 
			\parbox{35mm}{
				\begin{fmfgraph*}(90,30)
					\fmfleft{e1}
					\fmfright{e2}
					\fmf{fermion,tension=2}{v1,e1}
					\fmf{fermion,tension=1.5}{v2,v1}
					\fmf{fermion,tension=2}{e2,v2}
					\fmfv{label=$\Sigma$,label.dist=0,decor.shape=circle, decor.filled=empty, decor.size= 10thick}{v1}
					\fmfv{label=$\Sigma$,label.dist=0,decor.shape=circle, decor.filled=empty, decor.size= 10thick}{v2}
				\end{fmfgraph*} 
			} +
				\parbox{40mm}{
				\begin{fmfgraph*}(110,30)
					\fmfleft{e1}
					\fmfright{e2}
					\fmf{fermion,tension=2}{v1,e1}
					\fmf{fermion,tension=1.5}{v2,v1}
					\fmf{fermion,tension=1.5}{v3,v2}
					\fmf{fermion,tension=2}{e2,v3}
					\fmfv{label=$\Sigma$,label.dist=0,decor.shape=circle, decor.filled=empty, decor.size= 10thick}{v1}
					\fmfv{label=$\Sigma$,label.dist=0,decor.shape=circle, decor.filled=empty, decor.size= 10thick}{v2}
					\fmfv{label=$\Sigma$,label.dist=0,decor.shape=circle, decor.filled=empty, decor.size= 10thick}{v3}
				\end{fmfgraph*} 
			} + \dots
			\end{split}
		\end{equation}
which by treating the time (and potentially species) arguments in $R$ and $\Sigma$ as indices can be written in matrix form as
		\begin{equation}
				R = R_0 + R_0\Sigma R_0 +R_0\Sigma R_0 \Sigma R_0 + \dots = R_0 \left( \mathbb{I} - \Sigma R_0  \right)^{-1} = (R_0^{-1}-\Sigma)^{-1}
\end{equation}
or equivalently
\begin{equation}
				(R_0)^{-1} R = \mathbb{I} + \Sigma R
		\end{equation}
Using the inverse bare response $R_0^{-1}(\tau,\tau') = (\partial_{\tau} + k_2)\delta(\tau-\tau')$ we then have 
		\begin{equation}
			\partial_{\tau} R(\tau,\tau') = \delta(\tau-\tau') - k_2 R(\tau,\tau') +\int d\tau'' \,\Sigma(\tau,\tau'') R(\tau'',\tau') 
		\end{equation}
An exactly analogous relation also holds between the {\em physical dressed propagator} $R^*$ and the corresponding self-energy $\Sigma^*$: 
		\begin{equation}
			\partial_{\tau} R^*(\tau,\tau') = \delta(\tau-\tau') - k_2 R^*(\tau,\tau') +\int d\tau'' \, \Sigma^*(\tau,\tau'') R^*(\tau'',\tau') 
			\label{Feynman_star}
		\end{equation}
Both $R^*$ and $\Sigma^*$ are defined at the physical means $\mu=\mu^*$, i.e.\ with tadpoles included in the diagrams as explained in \cref{sec:internal_feynman}.

For the $A+A\rightarrow A$ reaction as pointed out earlier we have $\Sigma=0$ because there are no vertices with only one $\phi$ leg. Diagrams for the physical self-energy $\Sigma^*$ can also contain the $k_1$-vertex but as this has only one leg, at least one other internal vertex is required and we can include the three- and four-legged internal vertices from the interacting action to form diagrams for the self-energy. To organize these diagrams it is useful to recall the lowest order diagrams (to $O(\alpha^3)$) for the vertex functions $\Gamma^{1,m}$: 
		\begin{equation}
			\begin{split}
				&\parbox{25mm}{
					\begin{fmfgraph*}(70,40)
						\fmfleft{e1}
						\fmfright{e2}
						\fmf{end_arrow_dash}{v1,e1}
						\fmf{dashes}{e2,v1}
						\fmfv{label.angle=90,decor.shape=circle, decor.filled=full, decor.size= 4thick}{v1}
						\fmffreeze
						\fmfbottom{e3}
						\fmf{dashes,tension=1}{v1,e3}						
					\end{fmfgraph*}	
				} ,
				\parbox{38mm}{
					\begin{fmfgraph*}(105,30)
						\fmfleft{e1}
						\fmfright{e2}
						\fmf{end_arrow_dash,tension=1.5}{v1,e1}
						\fmf{fermion,left}{v2,v1}
						\fmf{fermion,right}{v2,v1}
						\fmf{dashes,tension=2}{e2,v2}
						\fmffreeze
						\fmfbottom{e3}
						\fmfforce{0.8w,0}{e3}
						\fmf{dashes}{e3,v2}
						\fmfv{label.angle=130,decor.shape=circle, decor.filled=full, decor.size= 4thick}{v1}
						\fmfv{label.angle=60,decor.shape=circle, decor.filled=full, decor.size= 4thick}{v2}						
					\end{fmfgraph*}	
				} ,
				\parbox{48mm}{
					\begin{fmfgraph*}(135,30)
						\fmfleft{e1}
						\fmfright{e2}
						\fmf{end_arrow_dash,tension=1.5}{v1,e1}
						\fmf{dashes,tension=2}{e2,v2}						
						\fmf{fermion,left}{v3,v1}
						\fmf{fermion,right}{v3,v1}
						\fmf{fermion,left}{v2,v3}
						\fmf{fermion,right}{v2,v3}
						\fmffreeze
						\fmfbottom{e3}
						\fmfforce{0.9w,0}{e3}
						\fmf{dashes}{e3,v2}
						\fmfv{label.angle=125,decor.shape=circle, decor.filled=full, decor.size= 4thick}{v1}
						\fmfv{label.angle=180,decor.shape=circle, decor.filled=full, decor.size= 4thick}{v2}
						\fmfv{label.angle=180,decor.shape=circle, decor.filled=full, decor.size= 4thick}{v3}						
					\end{fmfgraph*}
				},
			\parbox{60mm}{
				\begin{fmfgraph*}(145,50)
					\fmfleft{e1}
					\fmfright{e2,ex3,ex4}
					\fmf{end_arrow_dash,tension=1.}{v1,e1}
					\fmf{dashes,tension=1}{ex3,v2}
					\fmf{dashes,tension=1}{ex4,v2}					
					\fmf{fermion,right=0.4,tension=0.5}{v2,v1}
					\fmf{fermion,left=0.4,tension=0.2}{v2,v3,v1}
					\fmfforce{(0.5w,0.1h)}{v3}
					\fmfforce{(0.75w,0.5h)}{v2}
					\fmfforce{(0.25w,0.5h)}{v1}
					\fmffreeze
					\fmf{dashes,tension=5}{e2,v3}
					\fmfforce{(0.9w,-0.1h)}{e2}					
					\fmfv{label.angle=130,decor.shape=circle, decor.filled=full, decor.size= 4thick}{v1}
					\fmfv{label.angle=-45,decor.shape=circle, decor.filled=full, decor.size= 4thick}{v2}
					\fmfv{label.angle=60,decor.shape=circle, decor.filled=full, decor.size= 4thick}{v3}
				\end{fmfgraph*}
			} 		
			\end{split}
		\end{equation}
From these we can now construct diagrams for $\Sigma^*$, that is the self-energy with tadpoles, by replacing $m-1$ amputated $\phi$ legs by $m-1$ factors of $\mu$: this leaves one amputated $\tilde{\phi}$ leg and one amputated $\phi$ leg as required for self-energy diagrams. We obtain in this way
	\begin{equation}
		\begin{split}
			&\parbox{27mm}{
				\begin{fmfgraph*}(70,40)
					\fmfleft{e1}
					\fmfright{e2}
					\fmf{end_arrow_dash,tension=1.5}{v1,e1}
					\fmf{dashes,tension=2}{e2,v1}
					\fmfv{label=$\Sigma^*$,label.dist=0,decor.shape=circle, decor.filled=empty, decor.size= 7thick}{v1}
				\end{fmfgraph*}
			} = 
			\parbox{25mm}{
				\begin{fmfgraph*}(70,40)
					\fmfleft{e1}
					\fmfright{e2}
					\fmf{end_arrow_dash}{v1,e1}
					\fmf{dashes}{e2,v1}
					\fmfv{label.angle=90,decor.shape=circle, decor.filled=full, decor.size= 4thick}{v1}
					\fmffreeze
					\fmfbottom{e3}
					\fmf{dashes,tension=1}{v1,e3}
					\fmfv{decor.shape=circle, decor.filled=empty, decor.size= 7thick}{e3}
				\end{fmfgraph*}	
			} +
			\parbox{38mm}{
				\begin{fmfgraph*}(105,30)
					\fmfleft{e1}
					\fmfright{e2}
					\fmf{end_arrow_dash,tension=1.5}{v1,e1}
					\fmf{fermion,left}{v2,v1}
					\fmf{fermion,right}{v2,v1}
					\fmf{dashes,tension=2}{e2,v2}
					\fmffreeze
					\fmfbottom{e3}
					\fmfforce{0.8w,0}{e3}
					\fmf{dashes}{e3,v2}
					\fmfv{label.angle=130,decor.shape=circle, decor.filled=full, decor.size= 4thick}{v1}
					\fmfv{label.angle=60,decor.shape=circle, decor.filled=full, decor.size= 4thick}{v2}
					\fmfv{decor.shape=circle, decor.filled=empty, decor.size= 7thick}{e3}
				\end{fmfgraph*}	
			} +
			\parbox{48mm}{
				\begin{fmfgraph*}(135,30)
					\fmfleft{e1}
					\fmfright{e2}
					\fmf{end_arrow_dash,tension=1.5}{v1,e1}
					\fmf{dashes,tension=2}{e2,v2}					
					\fmf{fermion,left}{v3,v1}
					\fmf{fermion,right}{v3,v1}
					\fmf{fermion,left}{v2,v3}
					\fmf{fermion,right}{v2,v3}
					\fmffreeze
					\fmfbottom{e3}
					\fmfforce{0.9w,0}{e3}
					\fmf{dashes}{e3,v2}
					\fmfv{label.angle=125,decor.shape=circle, decor.filled=full, decor.size= 4thick}{v1}
					\fmfv{label.angle=180,decor.shape=circle, decor.filled=full, decor.size= 4thick}{v2}
					\fmfv{label.angle=180,decor.shape=circle, decor.filled=full, decor.size= 4thick}{v3}
					\fmfv{decor.shape=circle, decor.filled=empty, decor.size= 7thick}{e3}
				\end{fmfgraph*}
			} \\
			&+\parbox{60mm}{
				\begin{fmfgraph*}(145,60)
					\fmfleft{e1}
					\fmfright{e2,ex3,ex4}
					\fmf{end_arrow_dash,tension=1.}{v1,e1}
					\fmf{dashes,tension=1}{ex3,v2}
					\fmf{dashes,tension=1}{ex4,v2}					
					\fmf{fermion,right=0.4,tension=0.5}{v2,v1}
					\fmf{fermion,left=0.4,tension=0.2}{v2,v3,v1}
					\fmfforce{(0.5w,0.1h)}{v3}
					\fmfforce{(0.75w,0.5h)}{v2}
					\fmfforce{(0.25w,0.5h)}{v1}
					\fmffreeze
					\fmf{dashes,tension=5}{e2,v3}
					\fmfforce{(0.9w,-0.1h)}{e2}
					\fmfv{label.angle=130,decor.shape=circle, decor.filled=full, decor.size= 4thick}{v1}
					\fmfv{label.angle=-45,decor.shape=circle, decor.filled=full, decor.size= 4thick}{v2}
					\fmfv{label.angle=60,decor.shape=circle, decor.filled=full, decor.size= 4thick}{v3}
					\fmfv{decor.shape=circle, decor.filled=empty, decor.size= 7thick}{ex3}
					\fmfv{decor.shape=circle, decor.filled=empty, decor.size= 7thick}{ex4}
				\end{fmfgraph*}
			} 		
		+\parbox{60mm}{
			\begin{fmfgraph*}(145,60)
				\fmfleft{e1}
				\fmfright{e2,ex3,ex4}
				\fmf{end_arrow_dash,tension=1.}{v1,e1}
				\fmf{dashes,tension=1}{ex3,v2}
				\fmf{dashes,tension=1}{ex4,v2}				
				\fmf{fermion,right=0.4,tension=0.5}{v2,v1}
				\fmf{fermion,left=0.4,tension=0.2}{v2,v3,v1}
				\fmfforce{(0.5w,0.1h)}{v3}
				\fmfforce{(0.75w,0.5h)}{v2}
				\fmfforce{(0.25w,0.5h)}{v1}
				\fmffreeze
				\fmf{dashes,tension=5}{e2,v3}
				\fmfforce{(0.9w,-0.1h)}{e2}
				\fmfv{label.angle=130,decor.shape=circle, decor.filled=full, decor.size= 4thick}{v1}
				\fmfv{label.angle=-45,decor.shape=circle, decor.filled=full, decor.size= 4thick}{v2}
				\fmfv{label.angle=60,decor.shape=circle, decor.filled=full, decor.size= 4thick}{v3}
				\fmfv{decor.shape=circle, decor.filled=empty, decor.size= 7thick}{ex3}
				\fmfv{decor.shape=circle, decor.filled=empty, decor.size= 7thick}{e2}
			\end{fmfgraph*}} +\dots
		\end{split}
	\end{equation}
The value of these self-energy diagrams to $O(\alpha^2)$ is 
		\begin{equation}
			\begin{split}
				\Sigma^*(\tau,\tau')\Big|_{\alpha^2} = 2 \frac{\delta_{\tau_-,\tau'}}{\Delta t} (-\alpha k_3) \mu(\tau') +  4(-\alpha k_3)^2 R_0^2(\tau_-,\tau'_+) \mu(\tau') 			
\label{Sigma_star}
		\end{split}
		\end{equation}
		The equation for $R^*$ to $O(\alpha^2)$ in continuous time thus becomes
		\begin{equation}
			\partial_{\tau} R^*(\tau,\tau') = \delta(\tau-\tau') - k_2 R^*(\tau,\tau') -2\alpha k_3 \mu(\tau)R^*(\tau,\tau') +  4(-\alpha k_3)^2\int d\tau'' R_0^2(\tau,\tau'')\mu(\tau'') R^*(\tau'',\tau') 
		\end{equation} 

In the next section we will see how we can replace the bare response in our diagrams by the physical dressed response, as discussed in this section, which sums an infinite hierarchy of diagrams leading to the self-consistent response function approximation.

\subsection{Self-consistent response function approximation}

Up to this point, the equations for the mean copy number $\mu$ we have derived involve the bare response $R_0$. To implicitly include higher order diagrams, a standard approach is the Hartree-Fock (HF) approximation~\cite{hertz_path_2017}, where $R_0$ is replaced by the dressed response at zero mean, $R$. However, in the system we are considering, the two are identical as $\Sigma=0$ so nothing would be gained.

We will therefore proceed differently and self-consistently replace the bare propagator $R_0$ by the dressed {\em physical} propagator $R^*$, order by order in $\alpha$. In contrast to the standard HF approach we also do this not just to one-loop order but in all the diagrams that we keep, irrespective of the number of loops they have. Tadpole diagrams in $R^*$ then produce diagrams, in e.g.\ $\Omega^{1,2}$, that are not part of this quantity as originally defined, and we are effectively including contributions from $\Omega^{1,3}$ etc.\ in $\Omega^{1,2}$.

For notational simplicity and as we will no longer need to refer to the propagator at zero mean, we will from now on use $R$ to denote $R^*$, the dressed propagator at the physical copy number means. In the diagrams we will similarly use double lines with an arrow as in \cref{feynmann_dyson} to denote $R^*$.

Every self-consistent replacement of $R_0$ by $R$ sums up an infinite series of diagrams, which means we only have to explicitly consider a subset of diagrams in the diagrammatic expansion of the vertex functions. We will consider only the following series of diagrams for $\Gamma^{1,2}$ or equivalently $\Omega^{1,2}$,		
		\begin{equation}{\label{Gamma_12_2}}
			\setlength{\jot}{20pt}
			\begin{split}
				\Omega^{1,2} =  \parbox{20mm}{
					\begin{fmfgraph*}(55,30)
						\fmfleft{e1}
						\fmfright{e2,e3}
						\fmf{end_arrow_dash,tension=1}{v1,e1}
						\fmf{dashes,tension=1.5}{e2,v1}
						\fmf{dashes,tension=1.5}{e3,v1}
						\fmfv{decor.shape=circle, decor.filled=shaded, decor.size= 7thick}{v1}
						\fmfv{decor.shape=circle, decor.filled=empty, decor.size= 7thick}{e2}
						\fmfv{decor.shape=circle, decor.filled=empty, decor.size= 7thick}{e3}
					\end{fmfgraph*}
				}  
				&=  \parbox{20mm}{ \begin{fmfgraph*}(60,30)  
						\fmfleft{e1}              
						\fmfright{e2,e3} 
						\fmfv{label.angle=105,label.dist=6}{v1}
						\fmf{end_arrow_dash,tension=2}{v1,e1}    
						\fmf{dashes,tension=2}{e2,v1}
						\fmf{dashes,tension=2}{e3,v1}
						\fmfv{decor.shape=circle,decor.filled=full, decor.size=4thick}{v1 }	
						\fmfv{decor.shape=circle, decor.filled=empty, decor.size= 7thick}{e2}
						\fmfv{decor.shape=circle, decor.filled=empty, decor.size= 7thick}{e3}	
				\end{fmfgraph*}} + 
				\parbox{30mm}{ 	\begin{fmfgraph*}(105,30)
						\fmfleft{e1}
						\fmfright{e2,e3}
						\fmf{end_arrow_dash,tension=1.5}{v1,e1}
						\fmf{dashes}{e2,v2}
						\fmf{dashes}{e3,v2}
						\fmf{dbl_plain_arrow,left}{v2,v1}
						\fmf{dbl_plain_arrow,right}{v2,v1}
						\fmfv{label.angle=125,decor.shape=circle, decor.filled=full, decor.size= 4thick}{v1}
						\fmfv{label.angle=180,decor.shape=circle, decor.filled=full, decor.size= 4thick}{v2}
						\fmfv{decor.shape=circle, decor.filled=empty, decor.size= 7thick}{e2}
						\fmfv{decor.shape=circle, decor.filled=empty, decor.size= 7thick}{e3}
				\end{fmfgraph*}}
				+ 
				\parbox{45mm}{
					\begin{fmfgraph*}(140,30)
						\fmfleft{e1}
						\fmfright{e2,e3}
						\fmf{end_arrow_dash,tension=1.5}{v1,e1}
						\fmf{dashes}{e2,v2}
						\fmf{dashes}{e3,v2}
						\fmf{dbl_plain_arrow,left}{v3,v1}
						\fmf{dbl_plain_arrow,right}{v3,v1}
						\fmf{dbl_plain_arrow,left}{v2,v3}
						\fmf{dbl_plain_arrow,right}{v2,v3}
						\fmfv{label.angle=125,decor.shape=circle, decor.filled=full, decor.size= 4thick}{v1}
						\fmfv{label.angle=180,decor.shape=circle, decor.filled=full, decor.size= 4thick}{v2}
						\fmfv{label.angle=180,decor.shape=circle, decor.filled=full, decor.size= 4thick}{v3}
						\fmfv{decor.shape=circle, decor.filled=empty, decor.size= 7thick}{e2}
						\fmfv{decor.shape=circle, decor.filled=empty, decor.size= 7thick}{e3}
					\end{fmfgraph*}
				} + \dots			
			\end{split}
		\end{equation}
where we have replaced the bare $R_0$ by the dressed $R$, denoted by the double lines. To see the effect of this replacement we can look at the lowest order diagram that is affected, namely the $O(\alpha^2)$ diagram here. Expanding the dressed propagator also to $O(\alpha^2)$ -- by stringing together self-energy diagrams from \cref{Sigma_star} -- we get up to $O(\alpha^4)$
		 \begin{equation}
		 	\setlength{\jot}{20pt}
		 	\begin{split}
		 		\parbox{30mm}{ 	\begin{fmfgraph*}(100,30)
		 				\fmfleft{e1}
		 				\fmfright{e2,e3}
		 				\fmf{end_arrow_dash,tension=1.5}{v1,e1}
		 				\fmf{dashes}{e2,v2}
		 				\fmf{dashes}{e3,v2}
		 				\fmf{dbl_plain_arrow,left}{v2,v1}
		 				\fmf{dbl_plain_arrow,right}{v2,v1}
		 				\fmfv{label.angle=125,decor.shape=circle, decor.filled=full, decor.size= 4thick}{v1}
		 				\fmfv{label.angle=180,decor.shape=circle, decor.filled=full, decor.size= 4thick}{v2}
		 				\fmfv{decor.shape=circle, decor.filled=empty, decor.size= 7thick}{e2}
		 				\fmfv{decor.shape=circle, decor.filled=empty, decor.size= 7thick}{e3}
		 		\end{fmfgraph*}} &= 
		 		\parbox{30mm}{ 	\begin{fmfgraph*}(100,30)
		 				\fmfleft{e1}
		 				\fmfright{e2,e3}
		 				\fmf{end_arrow_dash,tension=1.5}{v1,e1}
		 				\fmf{dashes}{e2,v2}
		 				\fmf{dashes}{e3,v2}
		 				\fmf{fermion,left}{v2,v1}
		 				\fmf{fermion,right}{v2,v1}
		 				\fmfv{label.angle=125,decor.shape=circle, decor.filled=full, decor.size= 4thick}{v1}
		 				\fmfv{label.angle=180,decor.shape=circle, decor.filled=full, decor.size= 4thick}{v2}
		 				\fmfv{decor.shape=circle, decor.filled=empty, decor.size= 7thick}{e2}
		 				\fmfv{decor.shape=circle, decor.filled=empty, decor.size= 7thick}{e3}
		 		\end{fmfgraph*}} + 
		 		\parbox{50mm}{
		 			\begin{fmfgraph*}(135,30)
		 				\fmfleft{e1}
		 				\fmfright{e2,ex3,ex4}
		 				\fmf{end_arrow_dash,tension=1.}{v1,e1}
		 				\fmf{dashes,tension=1}{ex3,v2}
		 				\fmf{dashes,tension=1}{ex4,v2}
		 				\fmf{fermion,right=0.4,tension=0.5}{v2,v1}
		 				\fmf{fermion,left=0.4,tension=0.2}{v2,v3,v1}
		 				\fmfforce{(0.5w,0.1h)}{v3}
		 				\fmfforce{(0.75w,0.5h)}{v2}
		 				\fmfforce{(0.25w,0.5h)}{v1}
		 				\fmffreeze
		 				\fmf{dashes,tension=5}{e2,v3}
		 				\fmfforce{(0.75w,-0.25h)}{e2}
		 				\fmfv{label.angle=130,decor.shape=circle, decor.filled=full, decor.size= 4thick}{v1}
		 				\fmfv{label.angle=90,decor.shape=circle, decor.filled=full, decor.size= 4thick}{v2}
		 				\fmfv{label.angle=60,decor.shape=circle, decor.filled=full, decor.size= 4thick}{v3}
		 				\fmfv{decor.shape=circle, decor.filled=empty, decor.size= 7thick}{e2}
		 				\fmfv{decor.shape=circle, decor.filled=empty, decor.size= 7thick}{ex3}
		 				\fmfv{decor.shape=circle, decor.filled=empty, decor.size= 7thick}{ex4}
		 			\end{fmfgraph*}
		 		} 	+
	 		\parbox{45mm}{
	 			\begin{fmfgraph*}(145,40)
	 				\fmfleft{e1}
	 				\fmfright{ex3,ex4}
	 				\fmf{end_arrow_dash,tension=2.}{v1,e1}
	 				\fmf{dashes,tension=1}{ex3,v2}
	 				\fmf{dashes,tension=1}{ex4,v2}	 				
	 				\fmf{fermion,right=0.8,tension=0.5}{v2,v1}
	 				\fmf{fermion,left=0.1.,tension=0.7}{v2,v3,v4,v1}
	 				\fmf{fermion,right=0.8,tension=0.1}{v3,v4}	
	 				\fmffreeze
	 				\fmfbottom{e2}
	 				\fmf{dashes,tension=1.}{e2,v3}
	 				\fmfforce{(0.6w,-0.1h)}{e2}
	 				\fmfv{label.angle=130,decor.shape=circle, decor.filled=full, decor.size= 4thick}{v1}
	 				\fmfv{label.angle=0,decor.shape=circle, decor.filled=full, decor.size= 4thick}{v2}
	 				\fmfv{label.angle=90,decor.shape=circle, decor.filled=full, decor.size= 4thick}{v3}
	 				\fmfv{label.angle=-90,decor.shape=circle, decor.filled=full, decor.size= 4thick}{v4}
	 				\fmfv{decor.shape=circle, decor.filled=empty, decor.size= 7thick}{ex3}
	 				\fmfv{decor.shape=circle, decor.filled=empty, decor.size= 7thick}{ex4}
	 				\fmfv{decor.shape=circle, decor.filled=empty, decor.size= 7thick}{e2}
	 			\end{fmfgraph*}
	 		}\\
 		&+\parbox{50mm}{
 			\begin{fmfgraph*}(130,40)
 				\fmfleft{e1}
 				\fmfright{e2,ex3,ex4}
 				\fmfbottom{e5}
 				\fmf{end_arrow_dash,tension=0.2}{v1,e1}
 				\fmf{dashes,tension=1}{ex3,v2}
 				\fmf{dashes,tension=1}{ex4,v2} 				
 				\fmf{fermion,right=0.4,tension=0.5}{v2,v1}
 				\fmf{fermion,tension=1}{v2,v3}
 				\fmf{fermion,tension=1}{v3,v4}
 				\fmf{fermion,tension=1}{v4,v1}
 				\fmfforce{(0.6w,0.1h)}{v3}
 				\fmfforce{(0.4w,0.1h)}{v4}
 				\fmfforce{(0.75w,0.5h)}{v2}
 				\fmfforce{(0.25w,0.5h)}{v1}
 				\fmfforce{(0.25w,0h)}{e5}
 				\fmffreeze
 				\fmf{dashes,tension=5}{e2,v3}
 				\fmf{dashes,tension=5}{e5,v4}
 				\fmfv{label.angle=130,decor.shape=circle, decor.filled=full, decor.size= 4thick}{v1}
 				\fmfv{label.angle=-45,decor.shape=circle, decor.filled=full, decor.size= 4thick}{v2}
 				\fmfv{label.angle=-30,decor.shape=circle, decor.filled=full, decor.size= 4thick}{v3}
 				\fmfv{label.angle=60,decor.shape=circle, decor.filled=full, decor.size= 4thick}{v4}
 				\fmfv{decor.shape=circle, decor.filled=empty, decor.size= 7thick}{e2}
 				\fmfv{decor.shape=circle, decor.filled=empty, decor.size= 7thick}{ex3}
 				\fmfv{decor.shape=circle, decor.filled=empty, decor.size= 7thick}{ex4}
 				\fmfv{decor.shape=circle, decor.filled=empty, decor.size= 7thick}{e5}
 			\end{fmfgraph*}
 		}  +
 		\parbox{50mm}{
 			\begin{fmfgraph*}(150,60)
 				\fmfleft{e1}
 				\fmfright{e2,e3}
 				\fmftop{e4}
 				\fmfbottom{e5}
 				\fmf{end_arrow_dash,tension=1}{v1,e1}
 				\fmf{dashes,tension=1}{e4,v3}
 				\fmf{dashes,tension=1}{e2,v2}
 				\fmf{dashes,tension=1}{e3,v2}
 				\fmf{dashes,tension=1}{e5,v4}
 				\fmf{fermion,tension=0.4}{v2,v4,v1}
 				\fmf{fermion,tension=0.4}{v2,v3,v1}
 				\fmfv{label.angle=130,decor.shape=circle, decor.filled=full, decor.size= 4thick}{v1}
 				\fmfv{label.angle=0,decor.shape=circle, decor.filled=full, decor.size= 4thick}{v2}
 				\fmfv{label.angle=60,decor.shape=circle, decor.filled=full, decor.size= 4thick}{v3}
 				\fmfv{label.angle=90,decor.shape=circle, decor.filled=full, decor.size= 4thick}{v4}
 				\fmfv{decor.shape=circle, decor.filled=empty, decor.size= 7thick}{e2}
 				\fmfv{decor.shape=circle, decor.filled=empty, decor.size= 7thick}{e3}
 				\fmfv{decor.shape=circle, decor.filled=empty, decor.size= 7thick}{e4}
 				\fmfv{decor.shape=circle, decor.filled=empty, decor.size= 7thick}{e5}
 			\end{fmfgraph*}
 		}  	+ \dots 
		 	\end{split}
		 \end{equation}
Keeping the analogous diagrams in $\Sigma^*$ and making the same self-consistent approximation gives		 	
		 \begin{equation}{\label{Sigma_star_2}}
		 	\begin{split}
		 		\parbox{27mm}{
		 			\begin{fmfgraph*}(70,40)
		 				\fmfleft{e1}
		 				\fmfright{e2}
		 				\fmf{end_arrow_dash,tension=1.5}{v1,e1}
		 				\fmf{dashes,tension=2}{e2,v1}
		 				\fmfv{label=$\Sigma^*$,label.dist=0,decor.shape=circle, decor.filled=empty, decor.size= 8thick}{v1}
		 			\end{fmfgraph*}
		 		} = 
		 		\parbox{25mm}{
		 			\begin{fmfgraph*}(70,30)
		 				\fmfleft{e1}
		 				\fmfright{e2}
		 				\fmf{end_arrow_dash,tension=1.5}{v1,e1}
		 				\fmf{dashes,tension=1.5}{e2,v1}
		 				\fmfv{label.angle=90,decor.shape=circle, decor.filled=full, decor.size= 4thick}{v1}
		 				\fmffreeze
		 				\fmfbottom{e3}
		 				\fmf{dashes,tension=1}{v1,e3}
		 				\fmfv{decor.shape=circle, decor.filled=empty, decor.size= 7thick}{e3}
		 			\end{fmfgraph*}	
		 		} +
		 		\parbox{38mm}{
		 			\begin{fmfgraph*}(100,30)
		 				\fmfleft{e1}
		 				\fmfright{e2}
		 				\fmf{end_arrow_dash,tension=1.5}{v1,e1}
		 				\fmf{dbl_plain_arrow,left}{v2,v1}
		 				\fmf{dbl_plain_arrow,right}{v2,v1}
		 				\fmf{dashes,tension=2}{e2,v2}
		 				\fmffreeze
		 				\fmfbottom{e3}
		 				\fmfforce{0.8w,0}{e3}
		 				\fmf{dashes}{e3,v2}
		 				\fmfv{label.angle=130,decor.shape=circle, decor.filled=full, decor.size= 4thick}{v1}
		 				\fmfv{label.angle=60,decor.shape=circle, decor.filled=full, decor.size= 4thick}{v2}
		 				\fmfv{decor.shape=circle, decor.filled=empty, decor.size= 7thick}{e3}
		 			\end{fmfgraph*}	
		 		} +
		 		\parbox{45mm}{
		 			\begin{fmfgraph*}(125,30)
		 				\fmfleft{e1}
		 				\fmfright{e2}
		 				\fmf{end_arrow_dash,tension=1.5}{v1,e1}
		 				\fmf{dashes,tension=2}{e2,v2}		 				
		 				\fmf{dbl_plain_arrow,left}{v3,v1}
		 				\fmf{dbl_plain_arrow,right}{v3,v1}
		 				\fmf{dbl_plain_arrow,left}{v2,v3}
		 				\fmf{dbl_plain_arrow,right}{v2,v3}
		 				\fmffreeze
		 				\fmfbottom{e3}
		 				\fmfforce{0.9w,0}{e3}
		 				\fmf{dashes}{e3,v2}
		 				\fmfv{label.angle=125,decor.shape=circle, decor.filled=full, decor.size= 4thick}{v1}
		 				\fmfv{label.angle=180,decor.shape=circle, decor.filled=full, decor.size= 4thick}{v2}
		 				\fmfv{label.angle=180,decor.shape=circle, decor.filled=full, decor.size= 4thick}{v3}
		 				\fmfv{decor.shape=circle, decor.filled=empty, decor.size= 7thick}{e3}
		 			\end{fmfgraph*}
		 		} +\dots
		 	\end{split}
		 \end{equation}
Again one can expand the dressed response in each of these diagrams to generate the infinite series of diagrams that we have summed up. For example, expanding the propagator to $O(\alpha^2)$ in the $O(\alpha^2)$ diagram above yields to $O(\alpha^4)$
		\begin{equation}
			\begin{split}
				&\parbox{38mm}{
					\begin{fmfgraph*}(100,30)
						\fmfleft{e1}
						\fmfright{e2}
						\fmf{end_arrow_dash,tension=1.5}{v1,e1}
						\fmf{dbl_plain_arrow,left}{v2,v1}
						\fmf{dbl_plain_arrow,right}{v2,v1}
						\fmf{dashes,tension=2}{e2,v2}
						\fmffreeze
						\fmfbottom{e3}
						\fmfforce{0.8w,0}{e3}
						\fmf{dashes}{e3,v2}
						\fmfv{label.angle=130,decor.shape=circle, decor.filled=full, decor.size= 4thick}{v1}
						\fmfv{label.angle=60,decor.shape=circle, decor.filled=full, decor.size= 4thick}{v2}
						\fmfv{decor.shape=circle, decor.filled=empty, decor.size= 7thick}{e3}
					\end{fmfgraph*}	
				} =
			\parbox{38mm}{
				\begin{fmfgraph*}(100,30)
					\fmfleft{e1}
					\fmfright{e2}
					\fmf{end_arrow_dash,tension=1.5}{v1,e1}
					\fmf{fermion,left}{v2,v1}
					\fmf{fermion,right}{v2,v1}
					\fmf{dashes,tension=2}{e2,v2}
					\fmffreeze
					\fmfbottom{e3}
					\fmfforce{0.8w,0}{e3}
					\fmf{dashes}{e3,v2}
					\fmfv{label.angle=130,decor.shape=circle, decor.filled=full, decor.size= 4thick}{v1}
					\fmfv{label.angle=60,decor.shape=circle, decor.filled=full, decor.size= 4thick}{v2}
					\fmfv{decor.shape=circle, decor.filled=empty, decor.size= 7thick}{e3}
				\end{fmfgraph*}	
			} +
				\parbox{50mm}{
				\begin{fmfgraph*}(135,30)
					\fmfleft{e1}
					\fmfright{e2,ex3,ex4}
					\fmf{end_arrow_dash,tension=1.}{v1,e1}
					\fmf{dashes,tension=1}{ex3,v2}
					\fmf{dashes,tension=1}{ex4,v2}
					\fmf{fermion,right=0.4,tension=0.5}{v2,v1}
					\fmf{fermion,left=0.4,tension=0.2}{v2,v3,v1}
					\fmfforce{(0.5w,0.1h)}{v3}
					\fmfforce{(0.75w,0.5h)}{v2}
					\fmfforce{(0.25w,0.5h)}{v1}
					\fmffreeze
					\fmf{dashes,tension=5}{e2,v3}
					\fmfforce{(0.75w,-0.25h)}{e2}
					\fmfv{label.angle=130,decor.shape=circle, decor.filled=full, decor.size= 4thick}{v1}
					\fmfv{label.angle=90,decor.shape=circle, decor.filled=full, decor.size= 4thick}{v2}
					\fmfv{label.angle=60,decor.shape=circle, decor.filled=full, decor.size= 4thick}{v3}
					\fmfv{decor.shape=circle, decor.filled=empty, decor.size= 7thick}{e2}					
					\fmfv{decor.shape=circle, decor.filled=empty, decor.size= 7thick}{ex4}
				\end{fmfgraph*}
			} \\			&+
				 		\parbox{52mm}{
				\begin{fmfgraph*}(145,80)
					\fmfleft{e1}
					\fmfright{ex3}
					\fmf{end_arrow_dash,tension=2.}{v1,e1}
					\fmf{dashes,tension=2.}{ex3,v2}
					\fmf{dashes,tension=2.}{ex4,v2}
					\fmf{fermion,right=0.8,tension=0.5}{v2,v1}
					\fmf{fermion,left=0.1.,tension=0.7}{v2,v3,v4,v1}
					\fmf{fermion,right=0.8,tension=0.1}{v3,v4}
					\fmffreeze
					\fmfbottom{ext1,e2,ex4}
					\fmf{dashes,tension=2.}{e2,v3}
					\fmfforce{(0.7w,.35h)}{e2}
					\fmfforce{(0.9w,.35h)}{ex4}
					\fmfv{label.angle=130,decor.shape=circle, decor.filled=full, decor.size= 4thick}{v1}
					\fmfv{label.angle=45,decor.shape=circle, decor.filled=full, decor.size= 4thick}{v2}
					\fmfv{label.angle=90,decor.shape=circle, decor.filled=full, decor.size= 4thick}{v3}
					\fmfv{label.angle=-90,decor.shape=circle, decor.filled=full, decor.size= 4thick}{v4}					
					\fmfv{decor.shape=circle, decor.filled=empty, decor.size= 7thick}{ex4}
					\fmfv{decor.shape=circle, decor.filled=empty, decor.size= 7thick}{e2}
				\end{fmfgraph*}
			} 
			+\parbox{50mm}{
				\begin{fmfgraph*}(130,40)
					\fmfleft{e1}
					\fmfright{e2,ex3,ex4}
					\fmfbottom{e5}
					\fmf{end_arrow_dash,tension=0.2}{v1,e1}
					\fmf{dashes,tension=1}{ex3,v2}
					\fmf{dashes,tension=1}{ex4,v2}					
					\fmf{fermion,right=0.4,tension=0.5}{v2,v1}
					\fmf{fermion,tension=1}{v2,v3}
					\fmf{fermion,tension=1}{v3,v4}
					\fmf{fermion,tension=1}{v4,v1}
					\fmfforce{(0.6w,0.1h)}{v3}
					\fmfforce{(0.4w,0.1h)}{v4}
					\fmfforce{(0.75w,0.5h)}{v2}
					\fmfforce{(0.25w,0.5h)}{v1}
					\fmfforce{(0.25w,0h)}{e5}
					\fmffreeze
					\fmf{dashes,tension=5}{e2,v3}
					\fmf{dashes,tension=5}{e5,v4}
					\fmfv{label.angle=130,decor.shape=circle, decor.filled=full, decor.size= 4thick}{v1}
					\fmfv{label.angle=-45,decor.shape=circle, decor.filled=full, decor.size= 4thick}{v2}
					\fmfv{label.angle=-30,decor.shape=circle, decor.filled=full, decor.size= 4thick}{v3}
					\fmfv{label.angle=60,decor.shape=circle, decor.filled=full, decor.size= 4thick}{v4}
					\fmfv{decor.shape=circle, decor.filled=empty, decor.size= 7thick}{e2}					
					\fmfv{decor.shape=circle, decor.filled=empty, decor.size= 7thick}{ex4}
					\fmfv{decor.shape=circle, decor.filled=empty, decor.size= 7thick}{e5}
				\end{fmfgraph*}
			} + 
			 		\parbox{50mm}{
				\begin{fmfgraph*}(150,60)
					\fmfleft{e1}
					\fmfright{e2,e3}
					\fmftop{e4}
					\fmfbottom{e5}
					\fmf{end_arrow_dash,tension=1}{v1,e1}
					\fmf{dashes,tension=1}{e4,v3}
					\fmf{dashes,tension=1}{e2,v2}
					\fmf{dashes,tension=1}{e3,v2}
					\fmf{dashes,tension=1}{e5,v4}
					\fmf{fermion,tension=0.4}{v2,v4,v1}
					\fmf{fermion,tension=0.4}{v2,v3,v1}
					\fmfv{label.angle=130,decor.shape=circle, decor.filled=full, decor.size= 4thick}{v1}
					\fmfv{label.angle=0,decor.shape=circle, decor.filled=full, decor.size= 4thick}{v2}
					\fmfv{label.angle=60,decor.shape=circle, decor.filled=full, decor.size= 4thick}{v3}
					\fmfv{label.angle=90,decor.shape=circle, decor.filled=full, decor.size= 4thick}{v4}					
					\fmfv{decor.shape=circle, decor.filled=empty, decor.size= 7thick}{e3}
					\fmfv{decor.shape=circle, decor.filled=empty, decor.size= 7thick}{e4}
					\fmfv{decor.shape=circle, decor.filled=empty, decor.size= 7thick}{e5}
					\fmffreeze
					\fmfforce{(1w,0.5h)}{e2}
				\end{fmfgraph*}
			}  
			\end{split}
		\end{equation}
		
Both in the vertex functions and the self-energy, the self-consistent replacement procedure thus implicitly includes an infinite series of diagrams. To avoid double counting we must therefore exclude any diagrams containing self-energy parts, because these are automatically generated by the self-consistent replacement procedure. As any self-energy part can be separated from a diagram by cutting two lines, namely its incoming and outgoing response line, this can be formalized in the following {\em additional Feynman rule}:
		\begin{quote}
		For vertex functions and the self-energy, we keep all diagrams that do not separate when two lines are cut. Diagrams that do separate in this way are also kept, unless one of the parts has one incoming and one outgoing line, as this would form a self-energy contribution.	
		\end{quote} 
For example, all the ``bubble'' diagrams in \cref{Gamma_12_1} do separate if we cut two response lines in the same loop, but neither of the resulting pieces are self-energy components. On the other hand, if we cut the bottom two lines of the diagram in \cref{Gamma_13_1} then we get a self-energy component and hence that diagram must be excluded.

\subsection{Self-consistent bubble resummation (SBR)}{\label{section_resum}}

We are now ready to state the approximation we will use to determine copy number means as well as (physical) response functions. We retain in the exact equation of motion \cref{mu_star_equation} 
 only the $\Omega^{1,2}$ term and discard $\Omega^{1,m}$ with $m\geq 3$. Bearing in mind the interpretation that $\Omega^{1,2}$ gives all possible ways in which two $A$ particles interact in a time delayed fashion to form one particle, we are thus neglecting effective interactions between more than two particles. 
 This is physically reasonable as there are no direct reactions between more than two particles in the $A+A\to A$ system.

In the diagrammatic series \cref{Gamma_12_1} for $\Gamma^{1,2}$ we have only {\em bubble diagrams}. Writing out the corresponding $\Omega^{1,2}$ we have 
			\begin{align}
			\Omega^{1,2}(\tau)  &= -\alpha k_3  \mu^2(\tau_-) +   2(-\alpha k_3)^2  \Delta t \sum_{\tau'} R_0^2(\tau_-,\tau') \mu^2(\tau'_-) + 4(-\alpha k_3 )^3 (\Delta t)^2 \sum_{\tau', \tau''} R_0^2(\tau_-,\tau')R_0^2(\tau'_-,\tau'') \mu^2(\tau''_-)  \nonumber \\
		& + 8(-\alpha k_3)^4 ( \Delta t)^3 \sum_{\substack{\tau', \tau'' \\ \tau'''}} R_0^2(\tau_-,\tau')R^2(\tau'_-,\tau'')R_0^2(\tau''_-,\tau''') \mu^2(\tau'''_-) + \dots \label{Gamma_series_1_bare}
		\end{align}
As every order here acquires one extra factor of $(-2\alpha k_3) R_0^2$, this series of diagrams can be summed as a geometric series (see \cref{resum_geometric_appendix}), giving in the continuous time limit
			\begin{equation}{\label{Gamma_12_contTime_bare}}
				\Omega^{1,2}(\tau) =-\alpha k_3 \int_0^\tau d\tau' \left( \delta(\tau-\tau') + 2\alpha k_3 R_0^2(\tau,\tau') \right)^{-1} \mu^2(\tau')
			\end{equation}
where the inverse is now in the operator sense, generalizing the matrix inverse one has in the discrete time case. A similar geometric series summation has also been considered by Cardy~\cite{falkovich_john_2008} in the context of asymptotic vertex renormalization.

The resulting equation of motion for $\mu$, i.e\ \cref{mu_star_equation} is then simplified to
\begin{equation}{\label{resum_mu_bare}}
	(\partial_\tau + k_2)\mu(\tau) = \alpha k_1 + \Omega^{1,2}(\tau)
\end{equation}
This gives all quadratic contributions in $\mu$ on the r.h.s. We call this the {\em bubble resummation (BR)} approximation.
			 
Finally we consider the same series of diagrams for $\Omega^{1,2}$ but with the self-consistent replacement of the bare response $R_0$ by the physical dressed response $R$ as in \cref{Gamma_12_2}; the result is as in \cref{Gamma_series_1_bare} but with $R_0$ replaced by $R$, giving 
			\begin{equation}{\label{Gamma_12_contTime}}
				\Omega^{1,2}(\tau) = -\alpha k_3 \int_0^\tau d\tau' \left( \delta(\tau-\tau') + 2\alpha k_3 R^2(\tau,\tau') \right)^{-1} \mu^2(\tau')
		\end{equation}
As explained in the previous section, this effectively includes an infinite series of further diagrams that correspond to contributions from $\Omega^{1,m}$ with $m\geq 3$.

The full response $R$ then also needs to be determined via the self-energy $\Sigma^*$ and we use the analogous series of self-consistent bubble diagrams for this, as shown in \cref{Sigma_star_2}. These are the only diagrams containing only a single factor of $\mu$. One could interpret these as self-energy contributions where one particle interacts with an ``external'' particle to coagulate, possibly with some time delay, into a single particle. The resulting self-energy is
			\begin{align}{\label{sigma_series_1}}
				\Sigma^*(\tau,\tau')  &= 2 \delta_{\tau_-,\tau'} (-\alpha k_3) \mu(\tau') +  4(-\alpha k_3)^2 R^2(\tau_-,\tau'_+) \mu(\tau') + 8(-\alpha k_3 )^3\Delta t \sum_{\tau''} R^2(\tau_-,\tau'')R^2(\tau''_-,\tau'_+) \mu(\tau')  \nonumber \\
				&+ 16(-\alpha k_3)^4(\Delta t)^2 \sum_{\tau'',\tau'''} R^2(\tau_-,\tau'')R^2(\tau''_-,\tau''')R^2(\tau'''_-,\tau'_+) \mu(\tau') + \dots
		\end{align}
		Again this can be summed in the continuous time limit (see \cref{resum_geometric_appendix}) to yield
			\begin{equation}{\label{sigma_12_contTime}}
					\Sigma^*(\tau,\tau') =  (-2\alpha k_3) \left( \delta(\tau-\tau') + 2\alpha k_3 R^2(\tau,\tau') \right)^{-1} \mu(\tau')
			\end{equation}
The physical response is then given by the Feynman-Dyson equation
			\begin{equation}{\label{resum_R}}
					(\partial_t + k_2)R(\tau,\tau') = \delta(\tau-\tau') + \int_0^\tau d\tau'' \Sigma^*(\tau,\tau'') R(\tau'',\tau')
\end{equation}
and the last two equations have to be solved self-consistently, i.e.\ simultaneously, with \cref{resum_mu_bare} and \cref{Gamma_12_contTime}. We call this approach of summing infinitely many bubble diagrams and replacing the bare response by the physical dressed response, the {\em self-consistent bubble resummation (SBR)} method. As explained, the self-consistency captures in the equations for both means and response some but not all (see \cref{exclusion_section}) corrections of higher than quadratic order in $\mu$.

Eq.~(\ref{resum_R}) for the response $R$ is an integro-differential equation of Kadanoff-Baym type~\cite{baym_conservation_1961} as occurs in Quantum Field Theory, but in what would be imaginary time there. 
Within the scope of this paper we simply integrate it on a two-dimensional time grid with a fixed step size, but there are more efficient and adaptive methods available that could be used in future, see for example~\cite{stan_time_2009,meirinhos_adaptive_2022}. Even with our straightforward numerical approach, the results in \cref{numerics_AAA} will show superior performance and stability of SBR over the mean field solution and other common approximation methods.

 \subsection{Excluded diagrams}{\label{exclusion_section}}
 The series of bubble diagrams with the self-consistent physical dressed propagator replacement still leaves out some diagrams in the vertex functions $\Gamma^{1,m}$ with $m \geq 3$,  starting at $O(\alpha^4)$. These diagrams do separate if we cut two propagator lines, but the resulting pieces do not form a part of self-energy, that is they either have two incoming or two outgoing response lines. Up to $O(\alpha^5)$ the only omitted diagrams are contributions to $\Gamma^{1,3}$: 
\begin{equation}{\label{Gamma_13_neglect}}
 	\begin{split} 		
 				&\parbox{60mm}{
 			\begin{fmfgraph*}(160,45)
 				\fmfleft{e1}
 				\fmfright{ext1,ex3,ex4,ext2}
 				\fmf{end_arrow_dash,tension=1.3}{v1,e1}
 				\fmf{dashes,tension=1}{ex3,v2}
 				\fmf{dashes,tension=1}{ex4,v2} 				
 				\fmf{fermion,right=0.4,tension=0.5}{v2,v4,v1}
 				\fmf{fermion,left=0.4,tension=0.5}{v2,v3,v1}
 				\fmf{fermion,tenison=0.1}{v3,v4}
 				\fmfforce{(0.45w,0.7h)}{v4}
 				\fmfforce{(0.55w,0.3h)}{v3} 				
 				\fmffreeze
 				\fmfbottom{e2}
 				\fmf{dashes,tension=5}{e2,v3}
 				\fmfforce{(0.7w,-0.2h)}{e2}
 				\fmfv{label.angle=90,decor.shape=circle, decor.filled=full, decor.size= 4thick}{v1}
 				\fmfv{label.angle=-90,decor.shape=circle, decor.filled=full, decor.size= 4thick}{v2}
 				\fmfv{label.angle=-60,decor.shape=circle, decor.filled=full, decor.size= 4thick}{v3}
 				\fmfv{label.angle=30,decor.shape=circle, decor.filled=full, decor.size= 4thick}{v4}
		 		\end{fmfgraph*}} 
		 		+
		 		\parbox{60mm}{
		 			\begin{fmfgraph*}(160,60)
		 				\fmfleft{e1}
		 				\fmfright{e2,ex3,ex4}
		 				\fmf{end_arrow_dash,tension=1.3}{v1,e1}
		 				\fmf{dashes,tension=1}{ex3,v2}
		 				\fmf{dashes,tension=1}{ex4,v2}		 				
		 				\fmf{fermion,right=0.4,tension=0.5}{v2,v4,v1}
		 				\fmf{fermion,left=0.4,tension=0.5}{v2,v3,v5,v1}
		 				\fmf{fermion,tenison=0.1}{v3,v4}
		 				\fmf{fermion,tenison=0.1}{v4,v5}
		 				\fmfforce{(0.5w,0.7h)}{v4}
		 				\fmfforce{(0.65w,0.3h)}{v3}
		 				\fmfforce{(0.35w,0.3h)}{v5}	
		 				\fmffreeze
		 				\fmf{dashes,tension=5}{e2,v3}
		 				\fmfv{label.angle=130,decor.shape=circle, decor.filled=full, decor.size= 4thick}{v1}
		 				\fmfv{label.angle=-70,decor.shape=circle, decor.filled=full, decor.size= 4thick}{v2}
		 				\fmfv{label.angle=-45,decor.shape=circle, decor.filled=full, decor.size= 4thick}{v3}
		 				\fmfv{label.angle=90,decor.shape=circle, decor.filled=full, decor.size= 4thick}{v4}
		 				\fmfv{label.angle=-90,decor.shape=circle, decor.filled=full, decor.size= 4thick}{v5}
		 			\end{fmfgraph*}
		 		} \\
	 		&+ \parbox{60mm}{
		 			\begin{fmfgraph*}(175,60)
		 				\fmfleft{e1}
		 				\fmfright{ext1,ex3,ex4,ext2}
		 				\fmf{end_arrow_dash,tension=1.3}{v1,e1}
		 				\fmf{dashes,tension=1}{ex3,v2}
		 				\fmf{dashes,tension=1}{ex4,v2}		 				
		 				\fmf{fermion,left,tension=0.8}{v2,v5}
		 				\fmf{fermion,right,tension=0.8}{v2,v5}
		 				\fmf{fermion,right=0.4,tension=0.5}{v5,v4,v1}
		 				\fmf{fermion,left=0.4,tension=0.5}{v5,v3,v1}
		 				\fmf{fermion,tenison=0.1}{v3,v4}
		 				\fmfforce{(0.375w,0.7h)}{v4}
		 				\fmfforce{(0.425w,0.3h)}{v3} 
		 				\fmffreeze
		 				\fmfbottom{e2}
		 				\fmf{dashes,tension=5}{e2,v3}
		 				\fmfv{label.angle=130,decor.shape=circle, decor.filled=full, decor.size= 4thick}{v1}
		 				\fmfv{label.angle=-60,decor.shape=circle, decor.filled=full, decor.size= 4thick}{v2}
		 				\fmfv{label.angle=-90,decor.shape=circle, decor.filled=full, decor.size= 4thick}{v3}
		 				\fmfv{label.angle=90,decor.shape=circle, decor.filled=full, decor.size= 4thick}{v4}
		 				\fmfv{label.angle=180,decor.shape=circle, decor.filled=full, decor.size= 4thick}{v5}
		 			\end{fmfgraph*}
		 		} +\parbox{20mm}{
		 			\begin{fmfgraph*}(175,60)
		 				\fmfleft{e1}
		 				\fmfright{ext1,ex3,ex4,ext2}
		 				\fmf{end_arrow_dash,tension=1.3}{v1,e1}
		 				\fmf{fermion,left,tension=0.85}{v5,v1}
		 				\fmf{fermion,right,tension=0.85}{v5,v1}
		 				\fmf{dashes,tension=1}{ex3,v2}
		 				\fmf{dashes,tension=1}{ex4,v2}		 				
		 				\fmf{fermion,right=0.4,tension=0.5}{v2,v4,v5}
		 				\fmf{fermion,left=0.4,tension=0.5}{v2,v3,v5}
		 				\fmf{fermion,tenison=0.1}{v3,v4}
		 				\fmfforce{(0.575w,0.7h)}{v4}
		 				\fmfforce{(0.625w,0.3h)}{v3}
		 				\fmffreeze
		 				\fmfbottom{e2}
		 				\fmf{dashes,tension=5}{e2,v3}
		 				\fmfv{label.angle=130,decor.shape=circle, decor.filled=full, decor.size= 4thick}{v1}
		 				\fmfv{label.angle=-90,decor.shape=circle, decor.filled=full, decor.size= 4thick}{v2}
		 				\fmfv{label.angle=-90,decor.shape=circle, decor.filled=full, decor.size= 4thick}{v3}
		 				\fmfv{label.angle=90,decor.shape=circle, decor.filled=full, decor.size= 4thick}{v4}
		 				\fmfv{label.angle=180,decor.shape=circle, decor.filled=full, decor.size= 4thick}{v5}
		 			\end{fmfgraph*}
		 		}
		 	\end{split}
		 \end{equation}
	 	Starting at $O(\alpha^6)$ there are omitted diagrams from even higher order vertex functions like $\Gamma^{1,4}$, such as 
	 	\begin{equation}	 		
	 		\parbox{60mm}{
	 			\begin{fmfgraph*}(160,75)
	 				\fmfleft{e1}
	 				\fmfright{e2,ex3,ex4,e5}
	 				\fmf{end_arrow_dash,tension=1.3}{v1,e1}
	 				\fmf{dashes,tension=1}{ex3,v2}
	 				\fmf{dashes,tension=1}{ex4,v2}	 				
	 				\fmf{fermion,right=0.4,tension=0.5}{v2,v6,v1}
	 				\fmf{fermion,left=0.4,tension=0.5}{v2,v3,v5,v1}
	 				\fmf{fermion,tenison=0.1}{v3,v4}
	 				\fmf{fermion,tenison=0.1}{v4,v5}
	 				\fmf{fermion,tenison=0.1}{v6,v4}
	 				\fmfforce{(0.5w,0.6h)}{v4}
	 				\fmfforce{(0.65w,0.3h)}{v3}
	 				\fmfforce{(0.35w,0.3h)}{v5}
	 				\fmfforce{(0.55w,0.95h)}{v6}
	 				\fmffreeze
	 				\fmf{dashes,tension=5}{e2,v3}
	 				\fmf{dashes,tension=5}{e5,v6}
	 				\fmfforce{(0.8w,1.1h)}{e5}
	 				\fmfv{label.angle=130,decor.shape=circle, decor.filled=full, decor.size= 4thick}{v1}
	 				\fmfv{label.angle=-70,decor.shape=circle, decor.filled=full, decor.size= 4thick}{v2}
	 				\fmfv{label.angle=-45,decor.shape=circle, decor.filled=full, decor.size= 4thick}{v3}
	 				\fmfv{label.angle=0,decor.shape=circle, decor.filled=full, decor.size= 4thick}{v4}
	 				\fmfv{label.angle=135,decor.shape=circle, decor.filled=full, decor.size= 4thick}{v6}
	 				\fmfv{label.angle=-90,decor.shape=circle, decor.filled=full, decor.size= 4thick}{v5}
	 			\end{fmfgraph*}
	 		}
	 	\end{equation}
Similarly in the physical self-energy $\Sigma^*$ there are diagrams we have omitted, with at least two factors of $\mu$ and of $O(\alpha^3)$ or higher: 
		\begin{equation}
			\begin{split}				
				&\parbox{55mm}{
					\begin{fmfgraph*}(145,45)
						\fmfleft{e1}
						\fmfright{e2,ex3,ex4}
						\fmf{end_arrow_dash,tension=1.}{v1,e1}
						\fmf{dashes,tension=1}{ex3,v2}
						\fmf{dashes,tension=1}{ex4,v2}						
						\fmf{fermion,right=0.4,tension=0.5}{v2,v1}
						\fmf{fermion,left=0.4,tension=0.2}{v2,v3,v1}
						\fmfforce{(0.5w,0.1h)}{v3}
						\fmfforce{(0.75w,0.5h)}{v2}
						\fmfforce{(0.25w,0.5h)}{v1}
						\fmffreeze
						\fmf{dashes,tension=5}{e2,v3}
						\fmfforce{(0.9w,-0.1h)}{e2}
						\fmfv{label.angle=130,decor.shape=circle, decor.filled=full, decor.size= 4thick}{v1}
						\fmfv{label.angle=-45,decor.shape=circle, decor.filled=full, decor.size= 4thick}{v2}
						\fmfv{label.angle=60,decor.shape=circle, decor.filled=full, decor.size= 4thick}{v3}
						\fmfv{decor.shape=circle, decor.filled=empty, decor.size= 7thick}{ex3}
						\fmfv{decor.shape=circle, decor.filled=empty, decor.size= 7thick}{ex4}
					\end{fmfgraph*}
				} + 
			\parbox{60mm}{
				\begin{fmfgraph*}(160,45)
					\fmfleft{e1}
					\fmfright{ext1,ex3,ex4,ext2}
					\fmf{end_arrow_dash,tension=1.3}{v1,e1}
					\fmf{dashes,tension=1}{ex3,v2}
					\fmf{dashes,tension=1}{ex4,v2}
					\fmf{fermion,right=0.4,tension=0.5}{v2,v4,v1}
					\fmf{fermion,left=0.4,tension=0.5}{v2,v3,v1}
					\fmf{fermion,tenison=0.1}{v3,v4}
					\fmfforce{(0.45w,0.7h)}{v4}
					\fmfforce{(0.55w,0.3h)}{v3}
					\fmffreeze
					\fmfbottom{e2}
					\fmfforce{(0.7w,-0.2h)}{e2}
					\fmf{dashes,tension=5}{e2,v3}
					\fmfv{label.angle=90,decor.shape=circle, decor.filled=full, decor.size= 4thick}{v1}
					\fmfv{label.angle=-90,decor.shape=circle, decor.filled=full, decor.size= 4thick}{v2}
					\fmfv{label.angle=-60,decor.shape=circle, decor.filled=full, decor.size= 4thick}{v3}
					\fmfv{label.angle=30,decor.shape=circle, decor.filled=full, decor.size= 4thick}{v4}
					\fmfv{decor.shape=circle, decor.filled=empty, decor.size= 7thick}{ex3}
					\fmfv{decor.shape=circle, decor.filled=empty, decor.size= 7thick}{ex4}
			\end{fmfgraph*}} \\
		&+ \parbox{60mm}{
			\begin{fmfgraph*}(160,55)
				\fmfleft{e1}
				\fmfright{ext1,ex3,ex4,ext2}
				\fmf{end_arrow_dash,tension=1.3}{v1,e1}
				\fmf{dashes,tension=1}{ex3,v2}
				\fmf{dashes,tension=1}{ex4,v2}
				\fmf{fermion,right=0.4,tension=0.5}{v2,v4,v1}
				\fmf{fermion,left=0.4,tension=0.5}{v2,v3,v1}
				\fmf{fermion,tenison=0.1}{v3,v4}
				\fmfforce{(0.45w,0.7h)}{v4}
				\fmfforce{(0.55w,0.3h)}{v3}
				\fmffreeze
				\fmfbottom{e2}
				\fmf{dashes,tension=5}{e2,v3}
				\fmfforce{(0.7w,-0.2h)}{e2}
				\fmfv{label.angle=90,decor.shape=circle, decor.filled=full, decor.size= 4thick}{v1}
				\fmfv{label.angle=-90,decor.shape=circle, decor.filled=full, decor.size= 4thick}{v2}
				\fmfv{label.angle=-60,decor.shape=circle, decor.filled=full, decor.size= 4thick}{v3}
				\fmfv{label.angle=30,decor.shape=circle, decor.filled=full, decor.size= 4thick}{v4}
				\fmfv{decor.shape=circle, decor.filled=empty, decor.size= 7thick}{e2}
				\fmfv{decor.shape=circle, decor.filled=empty, decor.size= 7thick}{ex4}
			\end{fmfgraph*}} +
					\parbox{60mm}{
				\begin{fmfgraph*}(135,60)
					\fmfleft{e1}
					\fmfright{ex1,e2,e3}
					\fmf{end_arrow_dash,tension=2.}{v1,e1}
					\fmf{dashes}{e2,v2}
					\fmf{dashes}{e3,v2}
					\fmf{fermion,left}{v3,v1}
					\fmf{fermion,right}{v3,v1}
					\fmf{fermion,left=0.5,tension=1.}{v2,v4}
					\fmf{fermion,left=0.5,tension=2.}{v4,v3}
					\fmf{fermion,right}{v2,v3}
					\fmf{dashes,tension=0.5}{ex1,v4}
					\fmffreeze
					\fmfforce{(0.65w,0.2h)}{v4}
					\fmfv{label.angle=125,decor.shape=circle, decor.filled=full, decor.size= 4thick}{v1}
					\fmfv{label.angle=180,decor.shape=circle, decor.filled=full, decor.size= 4thick}{v2}
					\fmfv{label.angle=180,decor.shape=circle, decor.filled=full, decor.size= 4thick}{v3}
					\fmfv{label.angle=-90,decor.shape=circle, decor.filled=full, decor.size= 4thick}{v4}
					\fmfv{decor.shape=circle, decor.filled=empty, decor.size= 7thick}{e2}
					\fmfv{decor.shape=circle, decor.filled=empty, decor.size= 7thick}{e3}
				\end{fmfgraph*}
			}\\
			&+	\parbox{50mm}{
				\begin{fmfgraph*}(135,60)
					\fmfleft{e1}
					\fmfright{extra1,e2,e3,extra2}
					\fmf{end_arrow_dash,tension=2.}{v1,e1}
					\fmf{dashes}{e2,v2}
					\fmf{dashes}{e3,v2}
					\fmf{fermion,left=0.5,tension=0.5}{v3,v4,v1}
					\fmf{fermion,right,tension=1.5}{v3,v1}
					\fmf{fermion,left}{v2,v3}
					\fmf{fermion,right}{v2,v3}
					\fmfforce{(0.35w,0.1h)}{v4}
					\fmffreeze
					\fmfbottom{ex1}
					\fmf{dashes,tension=0.5}{ex1,v4}
					\fmfv{label.angle=125,decor.shape=circle, decor.filled=full, decor.size= 4thick}{v1}
					\fmfv{label.angle=180,decor.shape=circle, decor.filled=full, decor.size= 4thick}{v2}
					\fmfv{label.angle=180,decor.shape=circle, decor.filled=full, decor.size= 4thick}{v3}
					\fmfv{label.angle=-90,decor.shape=circle, decor.filled=full, decor.size= 4thick}{v4}
					\fmfv{decor.shape=circle, decor.filled=empty, decor.size= 7thick}{e2}
					\fmfv{decor.shape=circle, decor.filled=empty, decor.size= 7thick}{e3}
				\end{fmfgraph*}
			}			
			\end{split}
		\end{equation}

We have verified the diagrammatic calculations via a short-time expansion of the exact dynamics up to $O(t^5)$, using the hierarchy of coupled equations for the moments of the copy number distribution derived from the master equation. This can be compared to the corresponding expansion of the SBR~\cref{resum_R,resum_mu_bare,Gamma_12_contTime,sigma_12_contTime}. We find that the missing terms are accounted for precisely by the diagrams shown above. For brevity, these calculations are not shown in this paper.

		\subsection{Numerical results}{\label{numerics_AAA}}

		\begin{figure}
			\includegraphics[width=\linewidth]{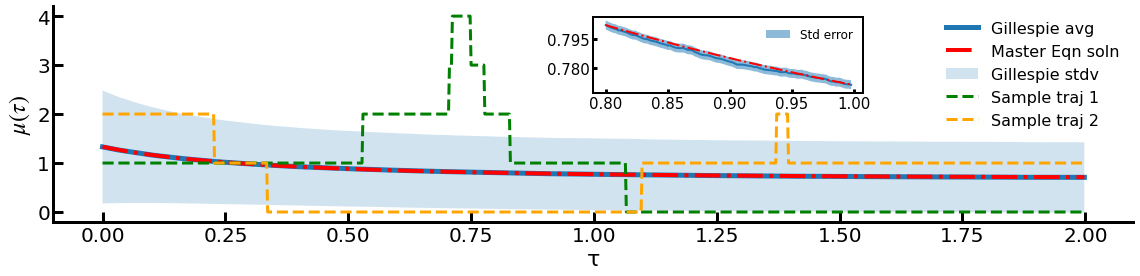}
			\caption{\small Average copy number dynamics 
				$\mu(\tau)$ for the $A+A\rightarrow A$ system with baseline creation and destruction. 
				The system was simulated with the stochastic Gillespie algorithm, two trajectories of which are shown as orange and green dashed lines. The average and standard deviation over $10^6$ such trajectories are plotted as the blue solid line and the shaded area, respectively. The mean from the numerical integration of the master equation is shown as the red dash-dotted line. Inset: enlarged part of the graph from $\tau=0.8$ to $\tau=1$ with the standard error as the shaded area, showing that master equation and Gillespie numerics are consistent. The simulation parameters are from the standard set given in \cref{numerics_AAA}.}
			\label{fig1}
		\end{figure}
		
		In this paper we focus on the challenging regime of small copy numbers of molecules. To demonstrate the power of the SBR method (see \cref{section_resum}), we compare the mean copy number and the two-time number-number correlator for the system considered so far, i.e.\ a single molecular species $A$ undergoing the reaction $A+A\xrightarrow{k_3} A$ with the baseline reactions $A \xrightarrow{k_2} \emptyset$ and $\emptyset \xrightarrow{k_1} A$ and with an initial Poisson distribution for the number of molecules of $A$. We compare our results and report the error relative to the numerical solution of the master equation. The latter is obtained by finite space projection~\cite{munsky_finite_2006}, i.e.\ by truncating the state space at a sufficiently large copy number (see \cref{app_ME}). For the SBR predictions, we use an Euler integration scheme to jointly integrate the discrete time equations for mean $\mu(\tau)$ and Response $R(\tau,\tau')$, performing the inversions required in \cref{Gamma_12_contTime,sigma_12_contTime} as matrix inversions of the corresponding discretized quantities. 
		
		Unless otherwise stated, the simulations for this system are carried out with the following {\em standard parameter set} $k_1 = 1$, $k_2=1$, $k_3 = 1$, $\alpha=1$, time step $\Delta t =0.002$, total integration time $t = 2$ and $\mu(0) = 4/3$, so that the initial copy number is Poisson with this mean and variance as shown in \cref{fig1}. As a consistency check on the numerical solution of the master equation, we also run $10^6$ stochastic Gillespie simulations~\cite{gillespie_exact_1977} and calculate the average and the standard deviation across these runs. As the dynamics proceeds from the initial time, the average copy number decreases to about $0.8$ in the steady state, with the standard deviation remaining of the same order as the mean throughout. We are therefore in the regime of large stochastic fluctuations as illustrated by the  two sample trajectories in \cref{fig1}. In the 
inset we show an enlarged section of the mean copy number time course from $\tau=0.8$ to $\tau=1$, which demonstrates that the Gillespie simulations and the master equation solution are consistent within their error bars as they should be. Because we are in the large fluctuation regime, note that even after averaging over a million stochastic trajectories, the Gillespie mean still has discernible fluctuations. This highlights the inefficiency of stochastic simulations, which becomes even more pronounced in the case of multiple species. Since we are interested in small errors, we will then rely on comparisons with the numerical solution of the master equation instead of stochastic simulations. 
The effects of the integration time step $\Delta t$ are discussed in \cref{time_step_errors_appendix}.		
		\begin{figure}
			\includegraphics[width=\linewidth]{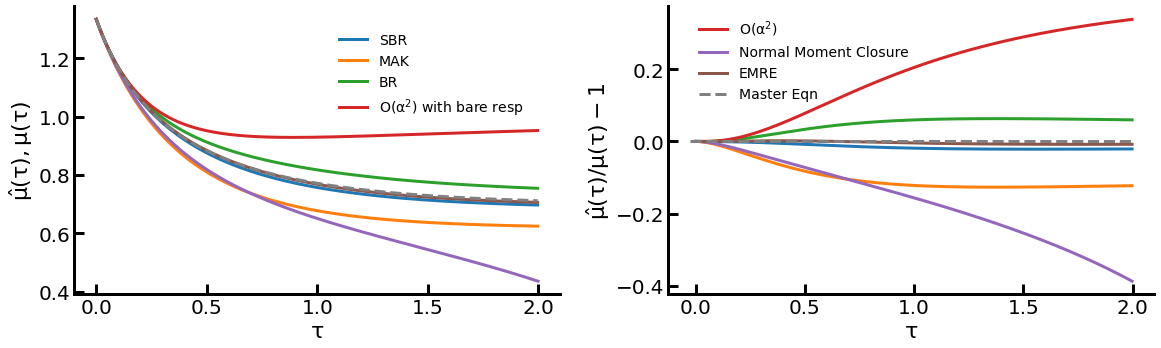}
			\caption{\small Dynamics of average copy number $\hat{\mu}$ (left) as predicted by different approximation methods and relative deviation (right) to the ground truth $\mu$ (grey dashed 
				line). The error of SBR is an order of magnitude lower than the MAK. The simulation parameters are from the standard set given in \cref{numerics_AAA}.}
			\label{fig2}
		\end{figure}
		
With the master equation solution as ground truth in place, we compare the performance of different approximation methods in \cref{fig2}. 
We plot the actual trajectories on the left and the relative deviation on the right. The latter is defined as $\frac{\hat{\mu}(\tau) - \mu(\tau)}{\mu(\tau)}$ where $\hat{\mu}$ is the estimator for the mean copy number as provided by the different methods and $\mu$ is the true mean. The SBR method 
has the second best performance, followed by the (bare) bubble resummation (BR), which is obtained by simply solving \cref{resum_mu_bare,Gamma_12_contTime_bare} with the bare response given by \cref{R_0}. The mean field approximation or mass action kinetics (MAK), obtained by integrating \cref{AAA_MAK}, is stable but has an error an order of magnitude worse than SBR. The results of \textit{normal moment closure} are also plotted. This method takes all cumulants of the copy number distribution of higher than second order as zero and was implemented using the package~\cite{sukys_momentclosurejl_2021}.
This works well only for small times, after which its copy number dynamics diverges and can even enter the negative copy number regime, a problem well known for moment closure methods. We also compare the results to the Effective Mesoscopic Rate Equations (EMRE)~\cite{grima_effective_2010} at unit volume, which for this system has the best performance. 
(The full power of SBR and its advantages over EMRE will become clear for reaction networks with multiple species, see~\cref{ABC_numerics}.) Finally, we also plot the results for a diagrammatic expansion to $O(\alpha^2)$ without resummation, thus keeping only the (bare) $\alpha^2$ term in \cref{bare_O_alpha2}. This gives poor performance 
because the bare response function used is not sensitive to relaxation effects from the $A+A\to A$ reaction (see \cref{bare_response_decay_appendix}). Comparing with the BR method, we conclude for this example that the resummation of an infinite series of bubble diagrams is more important for accuracy than the self-consistent replacement of the response function.

		\begin{figure}
			\includegraphics[width=\linewidth]{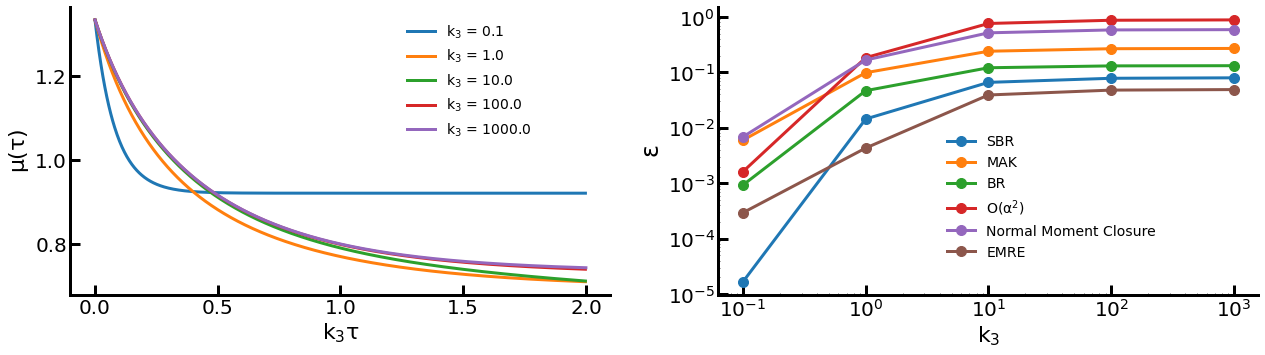}
			\caption{\small Average copy number dynamics from the master equation $\mu$ (left) and time-averaged absolute relative deviation $\epsilon$, \cref{epsilon}, of different approximation methods (right) for different binary interaction rate $k_3$. To measure errors systematically as we increase $k_3$, we decrease total integration time $t$ and time step $\Delta t$ by the same factor, keeping the number of time steps and $k_3 t$ the same for each copy number dynamics time course. From the error measurement on the right plot we see that the SBR method outperforms other common methods over a range of five orders of magnitude of $k_3$. Its also outperforms EMRE at small $k_3$, and has comparable performance at large $k_3$ values.}
			\label{fig3}
		\end{figure}
	
To understand the physical effects of $k_3$ we plot in \cref{fig3} (left) the actual (ground truth) mean copy number dynamics at different values of $k_3$ as a function of $k_3 \tau$. Note that as we increase $k_3$, the dynamics becomes faster due to the higher reaction rate for the $A+A\to A$ reaction. We therefore correspondingly decrease the total integration time $t$ and the time step $\Delta t$. The number of time steps (in our case $B=1000$) is then the same for each $k_3$, as is $k_3 t$. In \cref{fig3} (right) we use this to compare the performance of the different methods as we change the binary interaction rate $k_3$ as compared to the baseline rates. This clearly demonstrates that SBR not only works for large values of $k_3$ but outperforms other common methods over a large range of $k_3$. This is because even though we are starting from a perturbative expansion, the SBR method includes many non-perturbative effects both via the resummation and the self-consistent response function replacement. EMRE has a slightly better performance at large $k_3$ values, while SBR is much better at smaller $k_3$. The performance metric shown is the time-averaged relative deviation in mean copy number,
\begin{equation}
	\epsilon = \frac{1}{B}\sum_\tau \left|\frac{\hat{\mu}(\tau)-\mu(\tau)}{\mu(\tau)}\right|
	\label{epsilon}
\end{equation}  
where $B$ is the total number of time steps. As in our previous comparison at fixed $k_3$, the bare bubble resummation has the second lowest error behind SBR, and both outperform mass action kinetics and normal moment closure.
For small $k_3$ the $A+A\to A$ reaction is the slowest process and the steady state is reached within our time window. 
For large $k_3$, on the other hand, the decay of the mean copy number becomes largely independent of $k_3$ when plotted against $k_3\tau$: the fast $A+A\to A$ reaction dominates here rather than being a small perturbation, so it is encouraging that SBR still performs well. Note that outside the window shown there is then a slow (in relative terms) approach to the steady state determined primarily by the baseline creation and destruction rates $k_1$ and $k_2$ (see \cref{SBR_long_time_appendix}). 

	\begin{figure}
		\includegraphics[width=\linewidth]{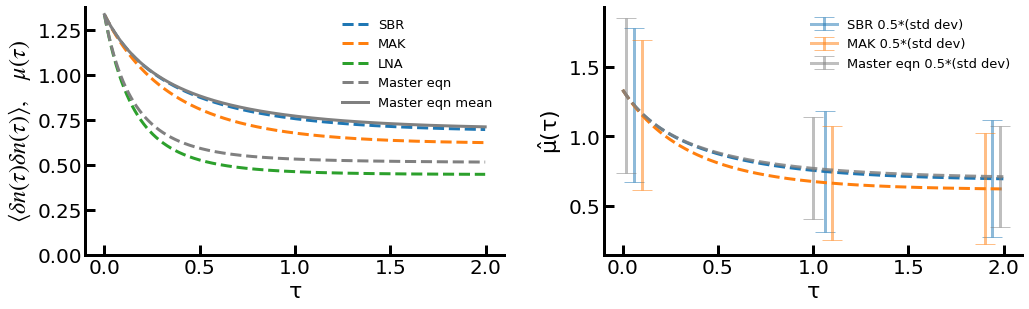}
		\caption{\small The equal time connected number correlator, i.e.\ copy number variance (left) and the copy numbers plus standard deviation error bars at three time points in the dynamics (right; error bars are scaled by $0.5$ for better visibility of the $y$-scale), for the master equation ground truth and SBR and MAK approximations as shown in the legend. On the left we additionally plot the Linear Noise Approximation (LNA) result. The standard parameter set is used. Both the SBR and MAK approximations work with the bare correlation functions $C_0\equiv 0$ and so predict a Poissonian variance $\langle \delta n(t) \delta n(t) \rangle = \hat{\mu}(t)$.
}
		\label{fig10}
	\end{figure}
	
Moving on to the second order statistics, in \cref{fig10}(left) we plot the copy number variance or equivalently the equal-time connected correlator $\langle \delta n(\tau)  \delta n(\tau) \rangle $ as a function of time $\tau$ for the ground truth (see \cref{app_ME}), MAK, SBR and the Linear Noise Approximation (LNA)~\cite{van_kampen_expansion_1976,elf_fast_2003} at unit volume. 
Both SBR and MAK assign to the correlation functions the bare value $C_0\equiv 0$, so from \cref{num_correlator_relation_1} predict a Poissonian variance $\langle \delta n(\tau)  \delta n(\tau) \rangle = \mu(\tau)$. In the ground truth one finds a smaller variance, i.e.\ anti-Poisson effects. This leads to MAK coincidentally predicting a variance closer to the ground truth (for $\tau >0$) than SBR, by a cancellation of two errors: on the one hand it underestimates the mean copy number, and on the other hand it ignores the anti-Poisson effects so overestimates the variance. The LNA makes the best prediction for the equal-time connected correlator (i.e.\ the variance), but we will see that in the comparison to SBR this 
is no longer the case for the two-time correlator, see~\cref{fig7}.

The SBR estimates for the mean copy numbers are more accurate so its predictions for the copy number error bars (plotted as mean $\pm$ half standard deviation) overlap well with the ground truth as shown in \cref{fig10}(right). There are deviations, which can be traced back to the fact that the true correlation functions $C(\tau,\tau)$ do not vanish, but we observe that these are quantitatively moderate. To avoid these deviations one could calculate the correlation functions in perturbation theory and include them in the variance calculation, as we briefly discuss in~\cref{section_discussions}.

		\begin{figure}
			\includegraphics[width=\linewidth]{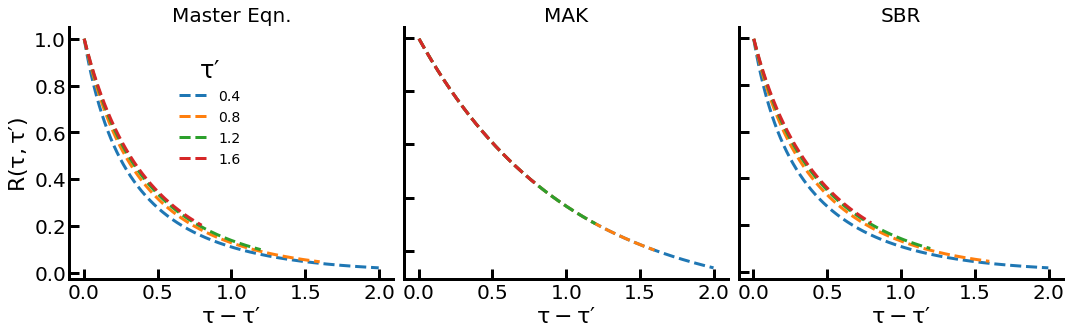}
			\caption{\small 
				 Response functions $R(\tau,\tau')$ obtained from the master equation (left) and the MAK (centre) and SBR (right) approximations, for the standard parameter set. The response functions are plotted against time lag $\tau-\tau'$ at several fixed values of the time $\tau'$ where the perturbation is applied. The different curves overlap for MAK because it predicts a time translation invariant response. The true response (left) is not invariant under time translation and SBR (right) correctly captures this effect.}
			\label{fig6}
		\end{figure}
	
	\begin{figure}
		\includegraphics[width=\linewidth]{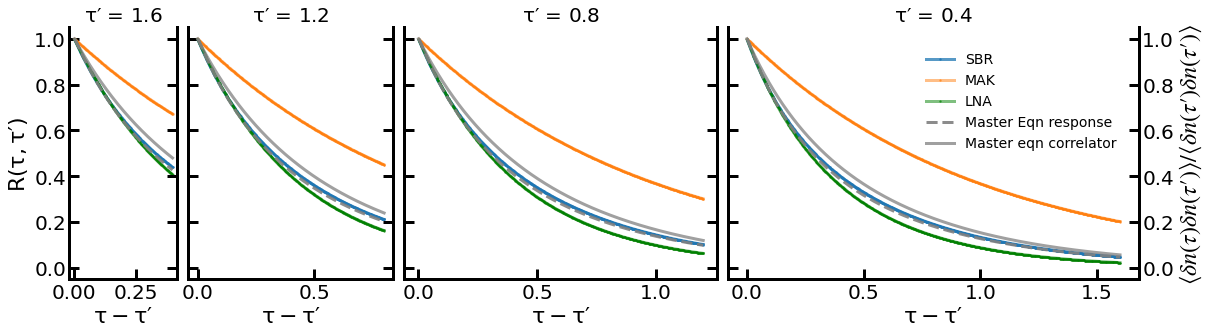}
		\caption{\small 
			Response functions $R(\tau,\tau')$ and two-time connected number correlator $\langle \delta n(\tau) \delta n(\tau')\rangle$ (normalized by the equal-time connected correlator $\langle \delta n(\tau') \delta n(\tau')\rangle$) as obtained from the master equation, the MAK, SBR and LNA methods, for the standard parameter set. The response functions are plotted as a function of the lag time $\tau-\tau'$ between perturbation and response, at fixed values of the perturbation time 
$\tau'=1.6, 1.2, 0.8, 0.4$ (left to right). 
			The MAK, SBR and the LNA predict that the response and the normalized correlator are identical to each other,
while
		in the true dynamics the two quantities are slightly different (solid and dashed grey lines). The SBR method very closely reproduces the true response function and correlator over the whole range of $\tau$ and $\tau'$, while the exponential MAK response decays too slowly and the LNA deviates in the other direction.}
		\label{fig7}
	\end{figure}

Now we look at the two-time quantities. We start with the response function and plot $R(\tau,\tau')$ in \cref{fig6} for fixed values of the time $\tau'$ at which the perturbation is created and as a function of the lag time $\tau-\tau'$ between perturbation and response. Since we are running our dynamics until the final time $t$, we can only calculate the response till $\tau-\tau' = t-\tau'$. From left to right the response functions from the master equation (\cref{app_ME}) and MAK and SBR methods are plotted. In the response from the master equation, the curves for different $\tau'$ do not overlap, consistent with the fact that we have transient dynamics that breaks time translation invariance: it matters when in absolute terms a perturbation is created, not just how long ago. SBR is able to capture this effect but MAK, which uses the bare response function and so is time translation invariant, does not.
		
From the response function, we can calculate the two time connected number number correlator $\langle \delta n(\tau) \delta n(\tau')\rangle$ using \cref{num_correlator_relation_2}. Both SBR and MAK replace the field correlation function $C$ by its bare counterpart $C_0$. This is zero, giving ${\langle \delta n(\tau) \delta n(\tau')\rangle }/{\langle \delta n(\tau')\delta n(\tau')\rangle} = R(\tau,\tau')$. In \cref{fig7} we plot the response and normalized correlator again as a function of the time lag, but now comparing the different methods at each fixed perturbation time. The procedure for calculating the correlator from the master equation is explained in \cref{app_ME}. The resulting true copy number correlator and response are slightly different due to non-Poissonian effects. SBR ignores these but nonetheless captures very closely the true response and correlation, demonstrating that the approximation of keeping only the bare field correlation $C_0$ is rather accurate. MAK, on the other hand, deviates substantially, giving decays in time that are too slow. The LNA also predicts identical response and correlation functions in the case of a single reacting chemical species, and underestimates these quantities when compared to their true values from the master equation.

\subsection{Inverse reaction volume expansion}
The propensity function in \cref{propensity_1} uses rates $k_\beta$ of physical dimension of inverse time, $1/t$, as so far we have worked with absolute molecule numbers. 
In chemical reaction kinetics when molecule numbers are large one typically works with concentrations, i.e.\ number densities. We now translate our formalism into this framework and make contact with earlier work on expansions using $1/V$ as a small parameter, where $V$ is the reaction volume. 
One such work by Thomas et al.~\cite{thomas_system_2014} also used diagrammatic perturbation theory, expanding in orders of $1/V$ around the (time translation invariant) steady state predicted by mass action kinetics. 
 
Starting with the binary reaction $A+B \rightarrow C$ as an example, the mass action description in terms of densities $\rho=n/V$ would be $\partial_\tau \rho_A = j_\beta \rho_A \rho_B$. 
Converting to an equation for the change in particle number $n=\rho V$ we get $\partial_\tau n_A = \frac{j_\beta}{V} n_A n_B$. In the general case one would accordingly write the propensity function as originally defined in \cref{propensity_1} as 
\begin{equation}
	f_\beta(\bm{n}) = j_\beta V \prod_i \frac{n_i !}{(n_i - r_i^\beta)! V^{r_i^\beta}}
\end{equation}
with the rate constant $j_\beta$ having dimension $1/(V^{\sum_i r_i^\beta-1}t)$.
This is again easiest to understand from examples:
\begin{enumerate}
	\item For a creation 
	reaction such as $\emptyset \rightarrow A$, $f_\beta(\bm{n}) = j_\beta V$
	\item For a unary destruction 
	reaction such as $A \rightarrow \emptyset$, $f_\beta(\bm{n}) = j_\beta n_A$
	\item For a 
	binary reaction such as $A+B \rightarrow C$, $f_\beta(\bm{n}) =\frac{j_\beta}{V} n_A n_B$
\end{enumerate}
With this change in notation to standard rate constants $j_\beta$, the MAK equation for $\rho$ in the $A + A \rightarrow A$ system together with the baseline creation and destruction reactions is 
\begin{equation}
	\partial_\tau \rho = j_1 - j_2 \rho - j_3 \rho^2
\end{equation}
with $j_1=k_1/V$, $j_2=k_2$ and $j_3=k_3 V$. 
On the r.h.s.\ we have here the terms up to $O(\alpha)$ from our expansion but we do not write powers of $\alpha$ explicitly. The next correction, of $O(\alpha^2)$, was of the form 
 $(-\alpha k_3)^2 \int_0^{\tau} d\tau' R_0^2(\tau,\tau') \mu^2(\tau')$, giving in terms of density 
\begin{equation}
	\partial_\tau \rho = j_1 - j_2 \rho - j_3 \rho^2 + \frac{2 j_3^2}{V} \int_0^{\tau}d\tau' R^2(\tau,\tau') \rho^2(\tau')
\end{equation}
Viewed as the beginning of an expansion in terms of $1/V$, the correction terms is $O(1/V)$ because the response function is of order unity. All further corrections like the two bubble diagram in \cref{Gamma_12_1} have additional $V^{-1}$ factors and thus give smaller corrections for large $V$. One could then be tempted to stop after the first correction in the equation for $\mu$ or equivalently density $\rho$, which is the one loop diagram in \cref{Gamma_12_1}. However, as we have seen, that correction only works for small values of $k_3 = j_3/V$, i.e.\ large volumes;  the SBR approach can be viewed as ``dressing'' the leading order correction by including all delayed two particle interactions. The final form of the SBR, \cref{Gamma_12_contTime,sigma_12_contTime}, is then stable even for large values of $k_3$. Notice that the SBR method is not a straightforward expansion in $1/V$ as there are diagrams that we have neglected that are of the same order as some that we have considered. For example, the two loop diagram that is included in SBR, of the form $k_3^3 \mu^2$ in \cref{Gamma_12_2}, makes a contribution of order $V^{-2}$ to the equation for the density. The diagram that we have neglected such as the first diagram of \cref{Gamma_13_neglect}, of the form $k_3^4 \mu^3$, is also of order $V^{-2}$. Thus SBR does include corrections of all orders in $V^{-1}$, but is not a systematic way to capture all corrections at each order.

We note finally that there are subtleties regarding the interpretation of the Doi-Peliti path integral in terms of density fluctuations~\cite{lefevre_dynamics_2007}. The $\phi$ field cannot be directly interpreted as a density. Nonetheless, performing a Cole-Hopf transformation on the $\phi,\tilde{\phi}$ in the action and terminating the expansion of the transformed action at $O(1/V)$ leads to a Langevin equation for the density~\cite{itakura_two_2010} that matches the system size expansion result by van Kampen~\cite{van_kampen_expansion_1976}. 
 In our approach, on the other hand, we calculate $\mu = \langle \phi \rangle$ directly and the mean density can be obtained afterwards simply by dividing by system volume $V$.
 				
	\subsection{Adding a back reaction: $A+A \rightleftharpoons A$}{\label{AAA_back_section}}
		
So far we have illustrated our approach with a system with baseline particle creation and destruction as well as coagulation $A+A\to A$. 
To generalize, we now add the back reaction  of particle branching, $A \xrightarrow{k'_3} A+ A$. We then have the following additional vertices in the interacting action: 		
			\begin{equation}{\label{AAA_back_vertices}}
			S_{\text{int}} = \dots + \Delta t\sum_\tau \left( \rule{0cm}{1cm}  \quad \quad
			\parbox{20mm}{
				\begin{fmfgraph*}(30,30)  
					\fmfleft{e1}
					\fmfright{e2}					
					\fmfv{label=$k_3',, \tau$, label.angle=105,label.dist=6}{v1}
					\fmfv{label=$\tilde{\phi}$ ,label.dist=0.5}{e1} 
					\fmfv{label=$\phi$ ,label.dist=0.8}{e2} 
					\fmf{fermion,tension=2}{v1,e1}					
					\fmf{plain,tension=2.5}{e2,v1}    
					\fmfv{decor.shape=circle,decor.filled=full, decor.size=2thick}{v1 }		
			\end{fmfgraph*}}	+ \quad \quad		\parbox{20mm}{\begin{fmfgraph*}(35,30)  
					\fmfleft{e1,e2}              
					\fmfright{e3}					
					\fmfv{label=$k_3',, \tau$, label.angle=45,label.dist=5}{v1}
					\fmfv{label=$ \tilde{\phi} $,label.dist=0.5}{e1} 
					\fmfv{label=$\tilde{\phi}$,label.dist=0.5}{e2} 
					\fmfv{label=$ \phi $,label.dist=0.5}{e3}   
					\fmf{fermion,tension=2}{v1,e1}    
					\fmf{fermion,tension=2}{v1,e2}					
					\fmf{plain,tension=3}{e3,v1}
					\fmfv{decor.shape=circle,decor.filled=full, decor.size=2thick}{v1 }		
			\end{fmfgraph*}} \quad\quad \right)
		\end{equation}
The first vertex is quadratic and could be absorbed in $S_0$, but we prefer to keep it in $S_{\text{int}}$ to avoid modifying the baseline. We now get non-vanishing contributions to the self-energy $\Sigma$ at zero mean, which without the back reaction was zero:
		 \begin{equation}
			\begin{split}
				\parbox{27mm}{
					\begin{fmfgraph*}(70,40)
						\fmfleft{e1}
						\fmfright{e2}
						\fmf{end_arrow_dash,tension=1.5}{v1,e1}
						\fmf{dashes,tension=2}{e2,v1}
						\fmfv{label=$\Sigma$,label.dist=0,decor.shape=circle, decor.filled=empty, decor.size= 8thick}{v1}
					\end{fmfgraph*}
				} = 
				\parbox{25mm}{
					\begin{fmfgraph*}(70,30)
						\fmfleft{e1}
						\fmfright{e2}
						\fmf{end_arrow_dash,tension=1.5}{v1,e1}
						\fmf{dashes,tension=1.5}{e2,v1}
						\fmfv{label.angle=90,decor.shape=circle, decor.filled=full, decor.size= 4thick}{v1}
					\end{fmfgraph*}	
				} +
				\parbox{38mm}{
					\begin{fmfgraph*}(100,30)
						\fmfleft{e1}
						\fmfright{e2}
						\fmf{end_arrow_dash,tension=1.5}{v1,e1}
						\fmf{fermion,left}{v2,v1}
						\fmf{fermion,right}{v2,v1}
						\fmf{dashes,tension=2}{e2,v2}
						\fmfv{label.angle=130,decor.shape=circle, decor.filled=full, decor.size= 4thick}{v1}
						\fmfv{label.angle=60,decor.shape=circle, decor.filled=full, decor.size= 4thick}{v2}
					\end{fmfgraph*}	
				} +
				\parbox{45mm}{
					\begin{fmfgraph*}(125,30)
						\fmfleft{e1}
						\fmfright{e2}
						\fmf{end_arrow_dash,tension=1.5}{v1,e1}
						\fmf{dashes,tension=2}{e2,v2}
						\fmf{fermion,left}{v3,v1}
						\fmf{fermion,right}{v3,v1}
						\fmf{fermion,left}{v2,v3}
						\fmf{fermion,right}{v2,v3}
						\fmfv{label.angle=125,decor.shape=circle, decor.filled=full, decor.size= 4thick}{v1}
						\fmfv{label.angle=180,decor.shape=circle, decor.filled=full, decor.size= 4thick}{v2}
						\fmfv{label.angle=180,decor.shape=circle, decor.filled=full, decor.size= 4thick}{v3}
					\end{fmfgraph*}
				} +\dots
			\end{split}
			\label{self_energy_back_reaction}
		\end{equation}	
This self-energy features in \cref{mu_star_equation} for the mean copy number. To $O(\alpha)$ it gives a correction term to \cref{AAA_MAK}, namely
		\begin{equation}
			\partial_\tau \mu(\tau) = \dots + \alpha k_3' \mu(\tau)
		\end{equation}
This is the MAK level of approximation and simply reflects the increase in particle number from each branching reaction.
Going to $O(\alpha^2)$, where previously we had \cref{AAA_alpha2}, we have the above MAK term and in addition 
		\begin{equation}
			\partial_\tau \mu(\tau) = \dots - 2\alpha^2 k_3 k_3' \int_0^{\tau} d\tau'\, R_0^2(\tau,\tau') \mu(\tau')
		\end{equation}
Continuing to higher orders one again has a geometric series that can be summed similarly to that in \cref{Gamma_12_contTime_bare}, giving
			\begin{equation}
			\Sigma(\tau,\tau') =   (\alpha k_3') \left( \delta(\tau-\tau') + 2\alpha k_3 R_0^2(\tau,\tau') \right)^{-1}
			\end{equation}
and the overall additional term in the equation of motion (\ref{resum_mu_bare},\ref{Gamma_12_contTime}) 
		\begin{equation}
			\partial_\tau \mu(\tau) = \dots + (\alpha k_3') \int_0^\tau d\tau' \left( \delta(\tau-\tau') + 2\alpha k_3 R^2(\tau,\tau') \right)^{-1} \mu(\tau')
		\end{equation}
	
Here we have again performed the self-consistent replacement of the bare response $R_0$ by the physical dressed response $R$. This implicitly includes diagrams that in principle contribute to the vertex functions $\Gamma^{1,m}$ with $m\geq 2$ but are not contained in our bubble diagram series.

For the physical self-energy $\Sigma^*$ we use the same approximation, i.e.\ we add to 
\cref{sigma_12_contTime} the sum of the diagrams in \cref{self_energy_back_reaction} with $R_0$ replaced by $R$, giving 
\begin{equation}
			\Sigma^*(\tau,\tau') = 
(-2\alpha k_3 \mu(\tau') + \alpha k_3')			\left( \delta(\tau-\tau') + 2\alpha k_3 R^2(\tau,\tau') \right)^{-1}
			\end{equation}
From this self-energy the physical dressed response is then determined by \cref{resum_R}.

Note that even with the self-consistency, there are some new $\Omega^{1,m \geq 2}$ made from the three legged vertex in \cref{AAA_back_vertices}  and other vertices of \cref{AAA_vertices} that are not accounted for within SBR.		
		
\begin{figure}[ht]
			\includegraphics[width=\linewidth]{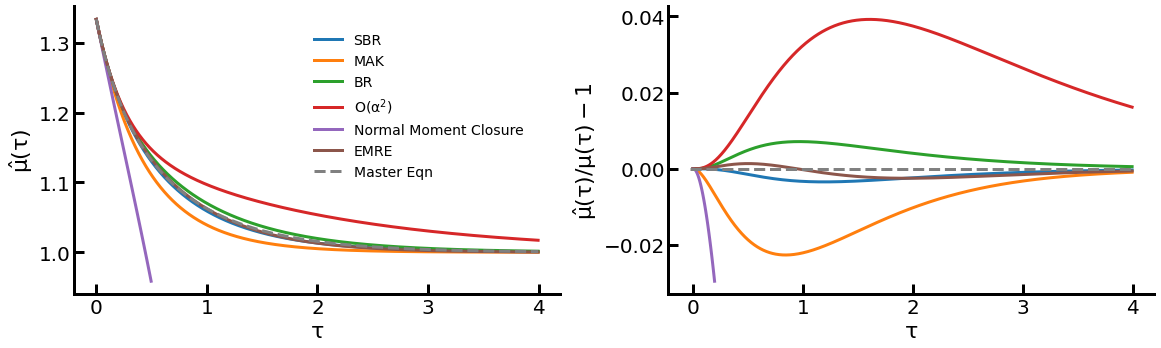}
			\caption{\small Dynamics of mean copy numbers $\mu$, $\hat{\mu}$ for the system with added branching reaction $A\to A+A$ 
				 (left); relative deviation (right) of different approximation methods $\hat{\mu}$ from the master equation solution $\mu$ (grey dashed line). SBR has a significantly lower error in the transient compared to the other methods. The normal moment closure method very quickly becomes unstable and predicts negative copy numbers. System parameters are given in \cref{AAA_back_section}.}
			\label{fig11}
		\end{figure}

\begin{figure}[ht]
			\includegraphics[width=\linewidth]{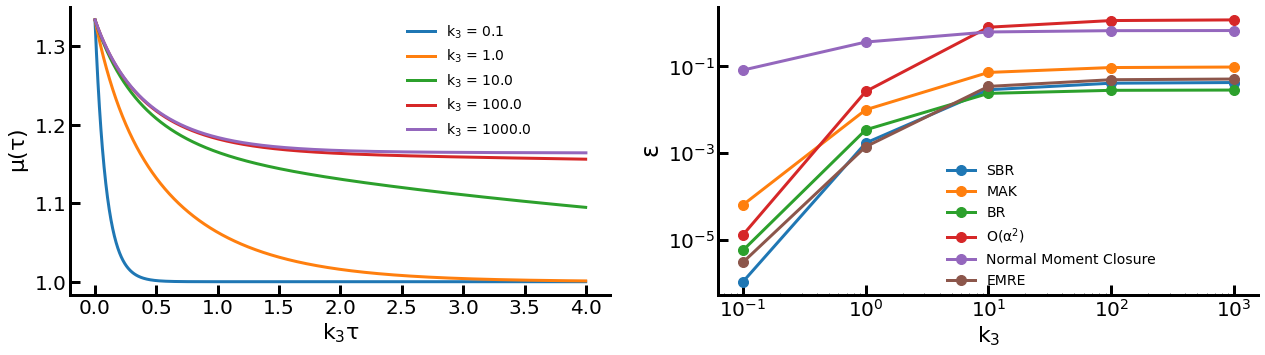}
			\caption{\small System with added branching reaction $A\to A+A$ for a range of coagulation and branching rates $k_3=k_3'$. 
				Mean copy number dynamics from the master equation (left) and time-averaged absolute relative deviation as defined in \cref{epsilon} of different approximation methods (right). Times and time ranges are scaled as in \cref{fig3}. The bubble resummed approaches (SBR and BR) perform similarly to EMRE and outperform all other methods over a large range of $k_3$ and $k_3'$. 
		}
			\label{fig12}
		\end{figure}

We now consider the above reaction system for the parameters $k_1 = 1$, $k_2=1$, $k_3 = 1$, $k_3'=1$, with time step $\Delta t =0.004$, total integration time $t = 4$ and $\mu(0) = 4/3$. As in \cref{numerics_AAA} we plot in \cref{fig11} the mean copy number dynamics as obtained from different approximation methods, and their respective relative deviations from the solution of the master equation. SBR again outperforms the other methods including MAK; normal moment closure diverges, and BR without self-consistency is second best to SBR. EMRE (at unit volume) has a similar performance as the SBR. These trends are confirmed in \cref{fig12} where we again change the perturbation parameter $k_3$, keeping $k_3=k_3'$, across four orders of magnitude, and plot the time-averaged absolute relative deviation as defined in \cref{epsilon} of the different approximation methods. Fig.~\ref{fig12}(left) shows the corresponding exact mean copy number dynamics as calculated from the master equation. Times and time ranges have been scaled as in \cref{fig3} and, as there, SBR and BR still perform well for large $k_3$ and $k_3'$ where the coagulation and branching reactions are dominant rather than small perturbations.
		
		\section{Multiple species interactions: $A+B \rightarrow C$}{\label{section_ABC}}
		
Having illustrated our approach so far for systems with a single molecular species, we now extend our considerations to multiple species. Unary reactions with conversion of one molecule to another like $A \rightarrow B$ will only produce two-legged vertices with one $\tilde{\phi}$ and one $\phi$ legs, like the first vertex of \cref{AAA_back_vertices}. These cannot appear in any vertex functions $\Gamma^{1,m}$ with $m\geq 2$ and just give rise to mean field ($O(\alpha)$) corrections that are straightforward to include. Binary reactions where two different molecules react, like $A+B \rightarrow A$, $A+B \rightarrow C$ or $A+B \rightarrow C+D$ require additional analysis that we will now present.

Focussing on the example $A+B\to C$ from now on, the interacting action as obtained from \cref{Hamiltonian_operator} after applying the Doi-Shift is 
		\begin{equation}
			\begin{split}
			S_{\text{int}} = \Delta t \sum_{\tau=\Delta t}^{t} &\left\lbrace k_{1A}\tilde{\phi}_A(\tau) + k_{1B}\tilde{\phi}_B(\tau) + k_{1C}\tilde{\phi}_C(\tau) + \right. \\
			 &\left. \phi_A(\tau_-) \phi_B(\tau_-) \left[ k_3 \tilde{\phi}_C(\tau) -k_3 \tilde{\phi}_A(\tau)  -k_3 \tilde{\phi}_B(\tau) -k_3 \tilde{\phi}_A(\tau)  \tilde{\phi}_B(\tau) \right] \right\rbrace	
		\end{split}
		\end{equation}
where the first line contains the particle creation reactions with rates $k_{1A}$, $k_{1B}$ and $k_{1C}$, respectively, and the second line the reaction $A+B\to C$ with rate $k_3$. 
Note that as in \cref{section_AAA} we treat the initial condition implicitly by considering a creation rate $k_{1i}(\tau) = k_{1i} + \bar{n}_{0i}\delta(\tau)$.
Diagrammatically we can represent the interacting action similarly to the previous single species case, but now using different colours to represent the field legs for different species, specifically $A \rightarrow \text{blue}$, $B \rightarrow \text{green}$, $C \rightarrow \text{red}$:
		\begin{equation}
			\begin{split}
			S_{\text{int}} = \Delta t\sum_\tau \left( \rule{0cm}{1cm}  
			\parbox{13mm}{
				\begin{fmfgraph*}(25,30)  
					\fmfleft{e1}
					\fmfright{v1}               
					\fmfv{label=$k_{1A},, \tau$, label.angle=105,label.dist=6}{v1}
					\fmf{end_arrow_two,tension=10,foreground=blue}{v1,e1}    
					\fmfv{decor.shape=circle,decor.filled=full, decor.size=4thick}{v1 }		
			\end{fmfgraph*}}	+  
			\parbox{13mm}{
			\begin{fmfgraph*}(25,30)  
				\fmfleft{e1}
				\fmfright{v1}               
				\fmfv{label=$k_{1B},, \tau$, label.angle=105,label.dist=6}{v1}
				\fmf{end_arrow_two,tension=10,foreground=green}{v1,e1}    
				\fmfv{decor.shape=circle,decor.filled=full, decor.size=4thick}{v1 }		
		\end{fmfgraph*}} +
		\parbox{13mm}{
		\begin{fmfgraph*}(25,30)  
			\fmfleft{e1}
			\fmfright{v1}               
			\fmfv{label=$k_{1C},, \tau$, label.angle=105,label.dist=6}{v1}
			\fmf{end_arrow_two,tension=10,foreground=red}{v1,e1}    
			\fmfv{decor.shape=circle,decor.filled=full, decor.size=4thick}{v1 }		
	\end{fmfgraph*}} + 
					\parbox{16mm}{\begin{fmfgraph*}(55,30)  
						\fmfleft{e1}              
						\fmfright{e2,e3} 
						\fmfv{label=$k_3,, \tau$, label.angle=90,label.dist=7}{v1}
						\fmf{end_arrow,tension=3,foreground=red}{v1,e1}    
						\fmf{plain,tension=2,foreground=blue}{e2,v1}
						\fmf{plain,tension=2,foreground=green}{e3,v1}
						\fmfv{decor.shape=circle,decor.filled=full, decor.size=4thick}{v1 }		
				\end{fmfgraph*}} +
				\parbox{16mm}{\begin{fmfgraph*}(55,30)  
					\fmfleft{e1}              
					\fmfright{e2,e3} 
					\fmfv{label=$-k_3,, \tau$, label.angle=90,label.dist=7}{v1}
					\fmf{end_arrow,tension=3,foreground=blue}{v1,e1}    
					\fmf{plain,tension=2,foreground=blue}{e2,v1}
					\fmf{plain,tension=2,foreground=green}{e3,v1}
					\fmfv{decor.shape=circle,decor.filled=full, decor.size=4thick}{v1 }		
			\end{fmfgraph*}} +
			\parbox{16mm}{\begin{fmfgraph*}(55,30)  
				\fmfleft{e1}              
				\fmfright{e2,e3} 
				\fmfv{label=$-k_3,, \tau$, label.angle=90,label.dist=7}{v1}
				\fmf{end_arrow,tension=3,foreground=green}{v1,e1}    
				\fmf{plain,tension=2,foreground=blue}{e2,v1}
				\fmf{plain,tension=2,foreground=green}{e3,v1}
				\fmfv{decor.shape=circle,decor.filled=full, decor.size=4thick}{v1 }		
		\end{fmfgraph*}} +	
		\parbox{20mm}{ \begin{fmfgraph*}(65,30)  
					\fmfleft{e1,e2}              
					\fmfright{e3,e4} 
					\fmfv{label=$-k_3,, \tau$, label.angle=90,label.dist=7}{v1}
					\fmf{end_arrow,tension=2,foreground=blue}{v1,e1}    
					\fmf{end_arrow,tension=2,foreground=green}{v1,e2} 
					\fmf{plain,tension=2,foreground=blue}{e3,v1}
					\fmf{plain,tension=2,foreground=green}{e4,v1}
					\fmfv{decor.shape=circle,decor.filled=full, decor.size=4thick}{v1 }		
			\end{fmfgraph*}}
		 \quad \right)
			\end{split}
		\end{equation}
	The vertex functions $\Gamma^{l,m}$ and their contractions $\Omega^{l,m}$ are now defined with extra lower indices that indicate which species the legs are associated with. For example, 
	 $\Gamma^{l,m}_{i_1 i_2 \dots i_l , j_1 j_2 \dots j_m}$ denotes that the vertex function $\Gamma$ has $l$ amputated $\tilde{\phi}$ legs associated with species $i_1,i_2,\dots,i_l$ and $m$ amputated $\phi$ legs associated with species $j_1,j_2,\dots,j_m$. 
	 $\Omega^{l,m}_{i_1 i_2 \dots i_l , j_1 j_2 \dots j_m}$ is similarly defined with the $\phi$ legs replaced by $\mu_{j_1}, \mu_{j_2},\dots \mu_{j_m}$ and summed over internal times.

\subsection{Vertex functions and equations of motion}
		
We begin with the bare propagator version of the theory. The baseline Hamiltonian $H_0$ decouples across species, so the bare response function $R_{0,ij} = 0$ when the two species indices $i$ and $j$ are different. This implies that in this flavour of approximation there are no mixed response functions, and $\phi$ and $\tilde{\phi}$ legs can connect only if they belong to the same species. We abbreviate the self-responses $R_{0,ii}$ as $R_{0i}$ below. 
		
The 1-point vertex function $\Gamma^{1,0}$ is the same as before, just for the different species, given by		
		\begin{equation}
			\Gamma^{1,0}_A = 
				\parbox{15mm}{
				\begin{fmfgraph*}(30,30)  
					\fmfleft{e1}
					\fmfright{v1}               
					\fmfv{label=$k_{1A}$, label.angle=105,label.dist=6}{v1}
					\fmf{end_arrow_dash_two,tension=10,foreground=blue}{v1,e1}    
					\fmfv{decor.shape=circle,decor.filled=full, decor.size=4thick}{v1 }		
			\end{fmfgraph*}}	\quad \quad
		\Gamma^{1,0}_B =
			\parbox{15mm}{
				\begin{fmfgraph*}(30,30)  
					\fmfleft{e1}
					\fmfright{v1}               
					\fmfv{label=$k_{1B}$, label.angle=105,label.dist=6}{v1}
					\fmf{end_arrow_dash_two,tension=10,foreground=green}{v1,e1}    
					\fmfv{decor.shape=circle,decor.filled=full, decor.size=4thick}{v1 }		
			\end{fmfgraph*}} \quad \quad
		\Gamma^{1,0}_C =
			\parbox{15mm}{
				\begin{fmfgraph*}(30,30)  
					\fmfleft{e1}
					\fmfright{v1}               
					\fmfv{label=$k_{1C}$, label.angle=105,label.dist=6}{v1}
					\fmf{end_arrow_dash_two,tension=10,foreground=red}{v1,e1}    
					\fmfv{decor.shape=circle,decor.filled=full, decor.size=4thick}{v1 }		
			\end{fmfgraph*}}
		\end{equation}
		The next vertex function $\Gamma^{1,2}$ is given by the following diagrams up to $O(\alpha^3)$: 
			\begin{equation}{\label{Gamma_A_AB}}
			\begin{split}
				\Gamma^{1,2}_{A,AB} = 	
				\parbox{20mm}{\begin{fmfgraph*}(55,30)  
						\fmfleft{e1}              
						\fmfright{e2,e3} 
						\fmfv{}{v1}
						\fmf{end_arrow_dash_two,tension=3,foreground=blue}{v1,e1}    
						\fmf{dashes,tension=2,foreground=blue}{e2,v1}
						\fmf{dashes,tension=2,foreground=green}{e3,v1}
						\fmfv{label=$-k_3$, label.angle=90,label.dist=6,decor.shape=circle,decor.filled=full, decor.size=4thick}{v1 }		
				\end{fmfgraph*}} +
				\parbox{30mm}{ 	\begin{fmfgraph*}(90,30)
						\fmfleft{e1}
						\fmfright{e2,e3}
						\fmf{end_arrow_dash_two,tension=2,foreground=blue}{v1,e1}
						\fmf{dashes,foreground=blue}{e2,v2}
						\fmf{dashes,foreground=green}{e3,v2}
						\fmf{plain_arrow,foreground=blue,left}{v2,v1}
						\fmf{plain_arrow,foreground=green,right}{v2,v1}
						\fmfv{label=$-k_3$, label.angle=90,label.dist=6,decor.shape=circle, decor.filled=full, decor.size= 4thick}{v1}
						\fmfv{label=$-k_3$, label.angle=90,label.dist=6,decor.shape=circle, decor.filled=full, decor.size= 4thick}{v2}
				\end{fmfgraph*}} +
				\parbox{45mm}{ 	\begin{fmfgraph*}(120,30)
						\fmfleft{e1}
						\fmfright{e2,e3}
						\fmf{end_arrow_dash_two,tension=2,foreground=blue}{v1,e1}
						\fmf{dashes,foreground=blue}{e2,v2}
						\fmf{dashes,foreground=green}{e3,v2}
						\fmf{plain_arrow,foreground=blue,left}{v2,v3}
						\fmf{plain_arrow,foreground=green,right}{v2,v3}
						\fmf{plain_arrow,foreground=blue,left}{v3,v1}
						\fmf{plain_arrow,foreground=green,right}{v3,v1}
						\fmfv{label=$-k_3$, label.angle=90,label.dist=6,decor.shape=circle, decor.filled=full, decor.size= 4thick}{v1}
						\fmfv{label=$-k_3$, label.angle=90,label.dist=6,decor.shape=circle, decor.filled=full, decor.size= 4thick}{v2}
						\fmfv{label=$-k_3$, label.angle=90,label.dist=6,decor.shape=circle, decor.filled=full, decor.size= 4thick}{v3}
				\end{fmfgraph*}} +\dots
			\end{split}
		\end{equation}

		\begin{equation}
		\begin{split}
			\Gamma^{1,2}_{B,AB} = 	
			\parbox{20mm}{\begin{fmfgraph*}(55,30)  
					\fmfleft{e1}              
					\fmfright{e2,e3} 
					\fmfv{}{v1}
					\fmf{end_arrow_dash_two,tension=3,foreground=green}{v1,e1}    
					\fmf{dashes,tension=2,foreground=blue}{e2,v1}
					\fmf{dashes,tension=2,foreground=green}{e3,v1}
					\fmfv{label=$-k_3$, label.angle=90,label.dist=6,decor.shape=circle,decor.filled=full, decor.size=4thick}{v1 }		
			\end{fmfgraph*}} +
			\parbox{30mm}{ 	\begin{fmfgraph*}(90,30)
					\fmfleft{e1}
					\fmfright{e2,e3}
					\fmf{end_arrow_dash_two,tension=2,foreground=green}{v1,e1}
					\fmf{dashes,foreground=blue}{e2,v2}
					\fmf{dashes,foreground=green}{e3,v2}
					\fmf{plain_arrow,foreground=blue,left}{v2,v1}
					\fmf{plain_arrow,foreground=green,right}{v2,v1}
					\fmfv{label=$-k_3$, label.angle=90,label.dist=6,decor.shape=circle, decor.filled=full, decor.size= 4thick}{v1}
					\fmfv{label=$-k_3$, label.angle=90,label.dist=6,decor.shape=circle, decor.filled=full, decor.size= 4thick}{v2}
			\end{fmfgraph*}} +
			\parbox{45mm}{ 	\begin{fmfgraph*}(120,30)
					\fmfleft{e1}
					\fmfright{e2,e3}
					\fmf{end_arrow_dash_two,tension=2,foreground=green}{v1,e1}
					\fmf{dashes,foreground=blue}{e2,v2}
					\fmf{dashes,foreground=green}{e3,v2}
					\fmf{plain_arrow,foreground=blue,left}{v2,v3}
					\fmf{plain_arrow,foreground=green,right}{v2,v3}
					\fmf{plain_arrow,foreground=blue,left}{v3,v1}
					\fmf{plain_arrow,foreground=green,right}{v3,v1}
					\fmfv{label=$-k_3$, label.angle=90,label.dist=6,decor.shape=circle, decor.filled=full, decor.size= 4thick}{v1}
					\fmfv{label=$-k_3$, label.angle=90,label.dist=6,decor.shape=circle, decor.filled=full, decor.size= 4thick}{v2}
					\fmfv{label=$-k_3$, label.angle=90,label.dist=6,decor.shape=circle, decor.filled=full, decor.size= 4thick}{v3}
			\end{fmfgraph*}} +\dots
		\end{split}
	\end{equation}	

	\begin{equation}
		\begin{split}
			\Gamma^{1,2}_{C,AB} = 	
			\parbox{20mm}{\begin{fmfgraph*}(55,30)  
					\fmfleft{e1}              
					\fmfright{e2,e3} 
					\fmfv{}{v1}
					\fmf{end_arrow_dash_two,tension=3,foreground=red}{v1,e1}    
					\fmf{dashes,tension=2,foreground=blue}{e2,v1}
					\fmf{dashes,tension=2,foreground=green}{e3,v1}
					\fmfv{label=$k_3$, label.angle=90,label.dist=6,decor.shape=circle,decor.filled=full, decor.size=4thick}{v1 }		
			\end{fmfgraph*}} +
			\parbox{30mm}{ 	\begin{fmfgraph*}(90,30)
					\fmfleft{e1}
					\fmfright{e2,e3}
					\fmf{end_arrow_dash_two,tension=2,foreground=red}{v1,e1}
					\fmf{dashes,foreground=blue}{e2,v2}
					\fmf{dashes,foreground=green}{e3,v2}
					\fmf{plain_arrow,foreground=blue,left}{v2,v1}
					\fmf{plain_arrow,foreground=green,right}{v2,v1}
					\fmfv{label=$k_3$, label.angle=90,label.dist=6,decor.shape=circle, decor.filled=full, decor.size= 4thick}{v1}
					\fmfv{label=$-k_3$, label.angle=90,label.dist=6,decor.shape=circle, decor.filled=full, decor.size= 4thick}{v2}
			\end{fmfgraph*}} +
			\parbox{45mm}{ 	\begin{fmfgraph*}(120,30)
					\fmfleft{e1}
					\fmfright{e2,e3}
					\fmf{end_arrow_dash_two,tension=2,foreground=red}{v1,e1}
					\fmf{dashes,foreground=blue}{e2,v2}
					\fmf{dashes,foreground=green}{e3,v2}
					\fmf{plain_arrow,foreground=blue,left}{v2,v3}
					\fmf{plain_arrow,foreground=green,right}{v2,v3}
					\fmf{plain_arrow,foreground=blue,left}{v3,v1}
					\fmf{plain_arrow,foreground=green,right}{v3,v1}
					\fmfv{label=$k_3$, label.angle=90,label.dist=6,decor.shape=circle, decor.filled=full, decor.size= 4thick}{v1}
					\fmfv{label=$-k_3$, label.angle=90,label.dist=6,decor.shape=circle, decor.filled=full, decor.size= 4thick}{v2}
					\fmfv{label=$-k_3$, label.angle=90,label.dist=6,decor.shape=circle, decor.filled=full, decor.size= 4thick}{v3}
			\end{fmfgraph*}} +\dots
		\end{split}
	\end{equation}
		Using these we can construct the elements of $\Omega^{1,2}$. For example $\Omega^{1,2}_{A,AB}$ is to leading order
		\begin{equation}
			\Omega^{1,2}_{A,AB}(\tau)\Big|_{\alpha} = -\alpha k_3 \mu_A(\tau_-) \mu_B(\tau_-)
		\end{equation} 
and $\Omega^{1,2}_{B,AB}$, $\Omega^{1,2}_{C,AB}$ are obtained analogously.
Neglecting $\Omega^{1,m}$ with $m\geq 3$, we have the mean copy number equation of motion
		\begin{equation}{\label{ABC_mean_general_eqn}}
			(\partial_\tau + k_{2i})\mu_i(\tau) = \alpha k_{1i} + \Omega^{1,2}_{i,AB}(\tau)
		\end{equation}
for each species $i=A,B,C$. To $O(\alpha)$ this gives the MAK equations as expected:
		\begin{equation}{\label{ABC_MAK}}
			\begin{split}
				\partial_\tau \mu_A(\tau) &= \alpha k_{1A} - k_{2A} \mu_A(\tau) - \alpha k_3 \mu_A(\tau) \mu_B(\tau) \\
				\partial_\tau \mu_B(\tau) &= \alpha k_{1B} - k_{2B} \mu_B(\tau) - \alpha k_3 \mu_A(\tau) \mu_B(\tau)  \\
				\partial_\tau \mu_C(\tau) &=  \alpha k_{1C} - k_{2C} \mu_C(\tau) + \alpha k_3 \mu_A(\tau) \mu_B(\tau) \\
			\end{split}
		\end{equation}
Looking then at the first non-trivial corrections, of $O(\alpha^2)$, we have from the one loop diagram of \cref{Gamma_A_AB}
		\begin{equation}
			\Omega^{1,2}_{A,AB}(\tau)\Big|_{\alpha^2} = (-\alpha k_3)^2 \Delta t \sum_{\tau'} R_{0A}(\tau_-,\tau') R_{0B}(\tau_-,\tau') \mu_A(\tau'_-) \mu_B(\tau'_-)
		\end{equation}
Together with the analogous results for other species this yields the continuous time equations for the means to $O(\alpha^2)$ as
		\begin{equation}{\label{ABC_bare_mean_eqn}}
			\begin{split}
				\partial_\tau \mu_A(\tau) &= \alpha k_{1A} - k_{2A} \mu_A(\tau) - \alpha k_3 \mu_A(\tau) \mu_B(\tau) + (-\alpha k_3)^2 \int_0^\tau d\tau' R_{0A}(\tau,\tau') R_{0B}(\tau,\tau')  \mu_A(\tau') \mu_B(\tau') \\
				\partial_\tau \mu_B(\tau) &= \alpha k_{1B} - k_{2B} \mu_B(\tau) - \alpha k_3 \mu_A(\tau) \mu_B(\tau) + (-\alpha k_3)^2 \int_0^\tau d\tau' R_{0A}(\tau,\tau') R_{0B}(\tau,\tau')  \mu_A(\tau') \mu_B(\tau') \\
				\partial_\tau \mu_C(\tau) &= \alpha k_{1C} - k_{2C} \mu_C(\tau) + \alpha k_3 \mu_A(\tau) \mu_B(\tau) - (\alpha k_3)^2 \int_0^\tau d\tau' R_{0A}(\tau,\tau') R_{0B}(\tau,\tau')  \mu_A(\tau') \mu_B(\tau') \\
			\end{split}
		\end{equation}
Following our approach for the single species $A+A \rightarrow A$ system, we will not explicitly include $\Omega^{1,m}$ terms with $m\geq 3$ and concentrate instead on the bubble series and its resummation.
		
		\subsection{Self-consistent response function approximation}
		
		\subsubsection{Using single species response functions}
		
The above expansion to $O(\alpha^2)$ can be improved following our previous strategy: we sum an infinite series of bubble diagrams, and we self-consistently replace the bare responses $R_{0i}$ by the physical dressed responses $R_i$. For $\Omega^{1,2}_{A,AB}$ this means
			\begin{equation}
			\begin{split}
				\Omega^{1,2}_{A,AB} = 	
				\parbox{20mm}{\begin{fmfgraph*}(55,30)  
						\fmfleft{e1}              
						\fmfright{e2,e3} 
						\fmfv{}{v1}
						\fmf{end_arrow_dash_two,tension=3,foreground=blue}{v1,e1}    
						\fmf{dashes,tension=2,foreground=blue}{e2,v1}
						\fmf{dashes,tension=2,foreground=green}{e3,v1}
						\fmfv{decor.shape=circle,decor.filled=full, decor.size=4thick}{v1 }	
						\fmfv{decor.shape=circle, decor.filled=empty, decor.size= 7thick,foreground=blue}{e2}
						\fmfv{decor.shape=circle, decor.filled=empty, decor.size= 7thick,foreground=green}{e3}
				\end{fmfgraph*}} +
				\parbox{30mm}{ 	\begin{fmfgraph*}(90,30)
						\fmfleft{e1}
						\fmfright{e2,e3}
						\fmf{end_arrow_dash_two,tension=2,foreground=blue}{v1,e1}
						\fmf{dashes,foreground=blue}{e2,v2}
						\fmf{dashes,foreground=green}{e3,v2}
						\fmf{dbl_plain_arrow,foreground=blue,left}{v2,v1}
						\fmf{dbl_plain_arrow,foreground=green,right}{v2,v1}
						\fmfv{decor.shape=circle, decor.filled=full, decor.size= 4thick}{v1}
						\fmfv{decor.shape=circle, decor.filled=full, decor.size= 4thick}{v2}
						\fmfv{decor.shape=circle, decor.filled=empty, decor.size= 7thick,foreground=blue}{e2}
						\fmfv{decor.shape=circle, decor.filled=empty, decor.size= 7thick,foreground=green}{e3}
				\end{fmfgraph*}} +
				\parbox{45mm}{ 	\begin{fmfgraph*}(120,30)
						\fmfleft{e1}
						\fmfright{e2,e3}
						\fmf{end_arrow_dash_two,tension=2,foreground=blue}{v1,e1}
						\fmf{dashes,foreground=blue}{e2,v2}
						\fmf{dashes,foreground=green}{e3,v2}
						\fmf{dbl_plain_arrow,foreground=blue,left}{v2,v3}
						\fmf{dbl_plain_arrow,foreground=green,right}{v2,v3}
						\fmf{dbl_plain_arrow,foreground=blue,left}{v3,v1}
						\fmf{dbl_plain_arrow,foreground=green,right}{v3,v1}
						\fmfv{decor.shape=circle, decor.filled=full, decor.size= 4thick}{v1}
						\fmfv{decor.shape=circle, decor.filled=full, decor.size= 4thick}{v2}
						\fmfv{decor.shape=circle, decor.filled=full, decor.size= 4thick}{v3}
						\fmfv{decor.shape=circle, decor.filled=empty, decor.size= 7thick,foreground=blue}{e2}
						\fmfv{decor.shape=circle, decor.filled=empty, decor.size= 7thick,foreground=green}{e3}
				\end{fmfgraph*}} +\dots
			\end{split}
		\end{equation}
and summing the geometric series gives 
	 	\begin{equation}
	 		{\label{omega_A_single_resum}}
	 		\Omega^{1,2}_{A,AB}(\tau) = (-\alpha k_3) \int_0^\tau d\tau' \left( \delta(\tau-\tau') + \alpha k_3 R_A(\tau,\tau')R_B(\tau,\tau') \right)^{-1} \mu_A(\tau') \mu_B(\tau')
	 	\end{equation}
The equation of motion for the mean copy number $\mu_A$ of species $A$ is then obtained by inserting this into \cref{ABC_mean_general_eqn}, with analogous expressions for the other species.

The dressed response functions at the nonzero, physical means are determined as before from the Feynman-Dyson equation using the self-energy $\Sigma^*_{ij}$. This now also carries two species indices.
Since we are considering only single species response functions, we only use $\Sigma_{ij}^*$ with $i=j$ and expand 
			\begin{equation}
				R_{i} = R_{0i} + R_{0i} \Sigma_{ii}^* R_{0i} + R_{0i} \Sigma_{ii}^* R_{0i} \Sigma_{ii}^* R_{0i} +\dots
				\label{R_0i_Dyson_series}
			\end{equation}
			which yields in continuous time
			\begin{equation}{\label{FD_eqn_single}}
					\partial_{\tau} R_{i}(\tau,\tau') = \delta(\tau-\tau') - k_{2i} R_{i}(\tau,\tau') +\int d\tau''\, \Sigma_{ii}^*(\tau,\tau'') R_{i}(\tau'',\tau') 
			\end{equation}
If we then express e.g.\ $\Sigma^*_{AA}$ using again self-consistently replaced dressed responses $R$ we obtain to $O(\alpha^3)$, 
			\begin{equation}
			\setlength{\jot}{20pt}
				\begin{split}
					\Sigma^*_{AA} &= 
						\parbox{27mm}{
						\begin{fmfgraph*}(70,40)
							\fmfleft{e1}
							\fmfright{e2}
							\fmf{end_arrow_dash,tension=1.5,foreground=blue}{v1,e1}
							\fmf{dashes,tension=2,foreground=blue}{e2,v1}
							\fmfv{label=$\Sigma^*$,label.dist=0,decor.shape=circle, decor.filled=empty, decor.size= 8thick}{v1}
						\end{fmfgraph*}} = 
						\parbox{27mm}{
						\begin{fmfgraph*}(70,40)
							\fmfleft{e1}
							\fmfright{e2}
							\fmf{end_arrow_dash,tension=1.5,foreground=blue}{v1,e1}
							\fmf{dashes,tension=1.5,foreground=blue}{e2,v1}
							\fmfv{decor.shape=circle, decor.filled=full, decor.size= 4thick}{v1}
							\fmffreeze
							\fmfbottom{e3}
							\fmf{dashes,tension=1,foreground=green}{v1,e3}
							\fmfv{decor.shape=circle, decor.filled=empty, decor.size= 7thick,foreground=green}{e3}
						\end{fmfgraph*}}	+
					\parbox{32mm}{
					\begin{fmfgraph*}(90,40)
						\fmfleft{e1}
						\fmfright{e2}
						\fmf{end_arrow_dash,tension=1.,foreground=blue}{v1,e1}
						\fmf{dbl_plain_arrow,tension=1.2,foreground=green}{v2,v1}
						\fmf{dashes,tension=1.5,foreground=blue}{v2,e2}
						\fmfv{decor.shape=circle, decor.filled=full, decor.size= 4thick}{v1}
						\fmfv{decor.shape=circle, decor.filled=full, decor.size= 4thick}{v2}
						\fmffreeze
						\fmfbottom{e4,e5}
						\fmfforce{(0.4w,0h)}{e4}
						\fmfforce{(0.75w,0h)}{e5}
						\fmf{dashes,tension=1,foreground=blue}{v1,e4}
						\fmfv{decor.shape=circle, decor.filled=empty, decor.size= 7thick,foreground=blue}{e4}
						\fmf{dashes,tension=1,foreground=green}{v2,e5}
						\fmfv{decor.shape=circle, decor.filled=empty, decor.size= 7thick,foreground=green}{e5}
					\end{fmfgraph*}	
					} +
				\parbox{35mm}{ 	\begin{fmfgraph*}(90,30)
						\fmfleft{e1}
						\fmfright{e2}
						\fmf{end_arrow_dash,tension=2,foreground=blue}{v1,e1}
						\fmf{dashes,tension=2,foreground=blue}{e2,v2}
						\fmf{dbl_plain_arrow,foreground=blue,left}{v2,v1}
						\fmf{dbl_plain_arrow,foreground=green,right}{v2,v1}
						\fmfv{decor.shape=circle, decor.filled=full, decor.size= 4thick}{v1}
						\fmfv{decor.shape=circle, decor.filled=full, decor.size= 4thick}{v2}
						\fmffreeze
						\fmfbottom{e3}
						\fmfforce{(0.8w,0h)}{e3}
						\fmf{dashes,foreground=green}{e3,v2}
						\fmfv{decor.shape=circle, decor.filled=empty, decor.size= 7thick,foreground=green}{e3}
				\end{fmfgraph*}} \\&+
				\parbox{40mm}{
				\begin{fmfgraph*}(110,40)
					\fmfleft{e1}
					\fmfright{e2}
					\fmf{end_arrow_dash,tension=1.,foreground=blue}{v1,e1}
					\fmf{dbl_plain_arrow,tension=1.2,foreground=green}{v2,v1}
					\fmf{dbl_plain_arrow,tension=1.2,foreground=green}{v3,v2}
					\fmf{dashes,tension=1.5,foreground=blue}{v3,e2}
					\fmfv{decor.shape=circle, decor.filled=full, decor.size= 4thick}{v1}
					\fmfv{decor.shape=circle, decor.filled=full, decor.size= 4thick}{v2}
					\fmfv{decor.shape=circle, decor.filled=full, decor.size= 4thick}{v3}
					\fmffreeze
					\fmfbottom{e4,e5,e6}
					\fmfforce{(0.3w,0h)}{e4}
					\fmfforce{(0.55w,0h)}{e5}
					\fmfforce{(0.81w,0h)}{e6}
					\fmf{dashes,tension=1,foreground=blue}{v1,e4}
					\fmfv{decor.shape=circle, decor.filled=empty, decor.size= 7thick,foreground=blue}{e4}
					\fmf{dashes,tension=1,foreground=blue}{v2,e5}
					\fmfv{decor.shape=circle, decor.filled=empty, decor.size= 7thick,foreground=blue}{e5}
					\fmf{dashes,tension=1,foreground=green}{v3,e6}
					\fmfv{decor.shape=circle, decor.filled=empty, decor.size= 7thick,foreground=green}{e6}
				\end{fmfgraph*}	
			} + 
			\parbox{45mm}{ 	\begin{fmfgraph*}(120,30)
					\fmfleft{e1}
					\fmfright{e2}
					\fmf{end_arrow_dash,tension=2,foreground=blue}{v1,e1}
					\fmf{dashes,tension=2,foreground=blue}{e2,v3}
					\fmf{dbl_plain_arrow,foreground=blue,left}{v2,v1}
					\fmf{dbl_plain_arrow,foreground=green,right}{v2,v1}
					\fmf{dbl_plain_arrow,foreground=blue,left}{v3,v2}
					\fmf{dbl_plain_arrow,foreground=green,right}{v3,v2}
					\fmfv{decor.shape=circle, decor.filled=full, decor.size= 4thick}{v1}
					\fmfv{decor.shape=circle, decor.filled=full, decor.size= 4thick}{v2}
					\fmfv{decor.shape=circle, decor.filled=full, decor.size= 4thick}{v3}
					\fmffreeze
					\fmfbottom{e3}
					\fmfforce{(0.9w,0h)}{e3}
					\fmf{dashes,foreground=green}{e3,v3}
					\fmfv{decor.shape=circle, decor.filled=empty, decor.size= 7thick,foreground=green}{e3}
			\end{fmfgraph*}} 
			+ 			\parbox{45mm}{ 	\begin{fmfgraph*}(120,40)
					\fmfleft{e1}
					\fmfright{e2}
					\fmf{end_arrow_dash,tension=2,foreground=blue}{v1,e1}
					\fmf{dashes,tension=2,foreground=blue}{e2,v3}
					\fmf{dbl_plain_arrow,foreground=green,tension=2}{v2,v1}
					\fmf{dbl_plain_arrow,foreground=blue,left}{v3,v2}
					\fmf{dbl_plain_arrow,foreground=green,right}{v3,v2}
					\fmfv{decor.shape=circle, decor.filled=full, decor.size= 4thick}{v1}
					\fmfv{decor.shape=circle, decor.filled=full, decor.size= 4thick}{v2}
					\fmfv{decor.shape=circle, decor.filled=full, decor.size= 4thick}{v3}
					\fmffreeze
					\fmfbottom{e4,e3}
					\fmfforce{(0.9w,0h)}{e3}
					\fmfforce{(0.25w,0h)}{e4}
					\fmf{dashes,foreground=green}{e3,v3}
					\fmf{dashes,foreground=blue}{e4,v1}
					\fmfv{decor.shape=circle, decor.filled=empty, decor.size= 7thick,foreground=green}{e3}
					\fmfv{decor.shape=circle, decor.filled=empty, decor.size= 7thick,foreground=blue}{e4}
			\end{fmfgraph*}} 
			\\&+ 	\parbox{45mm}{ 	\begin{fmfgraph*}(120,40)
					\fmfleft{e1}
					\fmfright{e2}
					\fmf{end_arrow_dash,tension=2,foreground=blue}{v1,e1}
					\fmf{dashes,tension=2,foreground=blue}{e2,v3}
					\fmf{dbl_plain_arrow,foreground=blue,left}{v2,v1}
					\fmf{dbl_plain_arrow,foreground=green,right}{v2,v1}
					\fmf{dbl_plain_arrow,foreground=green,tension=2}{v3,v2}
					\fmfv{decor.shape=circle, decor.filled=full, decor.size= 4thick}{v1}
					\fmfv{decor.shape=circle, decor.filled=full, decor.size= 4thick}{v2}
					\fmfv{decor.shape=circle, decor.filled=full, decor.size= 4thick}{v3}
					\fmffreeze
					\fmfbottom{e4,e3}
					\fmfforce{(0.6w,0h)}{e4}
					\fmfforce{(0.75w,0h)}{e3}
					\fmf{dashes,foreground=blue}{e4,v2}
					\fmf{dashes,foreground=green}{e3,v3}
					\fmfv{decor.shape=circle, decor.filled=empty, decor.size= 7thick,foreground=green}{e3}
					\fmfv{decor.shape=circle, decor.filled=empty, decor.size= 7thick,foreground=blue}{e4}
			\end{fmfgraph*}} 
			+\dots
				\end{split}
			\end{equation}
Note that because in the expansion \cref{R_0i_Dyson_series} we have only included response function factors for the {\em same} species $i$, we have to include here diagrams like the second and fourth that contain single response function links for {\em other} species. In other words, in the present approach a 1PI diagram for $\Sigma^*_{ii}$ has to be defined as one that cannot be split in two by cutting a response line of the same species $i$. The values of the first four diagrams of the above series term by term are 
		\begin{align}
			\Sigma_{AA}^*(\tau,\tau') &= -\frac{\delta_{\tau_-,\tau'}}{\Delta t} \alpha k_3 \mu_B(\tau') + (-\alpha k_3)^2 \mu_A(\tau_-) R_B(\tau_-,\tau'_+) \mu_B(\tau') + (-\alpha k_3)^2 R_A(\tau_-,\tau'_+) R_B(\tau_-,\tau'_+) \mu_B(\tau')  \nonumber
			\\&+ (-\alpha k_3)^3 \sum_{\tau''}  \mu_A(\tau_-) R_B(\tau_-,\tau'') \mu_A(\tau''_-) R_B(\tau''_-,\tau'_+) \mu_B(\tau')  + \dots
		\end{align} 
Following the approach in \cref{section_resum} this series can again be summed to all orders to obtain 
		 \begin{equation}
		 	{\label{sigma_AA_single_resum}}
		 	\begin{split}
		 	\Sigma^*_{AA}(\tau,\tau') &= (-\alpha k_3) \left[ \delta(\tau-\tau') + \alpha k_3 ( \mu_A(\tau) R_B(\tau,\tau') + R_A(\tau,\tau') )R_B(\tau,\tau')  \right]^{-1} \mu_B(\tau') \\
		 	\end{split}
	 	\end{equation}
Together with the corresponding expressions for $\Sigma^*_{BB}$ and $\Sigma^*_{CC}$ we then have a closed system of self-consistent equations for the means $\mu_i(\tau)$ and physical response functions $R_i(\tau,\tau')$ of all species $i$. 
For e.g.\ species $A$ one integrates \cref{ABC_mean_general_eqn} and \cref{FD_eqn_single} with $\Omega$ and $\Sigma$ given by \cref{sigma_AA_single_resum,omega_A_single_resum}. We call this approximation SBR-S where the additional S indicates that we are using single species responses only. Note that the $\Omega^{1,2}$ contribution is again a memory term: it integrates past values of the product $\mu_A\mu_B$, which is exactly the MAK term for the reaction $A+B\to C$, weighted by a self-consistently determined memory function.

\subsubsection{Including mixed response functions}	

The SBR-S approximation can be improved further by including mixed response functions, $R_{ij}$ for $i\neq j$. The resulting diagrammatic series for $\Omega^{1,2}_{A,AB}$ is, to $O(\alpha^3)$:
			\begin{equation}
				\setlength{\jot}{20pt}
				\begin{split}
					\Omega^{1,2}_{A,AB} &= 	
					\parbox{20mm}{\begin{fmfgraph*}(55,30)  
							\fmfleft{e1}              
							\fmfright{e2,e3} 
							\fmfv{}{v1}
							\fmf{end_arrow_dash_two,tension=3,foreground=blue}{v1,e1}    
							\fmf{dashes,tension=2,foreground=blue}{e2,v1}
							\fmf{dashes,tension=2,foreground=green}{e3,v1}
							\fmfv{decor.shape=circle,decor.filled=full, decor.size=4thick}{v1 }	
							\fmfv{decor.shape=circle, decor.filled=empty, decor.size= 7thick,foreground=blue}{e2}
							\fmfv{decor.shape=circle, decor.filled=empty, decor.size= 7thick,foreground=green}{e3}
					\end{fmfgraph*}} +
					\parbox{30mm}{ 	\begin{fmfgraph*}(90,30)
							\fmfleft{e1}
							\fmfright{e2,e3}
							\fmf{end_arrow_dash_two,tension=2,foreground=blue}{v1,e1}
							\fmf{dashes,foreground=blue}{e2,v2}
							\fmf{dashes,foreground=green}{e3,v2}
							\fmf{dbl_plain_arrow,foreground=blue,left}{v2,v1}
							\fmf{dbl_plain_arrow,foreground=green,right}{v2,v1}
							\fmfv{decor.shape=circle, decor.filled=full, decor.size= 4thick}{v1}
							\fmfv{decor.shape=circle, decor.filled=full, decor.size= 4thick}{v2}
							\fmfv{decor.shape=circle, decor.filled=empty, decor.size= 7thick,foreground=blue}{e2}
							\fmfv{decor.shape=circle, decor.filled=empty, decor.size= 7thick,foreground=green}{e3}
					\end{fmfgraph*}} +
					\parbox{30mm}{ 	\begin{fmfgraph*}(90,30)
							\fmfleft{e1}
							\fmfright{e2,e3}
							\fmf{end_arrow_dash_two,tension=2,foreground=blue}{v1,e1}
							\fmf{dashes,foreground=blue}{e2,v2}
							\fmf{dashes,foreground=green}{e3,v2}
							\fmf{mixed_response_BG,left}{v2,v1}
							\fmf{mixed_response_GB,right}{v2,v1}
							\fmfv{decor.shape=circle, decor.filled=full, decor.size= 4thick}{v1}
							\fmfv{decor.shape=circle, decor.filled=full, decor.size= 4thick}{v2}
							\fmfv{decor.shape=circle, decor.filled=empty, decor.size= 7thick,foreground=blue}{e2}
							\fmfv{decor.shape=circle, decor.filled=empty, decor.size= 7thick,foreground=green}{e3}
					\end{fmfgraph*}} +
					\parbox{45mm}{ 	\begin{fmfgraph*}(120,30)
							\fmfleft{e1}
							\fmfright{e2,e3}
							\fmf{end_arrow_dash_two,tension=2,foreground=blue}{v1,e1}
							\fmf{dashes,foreground=blue}{e2,v2}
							\fmf{dashes,foreground=green}{e3,v2}
							\fmf{dbl_plain_arrow,foreground=blue,left}{v2,v3}
							\fmf{dbl_plain_arrow,foreground=green,right}{v2,v3}
							\fmf{dbl_plain_arrow,foreground=blue,left}{v3,v1}
							\fmf{dbl_plain_arrow,foreground=green,right}{v3,v1}
							\fmfv{decor.shape=circle, decor.filled=full, decor.size= 4thick}{v1}
							\fmfv{decor.shape=circle, decor.filled=full, decor.size= 4thick}{v2}
							\fmfv{decor.shape=circle, decor.filled=full, decor.size= 4thick}{v3}
							\fmfv{decor.shape=circle, decor.filled=empty, decor.size= 7thick,foreground=blue}{e2}
							\fmfv{decor.shape=circle, decor.filled=empty, decor.size= 7thick,foreground=green}{e3}
					\end{fmfgraph*}} \\
					&+ \parbox{42mm}{ 	\begin{fmfgraph*}(120,30)
							\fmfleft{e1}
							\fmfright{e2,e3}
							\fmf{end_arrow_dash_two,tension=2,foreground=blue}{v1,e1}
							\fmf{dashes,foreground=blue}{e2,v2}
							\fmf{dashes,foreground=green}{e3,v2}
							\fmf{dbl_plain_arrow,foreground=blue,left}{v2,v3}
							\fmf{dbl_plain_arrow,foreground=green,right}{v2,v3}
							\fmf{mixed_response_BG,left}{v3,v1}
							\fmf{mixed_response_GB,right}{v3,v1}
							\fmfv{decor.shape=circle, decor.filled=full, decor.size= 4thick}{v1}
							\fmfv{decor.shape=circle, decor.filled=full, decor.size= 4thick}{v2}
							\fmfv{decor.shape=circle, decor.filled=full, decor.size= 4thick}{v3}
							\fmfv{decor.shape=circle, decor.filled=empty, decor.size= 7thick,foreground=blue}{e2}
							\fmfv{decor.shape=circle, decor.filled=empty, decor.size= 7thick,foreground=green}{e3}
					\end{fmfgraph*}} +
					\parbox{42mm}{ 	\begin{fmfgraph*}(120,30)
							\fmfleft{e1}
							\fmfright{e2,e3}
							\fmf{end_arrow_dash_two,tension=2,foreground=blue}{v1,e1}
							\fmf{dashes,foreground=blue}{e2,v2}
							\fmf{dashes,foreground=green}{e3,v2}
							\fmf{dbl_plain_arrow,foreground=blue,left}{v3,v1}
							\fmf{dbl_plain_arrow,foreground=green,right}{v3,v1}
							\fmf{mixed_response_BG,left}{v2,v3}
							\fmf{mixed_response_GB,right}{v2,v3}
							\fmfv{decor.shape=circle, decor.filled=full, decor.size= 4thick}{v1}
							\fmfv{decor.shape=circle, decor.filled=full, decor.size= 4thick}{v2}
							\fmfv{decor.shape=circle, decor.filled=full, decor.size= 4thick}{v3}
							\fmfv{decor.shape=circle, decor.filled=empty, decor.size= 7thick,foreground=blue}{e2}
							\fmfv{decor.shape=circle, decor.filled=empty, decor.size= 7thick,foreground=green}{e3}
					\end{fmfgraph*}} +
					\parbox{42mm}{ 	\begin{fmfgraph*}(120,30)
							\fmfleft{e1}
							\fmfright{e2,e3}
							\fmf{end_arrow_dash_two,tension=2,foreground=blue}{v1,e1}
							\fmf{dashes,foreground=blue}{e2,v2}
							\fmf{dashes,foreground=green}{e3,v2}
							\fmf{mixed_response_BG,foreground=blue,left}{v3,v1}
							\fmf{mixed_response_GB,foreground=green,right}{v3,v1}
							\fmf{mixed_response_BG,left}{v2,v3}
							\fmf{mixed_response_GB,right}{v2,v3}
							\fmfv{decor.shape=circle, decor.filled=full, decor.size= 4thick}{v1}
							\fmfv{decor.shape=circle, decor.filled=full, decor.size= 4thick}{v2}
							\fmfv{decor.shape=circle, decor.filled=full, decor.size= 4thick}{v3}
							\fmfv{decor.shape=circle, decor.filled=empty, decor.size= 7thick,foreground=blue}{e2}
							\fmfv{decor.shape=circle, decor.filled=empty, decor.size= 7thick,foreground=green}{e3}
					\end{fmfgraph*}}
					+\dots
				\end{split}
			\end{equation}
Here the half blue and half green double lines represent the mixed dressed response functions $R_{AB}$ and $R_{BA}$, with the species indices being in the same order as the colours of the double lines. The first additional term compared to the treatment in the previous section occurs at $O(\alpha^2)$ and is
			\begin{equation}
				\Omega^{1,2}_{A,AB}(\tau)\Big|_{\alpha^2} = \dots + (-\alpha k_3)^2 \Delta t \sum_{\tau'} R_{AB}(\tau_-,\tau') R_{BA}(\tau_-,\tau') \mu_A(\tau'_-) \mu_B(\tau'_-)
			\end{equation}
As before we can resum all higher orders in $\alpha$ to get, in continuous time,
\begin{equation}
{\label{omega_A_resum}}
	\Omega^{1,2}_{A,AB}(\tau) = (-\alpha k_3) \int_0^\tau d\tau' \left\{ \delta(\tau-\tau') + \alpha k_3  \left[ R_{AA}(\tau,\tau')R_{BB}(\tau,\tau') + R_{AB}(\tau,\tau')R_{BA}(\tau,\tau') \right] \right\}^{-1} \mu_A(\tau') \mu_B(\tau')
\end{equation}
which has to be inserted into \cref{ABC_mean_general_eqn} to get the final equation of motion for the mean copy number of species $A$; the other species $B, C$ can be treated in exactly the same manner.

To find the mixed response functions $R_{AB}$ and $R_{BA}$ that appear above, we now also need the general version of the Feynman-Dyson equation \cref{Feynman_star}. This is obtained by treating the species index analogously to the time indices, giving 
		\begin{equation}{\label{FD_eqn_mixed}}
			(\partial_\tau + k_{2i}) R_{ij}(\tau,\tau') = \delta(\tau-\tau') \delta_{ij} +\int d\tau''\, \sum_k \Sigma^*_{ik}(\tau,\tau'') R_{kj}(\tau'',\tau')
		\end{equation}
Looking at the diagrammatic expansion of the self-energies $\Sigma^*_{ij}$ one realizes that, because no internal vertex has an incoming $\phi_C$ leg ($C$ is not consumed in any reaction, only produced), $\Sigma_{AC}$, $\Sigma_{BC}$ and $\Sigma_{CC}$ are all zero. Of the remaining self-energy entries we only draw the diagrams for $\Sigma_{AA}^*$ and $\Sigma_{AB}^*$  below as the diagrams for $\Sigma_{BA}^*$, $\Sigma_{BB}^* $, $\Sigma_{CA}^*$ and $\Sigma^*_{CB}$ are analogous. To $O(\alpha^2)$ we have then 			
			\begin{equation}
				\begin{split}
					\Sigma^*_{AA} = 
					\parbox{27mm}{
						\begin{fmfgraph*}(70,40)
							\fmfleft{e1}
							\fmfright{e2}
							\fmf{end_arrow_dash,tension=1.5,foreground=blue}{v1,e1}
							\fmf{dashes,tension=2,foreground=blue}{e2,v1}
							\fmfv{label=$\Sigma^*$,label.dist=0,decor.shape=circle, decor.filled=empty, decor.size= 8thick}{v1}
					\end{fmfgraph*}} = 
					\parbox{25mm}{
						\begin{fmfgraph*}(70,40)
							\fmfleft{e1}
							\fmfright{e2}
							\fmf{end_arrow_dash,tension=1.5,foreground=blue}{v1,e1}
							\fmf{dashes,tension=1.5,foreground=blue}{e2,v1}
							\fmfv{decor.shape=circle, decor.filled=full, decor.size= 4thick}{v1}
							\fmffreeze
							\fmfbottom{e3}
							\fmf{dashes,tension=1,foreground=green}{v1,e3}
							\fmfv{decor.shape=circle, decor.filled=empty, decor.size= 7thick,foreground=green}{e3}
					\end{fmfgraph*}}	+
						\parbox{32mm}{ 	\begin{fmfgraph*}(90,30)
							\fmfleft{e1}
							\fmfright{e2}
							\fmf{end_arrow_dash,tension=2,foreground=blue}{v1,e1}
							\fmf{dashes,tension=2,foreground=blue}{e2,v2}
							\fmf{dbl_plain_arrow,foreground=blue,left}{v2,v1}
							\fmf{dbl_plain_arrow,foreground=green,right}{v2,v1}
							\fmfv{decor.shape=circle, decor.filled=full, decor.size= 4thick}{v1}
							\fmfv{decor.shape=circle, decor.filled=full, decor.size= 4thick}{v2}
							\fmffreeze
							\fmfbottom{e3}
							\fmfforce{(0.8w,0h)}{e3}
							\fmf{dashes,foreground=green}{e3,v2}
							\fmfv{decor.shape=circle, decor.filled=empty, decor.size= 7thick,foreground=green}{e3}
					\end{fmfgraph*}} 	+
				\parbox{32mm}{ 	\begin{fmfgraph*}(90,30)
						\fmfleft{e1}
						\fmfright{e2}
						\fmf{end_arrow_dash,tension=2,foreground=blue}{v1,e1}
						\fmf{dashes,tension=2,foreground=blue}{e2,v2}
						\fmf{mixed_response_BG,left}{v2,v1}
						\fmf{mixed_response_GB,right}{v2,v1}
						\fmfv{decor.shape=circle, decor.filled=full, decor.size= 4thick}{v1}
						\fmfv{decor.shape=circle, decor.filled=full, decor.size= 4thick}{v2}
						\fmffreeze
						\fmfbottom{e3}
						\fmfforce{(0.8w,0h)}{e3}
						\fmf{dashes,foreground=green}{e3,v2}
						\fmfv{decor.shape=circle, decor.filled=empty, decor.size= 7thick,foreground=green}{e3}
				\end{fmfgraph*}} +\dots
						\end{split}
			\end{equation}
which corresponds term by term to 
			\begin{equation}
				\begin{split}
				\Sigma_{AA}^*(\tau,\tau') &= -\frac{\delta_{\tau_-,\tau'}}{\Delta t} \alpha k_3 \mu_B(\tau') + (-\alpha k_3)^2 R_{AA}(\tau_-,\tau'_+) R_{BB}(\tau_-,\tau'_+) \mu_B(\tau')  \\
				& + (-\alpha k_3)^2 R_{AB}(\tau_-,\tau'_+) R_{BA}(\tau_-,\tau'_+) \mu_B(\tau')
				\end{split}
		\end{equation}
Similarly one has 
\begin{equation}
				\begin{split}
					\Sigma^*_{AB} = 
					\parbox{27mm}{
						\begin{fmfgraph*}(70,40)
							\fmfleft{e1}
							\fmfright{e2}
							\fmf{end_arrow_dash,tension=1.5,foreground=blue}{v1,e1}
							\fmf{dashes,tension=2,foreground=green}{e2,v1}
							\fmfv{label=$\Sigma^*$,label.dist=0,decor.shape=circle, decor.filled=empty, decor.size= 8thick}{v1}
					\end{fmfgraph*}} = 
					\parbox{25mm}{
						\begin{fmfgraph*}(70,40)
							\fmfleft{e1}
							\fmfright{e2}
							\fmf{end_arrow_dash,tension=1.5,foreground=blue}{v1,e1}
							\fmf{dashes,tension=1.5,foreground=green}{e2,v1}
							\fmfv{decor.shape=circle, decor.filled=full, decor.size= 4thick}{v1}
							\fmffreeze
							\fmfbottom{e3}
							\fmf{dashes,tension=1,foreground=blue}{v1,e3}
							\fmfv{decor.shape=circle, decor.filled=empty, decor.size= 7thick,foreground=blue}{e3}
					\end{fmfgraph*}}	+
					\parbox{32mm}{ 	\begin{fmfgraph*}(90,30)
							\fmfleft{e1}
							\fmfright{e2}
							\fmf{end_arrow_dash,tension=2,foreground=blue}{v1,e1}
							\fmf{dashes,tension=2,foreground=green}{e2,v2}
							\fmf{dbl_plain_arrow,foreground=blue,left}{v2,v1}
							\fmf{dbl_plain_arrow,foreground=green,right}{v2,v1}
							\fmfv{decor.shape=circle, decor.filled=full, decor.size= 4thick}{v1}
							\fmfv{decor.shape=circle, decor.filled=full, decor.size= 4thick}{v2}
							\fmffreeze
							\fmfbottom{e3}
							\fmfforce{(0.8w,0h)}{e3}
							\fmf{dashes,foreground=blue}{e3,v2}
							\fmfv{decor.shape=circle, decor.filled=empty, decor.size= 7thick,foreground=blue}{e3}
					\end{fmfgraph*}} 	+
					\parbox{32mm}{ 	\begin{fmfgraph*}(90,30)
							\fmfleft{e1}
							\fmfright{e2}
							\fmf{end_arrow_dash,tension=2,foreground=blue}{v1,e1}
							\fmf{dashes,tension=2,foreground=green}{e2,v2}
							\fmf{mixed_response_BG,left}{v2,v1}
							\fmf{mixed_response_GB,right}{v2,v1}
							\fmfv{decor.shape=circle, decor.filled=full, decor.size= 4thick}{v1}
							\fmfv{decor.shape=circle, decor.filled=full, decor.size= 4thick}{v2}
							\fmffreeze
							\fmfbottom{e3}
							\fmfforce{(0.8w,0h)}{e3}
							\fmf{dashes,foreground=blue}{e3,v2}
							\fmfv{decor.shape=circle, decor.filled=empty, decor.size= 7thick,foreground=blue}{e3}
					\end{fmfgraph*}} +\dots
				\end{split}
			\end{equation}
or 
		\begin{equation}
			\begin{split}
				\Sigma_{AB}^*(\tau,\tau') &= -\frac{\delta_{\tau_-,\tau'}}{\Delta t} \alpha k_3 \mu_A(\tau') + (-\alpha k_3)^2 R_{AA}(\tau_-,\tau'_+) R_{BB}(\tau_-,\tau'_+) \mu_A(\tau')  \\
				& + (-\alpha k_3)^2 R_{AB}(\tau_-,\tau'_+) R_{BA}(\tau_-,\tau'_+) \mu_A(\tau')
			\end{split}
		\end{equation}
These series can again be summed geometrically to all orders in $\alpha$ as we did in \cref{section_resum} to obtain
			\begin{equation}{\label{sigma_AA_resum}}
				\Sigma^*_{AA}(\tau,\tau') = (-\alpha k_3) \left\{ \delta(\tau-\tau') + \alpha k_3 \left[ R_{AA}(\tau,\tau')R_{BB}(\tau,\tau') + R_{AB}(\tau,\tau')R_{BA}(\tau,\tau') \right] \right\}^{-1} \mu_B(\tau')
			\end{equation}
			\begin{equation}{\label{sigma_AB_resum}}
				\Sigma^*_{AB}(\tau,\tau') = (-\alpha k_3) \left\{ \delta(\tau-\tau') + \alpha k_3 \left[ R_{AA}(\tau,\tau')R_{BB}(\tau,\tau') + R_{AB}(\tau,\tau')R_{BA}(\tau,\tau') \right] \right\}^{-1} \mu_A(\tau')
			\end{equation}
The resulting overall system of equations for copy number means $\mu_i$ and responses $R_{ij}$ defines the SBR-M approximation, where M indicates the inclusion of mixed responses. For e.g.\ species $A$, one  integrates \cref{ABC_mean_general_eqn} and \cref{FD_eqn_mixed} with $\Omega$ and $\Sigma$ given by \cref{sigma_AA_resum,sigma_AB_resum,omega_A_resum}.

\subsection{Numerical results}{\label{ABC_numerics}}
		
For numerical tests of the above two approximation methods we again focus on the regime of small copy numbers. We call that we are considering the $A+B\rightarrow C$ reaction system, with baseline creation and destruction reactions with rates $k_{1i}$ and $k_{2i}$, respectively.
Unless otherwise stated we use the standard parameter set $k_{1A} = 4$, $k_{1B} = 4$, $k_{1C} = 3$, $k_{2A}=3$, $k_{2B}=2$, $k_{2C}=3$, $k_3 = 1$, $\alpha=1$, with time step $\Delta t =0.001$ and total integration time $t = 1$. The initial condition is a product of independent Poisson distributions for each species, with means $\mu_i(0) = k_{1i}/k_{2i}$ for $i=A,B,C$. 
		
\begin{figure}
\includegraphics[width=\linewidth]{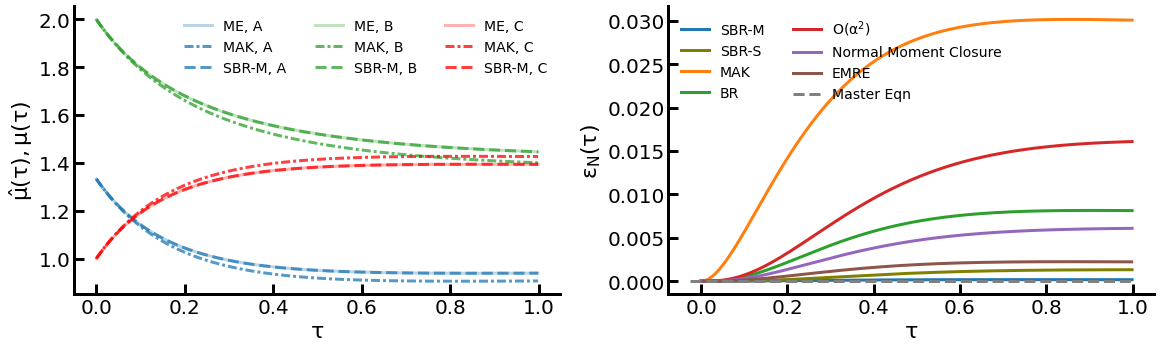}
			\caption{\small Dynamics of average copy numbers (left) 
				 for the three species $A$ (blue), $B$ (orange) and $C$ (green). 
				 Results are shown for the numerically exact solution of the master equation (ME, solid translucent lines), MAK (dash-dotted lines) and SBR-M (dashed lines). Absolute relative deviation (right) from the master equation solution averaged over the $N=3$ species, $\epsilon_N(\tau)$ as defined in \cref{epsilon_N}. SBR-M and SBR-S have significantly lower errors than most other methods. Standard system parameters are used as given in \cref{ABC_numerics}.}
			\label{fig13}
		\end{figure}

In \cref{fig13} (left) we plot the time courses of the mean copy numbers for the three species. As we start the system in what would be the steady state without the $A+B\to C$ reaction, it makes sense that the copy numbers of $A$ and $B$ decrease in time, while those of $C$ increase. All mean copy numbers are below 2 so we clearly are in the strongly fluctuating regime. We plot the time courses as obtained by numerical integration of the master equation (ME) (with an appropriately truncated state space) and by the MAK and SBR-M methods. The deviations of MAK from the true dynamics are evident, while the SBR-M predictions are essentially identical with the ground truth on the scale of the plot. For a more quantitative assessment we plot in \cref{fig13} (right) the absolute value of the relative deviations from the ground truth, averaged over the $N=3$ species, i.e.
\begin{equation}
\epsilon_N(\tau) = \frac{1}{N}\sum_i \left|\frac{\hat{\mu}_i(\tau) - \mu_i (\tau)}{\mu_i(\tau)}\right|
\label{epsilon_N}
\end{equation}
SBR-M is seen to perform best, followed closely by SBR-S, i.e.\ SBR with only single species response functions, and then EMRE. Normal moment closure is stable for this system and the next accurate method, followed by BR; the latter is obtained by integrating \cref{ABC_mean_general_eqn} with $\Omega$ given by \cref{omega_A_single_resum} but using the bare response functions. If we only keep the $O(\alpha^2)$ term in the dynamics with bare propagators (integrating \cref{ABC_bare_mean_eqn}), we get the next best approximation.

\begin{figure}[ht]
			\includegraphics[width=\linewidth]{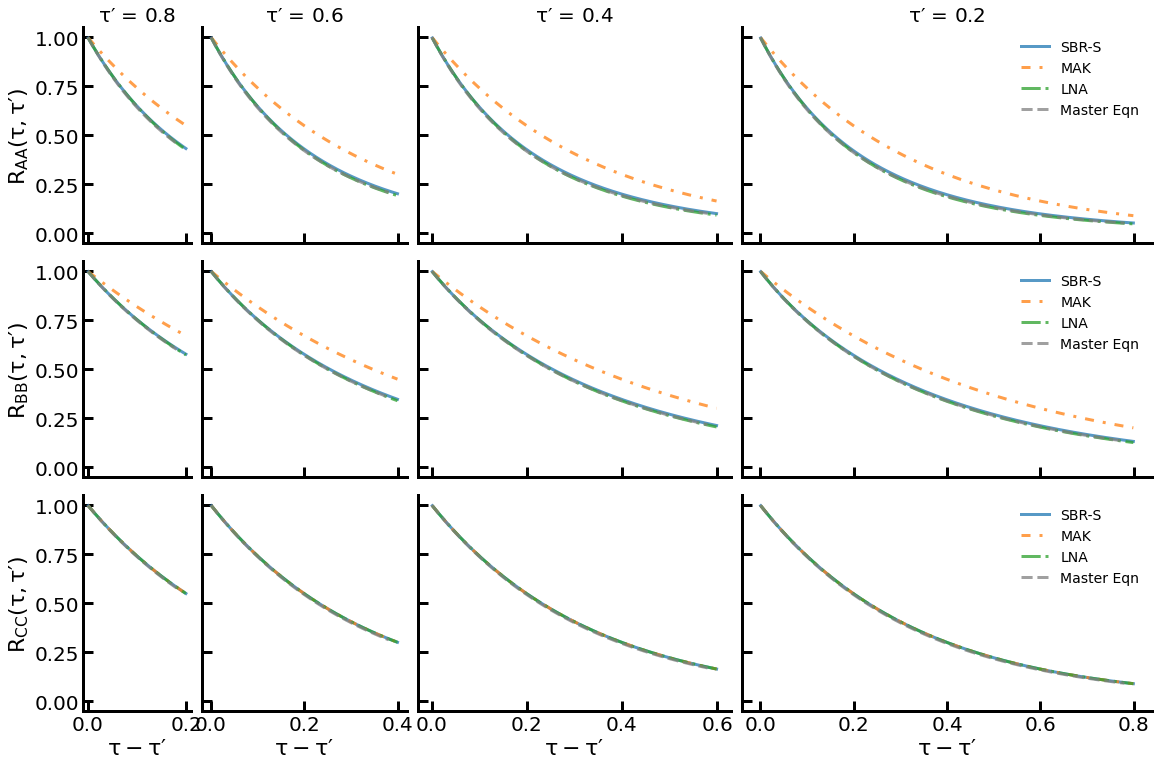}
			\caption{\small 
Single species response functions for the $A+B\to C$ system with the standard parameter set: $R_{AA}$ (top), $R_{BB}$ (middle) and $R_{CC}$ (bottom) as obtained from the master equation, and the MAK, LNA and SBR-S approximations.
The response functions are plotted at fixed values of the second time argument $\tau'$ at $\tau'=0.8, 0.6, 0.4, 0.2$ from left to the right, as a function of the lag $\tau-\tau'$. SBR-S and LNA very closely reproduce the true response function as obtained from the master equation across the whole temporal range while the exponentially decaying MAK response drops off too slowly. $R_{CC}$ is the same from all methods because the dynamics of the mean copy number $\langle n_C\rangle$ of $C$ is not dependent on higher moments involving $n_C$.}
			\label{fig14}
		\end{figure}

We next consider two-time quantities. In \cref{fig14} we show the single species response function $R_{ii}(\tau, \tau')$ for several fixed values of $\tau'$, the time at which the perturbation is applied, against time lag $\tau-\tau'$ between perturbation and response. The SBR-S method and the EMRE closely reproduce the true response functions calculated from the master equation, while MAK shows appreciable deviations. $R_{CC}$ is the same for all methods because the concentration of species $C$ does not appear in any propensity function and hence changing the creation rate of $C$ temporarily trivially changes its concentration, with the effect decaying exponentially only because of the baseline destruction reaction $C\to \emptyset$. 

\begin{figure}[ht]
		\includegraphics[width=\linewidth]{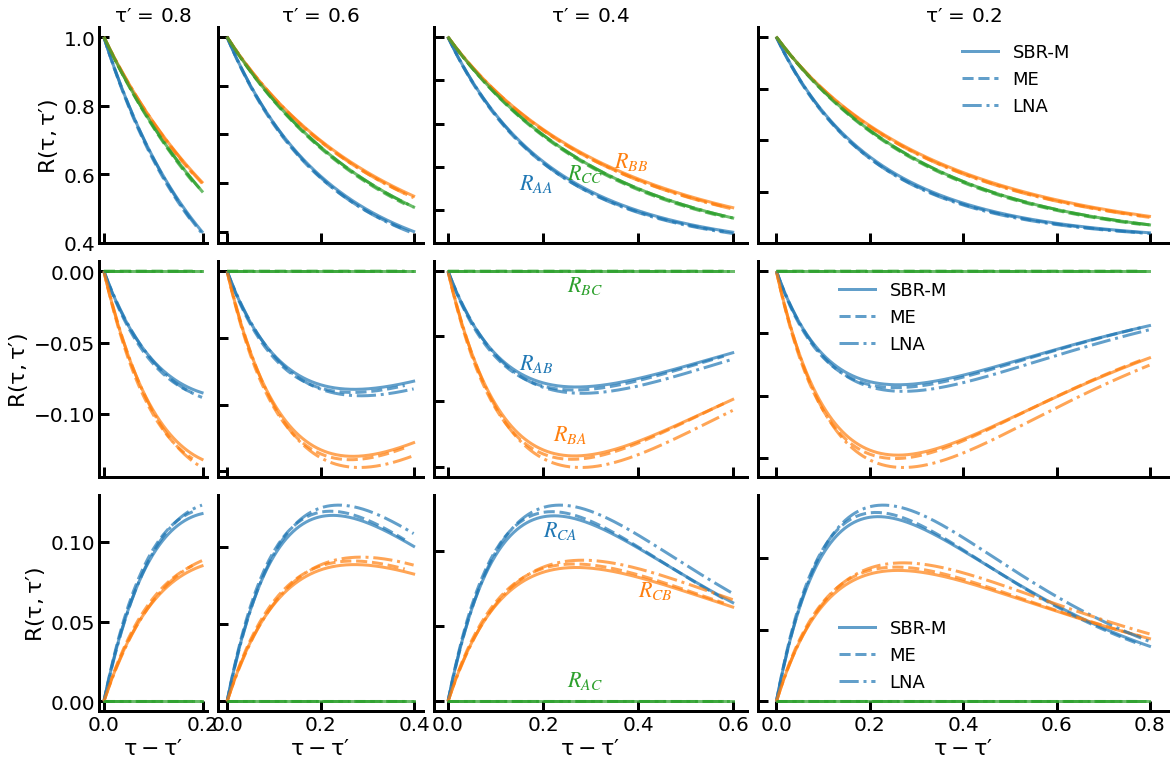}
			\caption{\small 
				Analog of~\cref{fig14} for single species and mixed response functions $R$ as obtained from the numerically exact master equation (dashed lines), the LNA (dashed-dotted lines) and the SBR-M approximation (solid lines). Single species responses $R_{AA}$, $R_{BB}$ and $R_{CC}$ 
				 are quite accurately reproduced by SBR-M and LNA (top). Mixed responses $R_{AB}$,  $R_{BA}$ and $R_{BC}$ (middle) and $R_{CA}$, $R_{CB}$ and $R_{AC}$ (bottom) are all reproduced by SBR-M including their non-monotonic behaviour,
				  while in simpler approximations like MAK all mixed responses would be predicted as identically zero. The LNA also reproduces the correct trends but has quantitative deviations at large time lags.
}
			\label{fig15}
		\end{figure}

With the more sophisticated SBR-M method we have access to all response functions, including the mixed responses. In \cref{fig15} we plot these in the same format as in \cref{fig14}, comparing to the ground truth from the solution of the master equation. The single species responses (top row) are quite accurately predicted by SBR-M and LNA. These responses decay from an initial value of unity as expected. In the middle and bottom rows the mixed responses $R_{ij}, i\neq j$ are plotted. $R_{BC}$ and $R_{AC}$ are zero throughout the dynamics in SBR-M, LNA and the master equation solution because perturbations of species $C$ affect neither $A$ nor $B$. In the middle row, $R_{AB}$ and $R_{BA}$ start from zero and become negative because increasing the creation rate of either $A$ or $B$ enables more $A+B\to C$ reactions to take place, thus decreasing the concentration of the other species. In the bottom row $R_{CA}$ and $R_{CB}$ start from zero and then take positive values because increasing the creation rate of either $A$ or $B$ {\em increases} the concentration of $C$, again because more $A+B\to C$ reactions take place. The non-trivial non-monotonic behaviour of these response functions is reproduced by SBR-M for the whole time range, while such cross-responses would be identically zero within e.g.\ MAK. Predictions for the mixed responses are possible only by including the corresponding mixed self-energies, and are stabilized by the bubble resummation procedure. This behaviour is also captured by the LNA, but deviates quantitatively from the true values at larger lag times.

\begin{figure}
			\includegraphics[width=\linewidth]{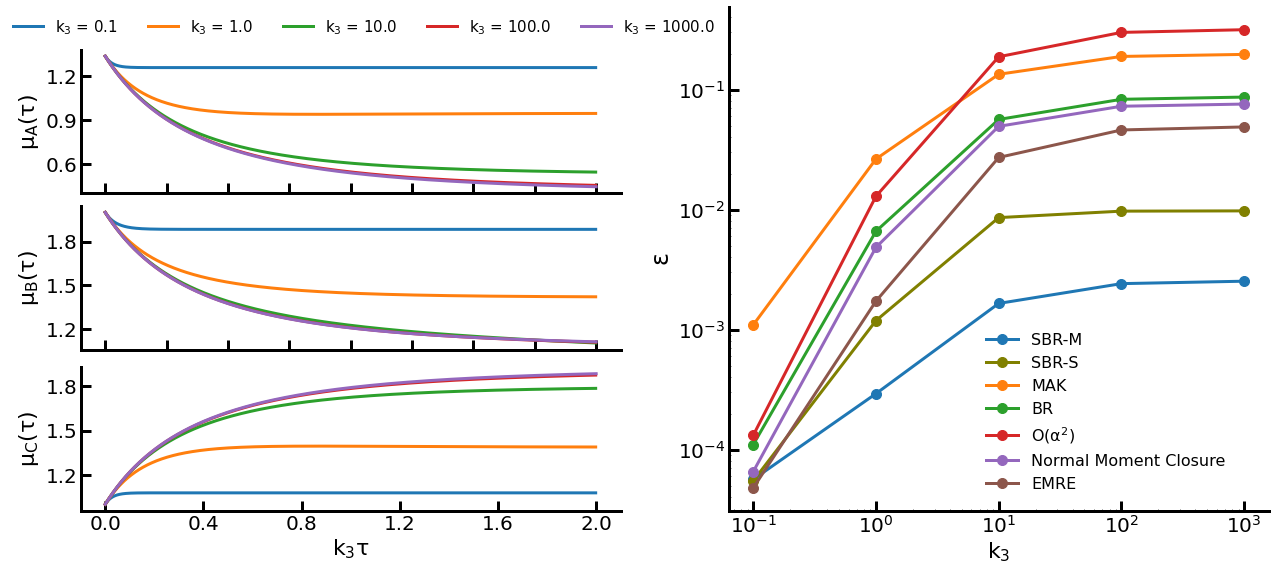}
			\caption{\small Time courses of mean copy numbers $\mu_A$, $\mu_B$ and $\mu_C$ (left) for different values of the interaction rate $k_3$ as 
				calculated from the master equation, and absolute relative deviation of the predictions from different methods (right) averaged over the three species and over time plotted against $k_3$. The total time and the time step have been rescaled analogously to \cref{fig3}. Both our SBR-M and SBR-S methods outperform every other approach over the entire range of $k_3$ shown.}
			\label{fig16}
		\end{figure}

Finally in \cref{fig16} we compare the performance of different methods as we change the rate of the non-trivial reaction $A+B\to C$ perturbation parameter $k_3$, that is changing the interaction rate while keeping the baseline rates constant. On the left we plot the numerically exact mean copy number time courses 
for the three species for different values of $k_3$. For $k_3=1$ we use total time  $t=2$ and time step $\Delta t=0.002$; for other $k_3$ we scale these as in \cref{fig3} keeping the number of time steps the same, and accordingly plot the results against $k_3 \tau$. On the right we plot the absolute relative error of the different methods from the master equation averaged over species and over time, i.e.\ $\epsilon = \frac{1}{B} \sum_\tau \epsilon_N(\tau)$ as a function of $\alpha$; $B=t/\Delta t = 1000$ is the total number of time steps as before. We observe that the SBR-M and SBR-S methods outperform all others across four orders of magnitude in $k_3$. In this multispecies case, it is worthy to note that the SBR methods have a clear advantage over the EMRE, thus highlighting the power and accuracy of the SBR to capture the dynamics of multispecies reaction networks with binary reactions in the challenging regime of small copy numbers.

		\section{Conclusions and discussion}{\label{section_discussions}}
		
We have used Doi-Peliti field theory methods in this paper to construct accurate approximations for the dynamics of chemical reaction networks in the challenging regime of large fluctuations. This approach leads to equations of motion for the mean copy numbers that involve {\em memory} to past (mean) copy numbers, and we determine this memory self-consistently via appropriate response functions.

Technically, we work with diagrammatic perturbation theory around a baseline that only has response ($\phi\tilde\phi$) lines in the bare propagator, while the bare field correlations ($\phi\phi$) are zero. By focussing on only means and response functions, we have managed to construct diagrams with a consistent flow of time. This significantly restricts the class of diagrams in the expansion for the $n$-point and vertex functions, and often the diagrams have simple physical interpretations. (One can of course also calculate the correlation function in this formalism and it does not vanish in general, but it also does not change the dynamics of the mean nor of the response.) By self-consistently replacing the bare response functions with the physical dressed responses in the vertex functions, even fewer diagrams need to be considered. 
We finally approximate by summing up a carefully chosen infinite sub-series of the diagrams for vertex functions and self-energies, giving overall what we call the ``self-consistent bubble resummation'' method (SBR). 

As an alternative we could have, in the diagrams, kept the bare $(\phi \phi)$ correlator, which is zero, and then self-consistently replaced it by its non-zero dressed counterpart. But this approach, which would be more analogous to the traditional diagrammatic approach for MSRJD path integrals where the bare  $(\phi \phi)$ correlator is non-zero, would involve many more diagrams. These diagrams also would not have a consistent flow of time, would contain self-loops, and offer no obvious route to resummation. 

For the binary reactions we consider, the SBR provides all $O(\mu)$ and $O(\mu^2)$ terms in the equations of motion for the mean copy numbers, which can be interpreted as time-delayed analogues of the terms appearing in standard mass-action kinetics. 
Higher order corrections in $\mu$ are included implicitly via the memory functions that weight these time-delayed terms.

We give explicitly the diagrammatic Feynman rules for a single-species example system but, as we demonstrate, these generalize to multi-species cases. In such cases one has a choice of whether or not to include (dressed, physical) mixed response functions between different species. If one does, then for $N$ species one has to consider $N^2$ response functions and their corresponding equations of motion, which for large $N$ becomes computationally expensive. As an alternative we propose a method where only dressed single species response functions are considered; in our numerical examples this is slightly less accurate, but of course also computationally less expensive as only $N$ response functions need to be tracked. 

We numerically demonstrate the superior performance of our SBR method in calculating mean copy numbers compared to other methods, over a very large range of parameter values and over the whole time range we consider. We also calculate the two-time response and correlation functions. Within our SBR approximation these quantities can easily be estimated and in numerical tests also prove to be rather accurate. The fact that we neglect field-field correlators implicitly means that all copy number variances are taken as equal to the corresponding means, as would be the case for Poisson statistics. Our results suggest that this is a reasonable approximation to the actual underlying copy number distributions, but of course it remains an approximation nonetheless.

 A key equation we derive and use extensively in this paper is \cref{mu_star_equation}, which gives the corrections to the mean coming from the past, i.e.\ the memory effects. Of course the underlying microscopic dynamics we consider is Markovian, with the rate for each chemical reaction depending only on the current state of the system via the propensity function. Nonetheless the equations for copy number means and responses do contain memory terms, and this is a natural consequence of the fact that a representation in terms of means and responses is effectively a lower-dimensional {\em projection} of the full microscopic dynamics of the system as specified by the joint distribution of the copy numbers for all species and times.

A number of interesting research perspectives arise from our work. Considering first numerical questions, the straightforwardly time-discretized implementation of the SBR method that we have used in this paper has a computational complexity (see \cref{resum_geometric_appendix}) of $O(B^3)$ where $B$ is the number of time steps. This can become expensive when long trajectories need to be simulated with small time steps, as could be the case when reaction rates in a system vary widely. More sophisticated techniques including adaptive time steps or integration order~\cite{meirinhos_adaptive_2022}, or low rank compression~\cite{kaye_low_2021} could be explored to address this. 
		 
We have also implicitly assumed that our reaction systems are well mixed rather than diffusion-limited, so that only overall copy numbers of each species affect the reaction rates. This assumption does not always hold, e.g.\ the binding and unbinding of transcription factors to DNA in gene regulation networks can be significantly influenced by the spatial arrangement of the DNA~\cite{pai_spatial_2010,hafner_spatial_2023}. It is, however, straightforward to include space in our analysis by considering a spatial grid of compartments, with molecules in different compartments treated effectively as different species. Diffusion is then just a unary conversion reaction from a species in one compartment to one in a neighbour compartment. This introduces terms quadratic in $\tilde{\phi}, \phi$ in the Hamiltonian and does not give rise to memory corrections. In the continuous time and space limit, the Hamiltonian simply acquires an additional term $-D\tilde{\phi}\nabla^2\phi$, with $D$ the diffusion coefficient and $\phi,\tilde{\phi}$ now also dependent on position. Doi-Peliti field theory with diffusing particles preserves their discrete particle identities~\cite{bothe_particle_2022}, i.e.\  the field theory maintains the particle nature of the degrees of freedom. This approach is therefore well suited to the study of chemical reactions especially at low copy numbers, because it does not require any prior spatial coarse graining.
		
Our formalism also allows us to consider time-dependent reaction rates such as $k_1(t)$, $k_2(t)$ or $k_3(t)$. These modify the equations of motion in relatively simple ways, with corrections from integrals over the past then constructed with reaction rates at the corresponding times, see e.g.~\cref{omega_time_dependent,sigma_time_dependent}. This extension could be deployed in situations where the relaxation time scale of the system is comparable to that of any variation in the interactions or the time period of periodic driving, as in time-dependent branching processes~\cite{assaf_population_2008}. It is also straightforward to treat non-Poisson initial conditions using our method. This will change the initial overlap from \cref{initial_overlap_path_integral} and introduce different $t=0$ boundary terms in the action in \cref{action_continuous}.
		
At the more technical level, it will be interesting to explore whether our approach can be used also for Martin-Siggia-Rose-Jansen-de Dominicis (MSRJD)~\cite{martin_statistical_1973,dominicis_techniques_1976,janssen_lagrangean_1976} path integrals for classical interacting particle systems with Langevin dynamics. These share some similarities with the Doi-Peliti approach and a diagrammatic expansion can be constructed with only response functions if noise sources are treated as perturbations~\cite{chow_path_2015} so that approaches analogous to the ones developed here should be applicable. We also intend to investigate the connections between the formalism we have presented here and the two-particle irreducible (2PI) effective action known from field theory~\cite{cornwall_effective_1974}. 

Finally, since our method predicts marginal Poisson distributions for all species, we have access to the entire time trajectory of the copy number distribution. It will be interesting to deploy this for {\em inference}, by extending it to approximations for the likelihood of a time series of observed copy numbers of molecular species, especially in biochemical reaction networks. This inference problem becomes more challenging in the not uncommon situation when not all species can be observed, and one then expects further memory effects from the projection onto the observable part of the reaction network~\cite{rubin_memory_2014,rubin_michaelis-menten_2016,bravi_statistical_2017,bravi_inferring_2017,bravi_inference_2017,bravi_systematic_2020,herrera-delgado_memory_2018}. 
 
		\newpage
	
	\printbibliography
	
	\newpage
	
	\appendix
	\section{Appendix}
	\renewcommand\thefigure{\thesection.\arabic{figure}}
	\setcounter{figure}{0}    

\subsection{Time step dependence of errors}\label{time_step_errors_appendix}

\begin{figure}[ht]
		\includegraphics[width=\linewidth]{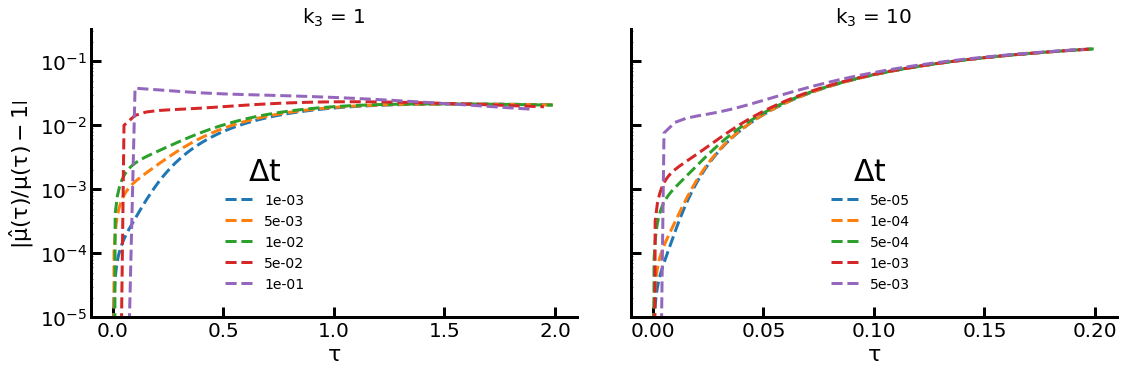}
		\caption{\small Dependence on time step $\Delta t$ of the prediction error of SBR for the $A+A\rightarrow A$ system for 
			$k_3=1$ (left) and $k_3=10$ (right), with the standard parameter set otherwise (see \cref{numerics_AAA}). Smaller step sizes and a smaller time range are used on the right because of the larger $k_3$. A limit is approached for small $\Delta t$, with convergence significantly faster near the steady state regime. 
		}
		\label{fig5}
	\end{figure}

Numerically, an important parameter to consider is the time step $\Delta t$ we use to evaluate predictions of the various approximations. We plot the dependence of the absolute value of the relative deviation from the ground truth (master equation solution) as a function of time for different step sizes for $k_3=1$ and $k_3=10$ in \cref{fig5}. Different step sizes lead to the same error in the steady state while in the transient the errors decrease with decreasing step size. Note that the scaling in the limit $\Delta t\to 0$ is not as simple here as for a standard Euler integration scheme for a differential equation: 
as $\Delta t$ is decreased, the number of time steps and thus the size of the response matrix increases, which increases the potential for numerical errors in the matrix inversion required in \cref{Gamma_12_contTime,sigma_12_contTime}.
	
\subsection{Bare response functions}\label{bare_response_decay_appendix}

\begin{figure}[ht]
	\includegraphics[width=\linewidth]{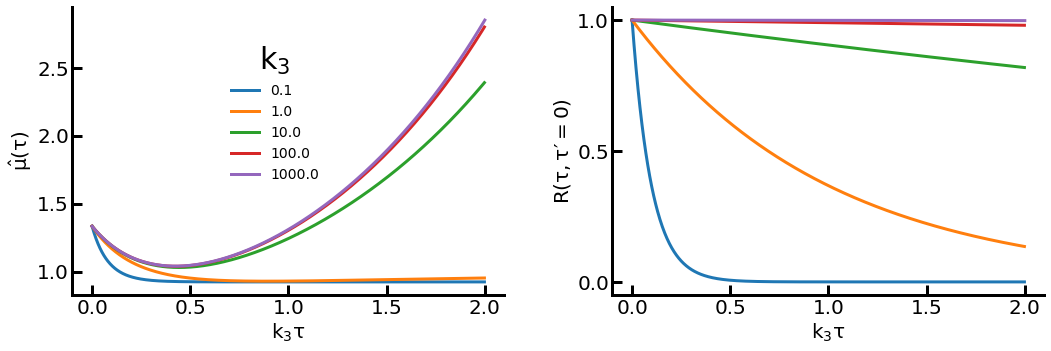}
	\caption{\small Mean copy numbers (left) and response functions (right) obtained by including only $O(\alpha^2)$ terms in the diagrammatic expansion. Proceeding as in \cref{fig3}, we scale down integration time $t$ and time step $\Delta t$ as we increase $k_3$. The bare response functions $R_0(\tau,\tau'=0)$ used in this method are plotted as a function of $k_3 \tau$ for different $k_3$. We see that the method is unstable at large $k_3$ because the bare response functions do not decay fast enough at large $k_3$, that is a perturbation created at $\tau'=0$ in the system has an apparent effect for a time that is much longer than the timescale of the $A+A\to A$ reaction.}
	\label{fig4}
\end{figure}

As briefly touched upon in \cref{numerics_AAA}, including only the $O(\alpha^2)$ correction to the diagrammatic expansion for the mean copy number leads to instabilities in the predicted time evolution, especially at larger values of $k_3$, because the response functions do not decay fast enough; they are not sensitive to the relaxation from the binary reaction and its rate $k_3$. In \cref{fig4}(left) we plot the mean copy number obtained by this method as a function of $k_3 \tau$ (i.e.\ with time rescaling as before, see \cref{fig3}) for different $k_3$. For $k_3>1$, the mean copy numbers quickly separate from the ground truth and would diverge if we ran the dynamics to longer times. In \cref{fig4}(right) we plot the response function $R_0(\tau,\tau')$ for $\tau'=0$ as a function of $k_3 \tau$. This measures the change in $\mu$ at time $k_3 \tau$ from a small perturbation in the creation rate $k_1$ at time $\tau'=0$. For large $k_3$ this effect should decay rapidly because of the fast $A+A\to A$ reaction, but the approximation does not capture this because the bare response always decays on a timescale of $O(1)$.

\subsection{Long time behaviour of SBR}\label{SBR_long_time_appendix}

\begin{figure}
		\includegraphics[width=\linewidth]{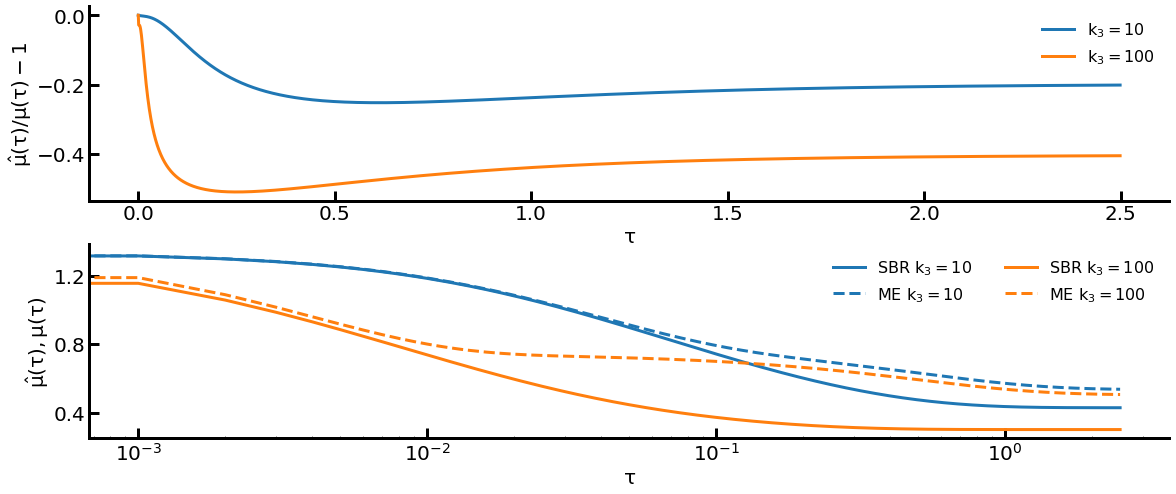}
		\caption{\small (Top) Relative deviations as a function of time $\tau$ between the self-consistent bubble resummed (SBR) approximation and the master equation (ME) for the $A+A\rightarrow A$ system with baseline reactions, with the standard parameter set but larger $k_3$ as shown. These results are calculated for a longer total time $t=2.5$ (with time step $\Delta t=0.001$) to show the approach to the final steady state where the copy numbers and deviations become constant. (Bottom) Corresponding dynamics of the mean copy numbers, plotted on a logarithmic time scale to show the fast $O(1/k_3)$ and slow $O(1)$ time regimes. }
		\label{fig_app1}
	\end{figure}

In \cref{fig_app1} we show some additional information regarding the long-time behaviour of the dynamics of the $A+A\rightarrow A$ system with the baseline reactions, for the standard parameter set but with larger rates $k_3=10$ and $k_3=100$ for the coagulation reaction. We show the predictions of SBR and compare to the solution of the master equation. The initial decay in the trajectories is driven by the fast coagulation reaction with rate $k_3$ and the second decay by the baseline destruction rate $k_2$, which is of $O(1)$. From these plots we see that, for short times of the order of $1/k_3$, 
 the SBR closely reproduces the ME trajectories. After that time, there are deviations between SBR and ME driven by the fact that SBR leaves out $\Omega^{1,m}$ terms with $m>2$, so SBR does not predict the correct steady state and after times of $O(1)$, its deviation from the master equation solution is constant.
		
	\subsection{Construction of the path integral}{\label{app_path_integral}}

	We want to be able to express averages in terms of the formalism introduced in the main text. The construction of the path integral is closely inspired by \cite{tauber_applications_2005}. 
	
	We define the associated bra states of the kets defined earlier, namely  $\langle \bm{n}|  = \langle n_1, \dots, n_N |$ which is generated by the action of $a_i$ on the state $\langle \bm{0}|$:
	\begin{equation}
		\langle \bm{n} | =	\langle \bm{0} | \prod_i \frac{a_i^{n_i}}{n_i!} 
		\label{bra}  
	\end{equation}
The difference from the standard quantum mechanics normalization is explicit here and comes from demanding that the bra states are orthonormal to the ket states, $\langle \bm{n} | \bm{n}'\rangle = \delta_{\bm{n},\bm{n}'}$.

We now define a uniform state or  ``sum state'', 
	\begin{equation}
		\langle \bm{1} | = \sum_{\bm{n}} \langle \bm{n}| = \langle \bm{0} | e^{\sum_i \hat{a}_i }
	\end{equation}
where the second expression follows from \cref{bra}. Because bra and ket states are orthonormal we have then $\langle \bm{1}| \bm{n}\rangle = 1$, and so 
	\begin{equation}{\label{prob_cons_proj}}
		\langle \bm{1} | P(\tau) \rangle = \sum_{\bm{n}} P(\bm{n},\tau) = 1 \qquad \qquad \forall \text{ } 0<\tau<t 
	\end{equation}
which expresses conservation of probability. More generally, the mean value of any observable $A$ at time $\tau$, denoted by $\bar{A}(\tau)$, is given by
	\begin{equation}
		\begin{split}
			\bar{A}(\tau) &= \sum_{\bm{n}} P(\bm{n},\tau) A(\bm{n}) \\
			&=  \sum_{\bm{n}} P(\bm{n},\tau)  \langle \bm{1} | \hat{A} 			
			|\bm{n} \rangle \\
			&= \langle  \bm{1} |\hat{A}| P(\tau) \rangle
		\end{split}
	\end{equation}	
	where $\hat{A}$ is obtained by replacing $n_i \rightarrow \hat{a}_i^\dagger \hat{a}_i$ in $A(\bm{n})$. 
	Using the explicit expression for $|P(\tau)\rangle$ we can also write
	\begin{equation}
		\bar{A}(\tau) = \langle \bm{1} | \hat{A} e^{\hat{H}\tau} | P(0) \rangle  
	\end{equation}
where the initial state $|P(0)\rangle$ is the Poisson mixture of species with initial means $\bar{n}_{0i}$ as defined in \cref{Poisson_initial_state}. 

	We introduce the coherent states $| \psi \rangle, \langle \psi|$, which are eigenstates of the creation and annihilation operators $\hat{a}$ and $\hat{a}^\dagger$ with eigenvalues $\psi$ and $\psi^*$, respectively:
	\begin{eqnarray}
		\hat{a} | \psi \rangle = \psi | \psi \rangle  \quad & \langle \psi | \hat{a}^\dagger = \langle \psi | \psi^*
	\end{eqnarray}
(For simplicity we treat a single species for now and so omit the species index $i$.) Explicitly, these eigenstates can be written as 
	\begin{equation}
		| \psi \rangle = \exp \left( -\frac{1}{2} |\psi|^2 + \psi \hat{a}^\dagger \right) |0 \rangle, \qquad
				\langle \psi | = \langle 0| \exp \left( -\frac{1}{2} |\psi|^2 + \psi^* \hat{a} \right)
	\end{equation}
	with overlap given by
	\begin{equation}
		\langle \psi_1| \psi_2 \rangle = \exp \left( -\frac{1}{2} |\psi_1|^2 -\frac{1}{2} |\psi_2|^2 + \psi_1^* \psi_2 \right)
	\end{equation}
From these states one can construct an overcomplete resolution of the identity, 
	\begin{equation}
		\mathbb{I} = \int \frac{d\phi^* d\phi}{\pi} | \phi \rangle \langle \phi | \quad \text{with} \quad d\phi^* d\phi =  d(\text{Re} \phi) d(\text{Im} \phi)
	\end{equation}
	We now insert such a resolution of the identity after every time step $\Delta t$, using one field for each species, i.e.\ $\bm{\phi}(\tau) \equiv \bm{\phi}_\tau = (\phi_1(\tau), \phi_2(\tau), \dots, \phi_N(\tau))$. 
Doing this in the normalization \cref{prob_cons_proj} at time $t$ gives
	\begin{equation}
		1 = \mathcal{N}^{-1} \lim_{\Delta t \rightarrow 0} \int \prod_{\tau=0}^t d \bm{\phi}^*_\tau  d \bm{\phi}_\tau \langle \bm{1} | 
		\bm{\phi}_t \rangle \langle \bm{\phi}_t | e^{\hat{H} \Delta t} | \bm{\phi}_{t - \Delta t}\rangle \dots \langle \bm{\phi}_\tau | e^{\hat{H} \Delta t} | \bm{\phi}_{\tau - \Delta t}\rangle  \dots \langle \bm{\phi}_0 | P(0) \rangle
		\label{normalization_one}
	\end{equation}
with a normalization factor $\mathcal{N}=\pi^{(t/\Delta t)+1}$ that will not play a role in the following. The matrix element factors can be evaluated to $O(\Delta t)$ as
	\begin{equation}
	\langle \bm{\phi}_\tau | e^{\hat{H} \Delta t} | \bm{\phi}_{\tau-\Delta t} \rangle = 
	\langle \bm{\phi}_\tau | \bm{\phi}_{\tau-\Delta t} 	\rangle  
=\langle \bm{\phi}_\tau | \bm{\phi}_{\tau-\Delta t} \rangle  e^{H(\bm{\phi}^*_\tau, \bm{\phi}_{\tau-\Delta t})\Delta t}
\label{matrix_element}
	\end{equation}
because the Hamiltonian in \cref{Hamiltonian_operator} is \textit{normal ordered}, i.e.\ with creation operators $\hat{a}_i^\dagger$ always to the left of annihilation operators $\hat{a}_i$; these operators then simply act on their eigenvectors to the left and right, respectively, giving the relevant factors of $\phi^*_i(\tau)$ or $\phi_i(\tau)$:
	\begin{equation}
		H(\bm{\phi}^*_\tau, \bm{\phi}_{\tau-\Delta t}) = \langle \bm{\phi}_\tau | \hat{H}| \bm{\phi}_{\tau - \Delta t} \rangle
	\end{equation}
	Note that the conjugate field $\phi^*$ is always a time step $\Delta t$ ahead of $\phi$. The remaining overlap factor in \cref{matrix_element} is
	\begin{equation}{\label{overlap_app1}}
		\langle \bm{\phi}_\tau | \bm{\phi}_{\tau_-}\rangle = \exp\left[ -(\bm{\phi}^*_\tau)^{\rm T}( \bm{\phi}_\tau - \bm{\phi}_{\tau_-}) \right] \exp\left(\frac{1}{2} | \bm{\phi}_\tau|^2 - \frac{1}{2} |\bm{\phi}_{\tau_-} |^2 \right)
	\end{equation}
After multiplication across $\tau=0,\Delta t, \ldots t$ the second exponential cancels for all times other than the initial and the final ones. The first exponential can be expressed in terms of the discrete derivative $\Delta_\tau \phi_i(\tau) = \frac{1}{\Delta t} (\phi_i(\tau) - \phi(\tau_-))$:
	\begin{equation}
	\exp\left[ -(\bm{\phi}^*_\tau)^{\rm T} ( \bm{\phi}_\tau - \bm{\phi}_{\tau_-}) \right]	
	= \exp\left[ -(\bm{\phi}^*_\tau)^{\rm T} \Delta_\tau \bm{\phi}(\tau) \Delta t \right]		
	\end{equation}
	The final term in \cref{normalization_one} including the final surviving squared exponential can be evaluated by noticing that the sum state $\langle \bm{1}|$ is, up to normalization, a coherent state with eigenvalue 1, such that 
	\begin{equation}{\label{overlap_app2}}
		\langle \bm{1} | \bm{\phi}_t \rangle \exp\left(\frac{1}{2}|\bm{\phi}_t|^2\right)  =  \exp\left(\sum_i \phi_i(t)\right)
	\end{equation}
For the initial term one has similarly, using that $|P(0)\rangle$ is proportional to an eigenvector of $a_i$ with eigenvalue $\bar{n}_{0i}$,
	\begin{equation}{\label{initial_overlap_path_integral}}
		\langle \bm{\phi}_0 | P(0) \rangle \exp\left(-\frac{1}{2}|\bm{\phi}_0|^2\right) = \exp\left(\sum_i \left[\bar{n}_{0i} (\phi_i^*(0) - 1) - |\bm{\phi}_i(0)|^2\right]\right)
	\end{equation}
	
	We now generalize the above construction by inserting instead of an identity $\mathbb{I}$, a factor
	\begin{equation}
e^{ \sum_i \theta_i(\tau)\Delta t\,\hat{a}_i} \, \mathbb{I} \, 	e^{ \sum_i \tilde\theta_i(\tau)\Delta t\,(\hat{a}^\dagger_i -1 )}
	\end{equation}
	 at each discretized time $\tau$. Differentiation w.r.t.\ the generating fields $\theta_i(\tau), \tilde{\theta}_i(\tau)$, defined for each species at each time step, then allows us to generate averages such as mean copy numbers, correlation functions etc.\ as explained in the main text. This leads to following path integral representation of the generating function:
	\begin{equation}{\label{path_integral1}}
		\mathcal{Z}( \tilde{\theta}  , \theta ) = \lim_{\Delta t \rightarrow 0} \mathcal {N}^{-1} \int \prod_{\tau} d \bm{\phi}_\tau^* d\bm{\phi}_\tau  \langle \bm{1} |\bm{\phi}_t \rangle \prod_{\tau=\Delta t}^{t} 	\langle \bm{\phi}_\tau | e^{\sum_i (\hat{a}_i^\dagger - 1 )\tilde{\theta}_i(\tau) \Delta t} e^{\hat{H} \Delta t} e^{\sum_i \hat{a}_i \theta_i(\tau-\Delta t) \Delta t} | \bm{\phi}_{\tau - \Delta t} \rangle \langle \bm{\phi}_0 | P(0) \rangle
	\end{equation}
	The matrix element for the propagation is then given by
	\begin{equation}
		\begin{split}
		\langle \bm{\phi}_\tau | e^{\sum_i (\hat{a}_i^\dagger-1) \tilde{\theta}_i(\tau) \Delta t} e^{\hat{H} \Delta t} e^{\sum_i \hat{a}_i \theta_i(\tau-\Delta t) \Delta t} | \bm{\phi}_{\tau - \Delta t} \rangle = \\
		e^{H(\bm{\phi}^*_\tau, \bm{\phi}_{\tau-\Delta t})\Delta t} \langle \bm{\phi}_\tau | \bm{\phi}_{\tau-\Delta t} \rangle  \prod_i \left\lbrace e^{\tilde{\theta}_i(\tau) (\phi_i^*(\tau)-1)  \Delta t + \theta_i(\tau-\Delta t) \phi_i(\tau-\Delta t) \Delta t} \right\rbrace 
		\end{split}
	\end{equation}
	Putting all the elements together with the initial and final overlap, that is \cref{overlap_app1,overlap_app2,initial_overlap_path_integral}, the partitition function can be written in terms of an action, $S$, as
	\begin{equation}
		\mathcal{Z}( \tilde{\theta}  , \theta ) = \lim_{\Delta t \rightarrow 0} \mathcal{N}^{-1} \int \prod_\tau d \bm{\phi}_\tau^* d\bm{\phi}_\tau 
		e^{S\left(  \phi^* ,  \phi  \right)}
	\end{equation}
	where $S$ and its subsequent forms are given in the main text in \cref{discrete_action} and below.
	
	\subsection{Matrix form of baseline action and discrete time response function }{\label{matrix_action}}
		In this section we cast the baseline action in a matrix form for a multivariate Gaussian distribution and obtain the response function by inverting the precision matrix of the Gaussian. We start by defining the following vectors, 
		\begin{equation}
			\begin{split}
				\boldsymbol{\phi}_i &= (\phi_i(0),\phi_i(\Delta t),\dots,\phi_i(t-\Delta t),\phi_i(t)) \\
				\boldsymbol{\tilde{\phi}}_i &= (\tilde{\phi}_i(0),\tilde{\phi}_i(\Delta t),\dots,\tilde{\phi}_i(t-\Delta t),\tilde{\phi}_i(t)) \\
				\boldsymbol{b}_i &= (\theta_i(0), \theta_i(\Delta t), \dots, \theta_i(t-\Delta t) ,\theta_i(t) ) \Delta t \\
				\boldsymbol{\tilde{b}}_i &= (\tilde{\theta}_i(0)+ k_{1i},  \tilde{\theta}_i(\Delta t)+ k_{1i} ,\dots, \tilde{\theta}_i(t-\Delta t)+ k_{1i}, \tilde{\theta}_i(t) + k_{1i}) \Delta t 
			\end{split}
		\end{equation}
	
		 Then one has the quadratic action, 
	\begin{equation}{\label{Gaussian_action}}
		S_0\left(  \tilde{\phi} ,  \phi  \right)  = \sum_i \left[ \boldsymbol{\tilde{\phi}}_i^{\rm T} \boldsymbol{\tilde{b}}_i + \boldsymbol{\phi}_i^{\rm T} \boldsymbol{b}_i  
		- \boldsymbol{\phi}_i^{\rm T} \bm{J}_i^{\rm T} \boldsymbol{\tilde{\phi}}_i 
		\right] + \mbox{const.}
	\end{equation}
	where $\bm{J}_i$ is a lower triangular matrix defined as
	\begin{equation}
		\bm{J}_i = 
		\begin{pmatrix}
			1 & 0 & 0 & 0 &\dots & 0\\
			k_{2i}\Delta t - 1 & 1 & 0 &0 & \dots & 0 \\
			0& k_{2i}\Delta t - 1  & 1 &0 & \dots & 0 \\
			0 & 0 & k_{2i}\Delta t - 1  & 1  & \dots & 0 \\
			\vdots & \vdots & \vdots  & \vdots  & \dots & 0 \\
			0 & 0 & \dots &0 & k_{2i}\Delta t -1 & 1
		\end{pmatrix}
	\end{equation}
The corresponding path integral is Gaussian and decoupled across molecular species.

 To find the response and correlation functions one writes the quadratic part of the action (for one species) as
\begin{equation}
-\frac{1}{2}
\begin{pmatrix}
\bm{\phi}_i \\
\bm{\tilde{\phi}}_i
\end{pmatrix}^{\rm T}
\begin{pmatrix}
0 & \bm{J}_i^{\rm T} \\
\bm{J}_i & 0
\end{pmatrix}
\begin{pmatrix}
\bm{\phi}_i \\
\bm{\tilde{\phi}}_i
\end{pmatrix}
\end{equation}
Because the fields have a Gaussian distribution, the matrix that appears here is the inverse covariance or precision matrix. Inverting gives the covariance matrix in block form:
\begin{equation}
\begin{pmatrix}
\langle \delta\bm{\phi}_i \delta\bm{\phi}_i^{\rm T}\rangle & \langle \delta\bm{\phi}_i \delta\bm{\tilde{\phi}}_i^{\rm T}\rangle \\  
\langle \delta\bm{\tilde{\phi}}_i \delta\bm{\phi}_i^{\rm T}\rangle & \langle \delta\bm{\tilde{\phi}}_i \delta\bm{\tilde{\phi}}_i^{\rm T}\rangle
\end{pmatrix}
=\begin{pmatrix}
0 & \bm{J}_i^{\rm T} \\
\bm{J}_i & 0
\end{pmatrix}^{-1}
=\begin{pmatrix}
0 & \bm{J}_i^{-1} \\
(\bm{J}_i^{\rm T})^{-1} & 0
\end{pmatrix}
\end{equation}
As expected on general grounds of causality, the bottom right block with its averages of products of $\tilde\phi$-factors vanishes. The correlation block (top left) also vanishes because the precision matrix has no $\tilde\phi\tilde\phi$-couplings. The only nonzero covariance block is the response function matrix 
$\langle \delta\bm{\phi}_i \delta\bm{\tilde{\phi}}_i^{\rm T}\rangle = \bm{J}_i^{-1}$,
with entries $R_{0i}(\tau,\tau') = \langle \delta \phi_i(\tau) \delta \tilde{\phi}_i(\tau') \rangle$.
As $\bm{J}_i$ is lower triangular with unit diagonal entries, it follows directly that the response function matrix has the same structure: 
the equal-time response (diagonal elements) is $R_{0i}(\tau,\tau)=1$ and the response is causal in that the elements above the diagonal vanish,  
	\begin{equation}
		R_{0i} (\tau,\tau') = 0 \text{ if } \tau < \tau' 
	\end{equation}
This is of course consistent with the interpretation discussed in the main text, of $\tilde\phi_i(\tau')$ representing a perturbation (increase in creation rate of species $i$) and $R_{0i}(\tau,\tau')$ measuring the response of the ``real'' field $\phi_i(\tau)$ at $\tau\geq \tau'$.

The non-trivial elements of the response can be found from the condition for it to be the inverse of $\bm{J}_i$:
	\begin{equation}
		\sum_{\tau''} J_i(\tau,\tau'') R_{0i}(\tau'',\tau') = \delta_{\tau,\tau'}
	\end{equation}
	which for $\tau > \tau'$ implies 
	\begin{equation}
		R_{0i}(\tau,\tau') - R_{0i}(\tau-\Delta t,\tau') = -k_{2i} R_{0i}(\tau-\Delta t,\tau')
	\end{equation}
Solving iteratively starting from $R(\tau',\tau')=1$ gives
\begin{equation}
R_{0i}(\tau,\tau')=(1-k_{2i}\Delta t)^{(\tau-\tau')/\Delta t}
\end{equation}
For reference, we give the corresponding result in the continuous time limit $\Delta t\to 0$:
\begin{equation}
		R_{0i}(\tau,\tau') = e^{-k_{2i}(\tau-\tau')} \Theta(\tau-\tau')
\end{equation}
where $\Theta(\cdot)$ is the Heaviside step function. This is the solution of the inversion condition, which in continuous time becomes a differential equation:
	\begin{equation}
		\partial_\tau R_{0i}(\tau,\tau') = -k_{2i} R_{0i}(\tau,\tau') \quad \text{for } \tau>\tau'
	\end{equation}
Writing $\bm{J}_i$ in continuous time form also shows that
\begin{equation}
		R_{0i}^{-1}(\tau,\tau') = (\partial_{\tau} + k_{2i}) \delta(\tau-\tau')
\end{equation}
which as expected is the same result as \cref{R_inv_defn} in the main text.

	\subsection{Resumming bubble diagrams as a geometric series}{\label{resum_geometric_appendix}}
	In this section we show how to perform the bubble summation of \cref{section_resum} in discrete time by considering the series of matrix products as a geometric series. We start by summing the series in \cref{Gamma_series_1_bare} 
	To do so we define a new matrix $\chi(\tau,\tau')$ such that
	\begin{equation}{\label{chi_defn}}
		\chi(\tau,\tau') = R_0^2(\tau,\tau') \quad \text{element-wise}
	\end{equation}
and a column vector $M(\tau)$ as $M(\tau) = \mu^2(\tau)$. The matrix $\chi$ is a lower triangular matrix because of the causality of the response function, with all entries equal to unity on the diagonal. This matrix is thus not diagonalizable and powers of the matrix $\chi^2$, $\chi^3$, $\chi^4$, $\dots$ need to be explicitly calculated. The dimension of the matrix $\chi$ (as for $R_0$) is $(B+1,B+1)$, where $B=t/\Delta t$ is the number of time steps.

For the following we include a factor $\Delta t$ in each matrix product, i.e.\ we define the matrix product $PQ$ to have $(\tau,\tau')$ entry $\Delta t \sum_{\tau''} P(\tau,\tau'')Q(\tau'',\tau')$. This is useful because such matrix products become operator products in the limit $\Delta t\rightarrow 0$. We also define a lower shift matrix $L$ with entries $L(\tau,\tau')=\delta_{\tau_-,\tau'}/\Delta t$. Multiplying this from the left with a matrix shifts the first time index, e.g.\ $L\chi$ has $(\tau,\tau')$ entry
\begin{equation}
\Delta t \sum_{\tau''} L(\tau,\tau'')\chi(\tau'',\tau') = \chi(\tau_-,\tau')
\end{equation} 

In the continuous time limit this will become a delta function, $L(\tau,\tau') = \delta(\tau -\tau')$, and so drop out of the expressions. The first sum in \cref{Gamma_series_1_bare} is then
\begin{equation}
\Delta t \sum_{\tau'} R_0^2(\tau_-,\tau')\mu^2(\tau'_-) = \Delta t \sum_{\tau'} (L\chi)(\tau,\tau')(LM)(\tau') = (L\chi LM)(\tau)
\end{equation}
and the second one similarly
\begin{equation}
(\Delta t)^2 \sum_{\tau'} R_0^2(\tau_-,\tau')R_0^2 (\tau'_-,\tau'')\mu^2(\tau''_-) = (\Delta t)^2 \sum_{\tau'} (L\chi)(\tau,\tau')(L\chi)(\tau',\tau'')(LM)(\tau'') = (L\chi L\chi LM)(\tau)
\end{equation}

We can then write the full sum \cref{Gamma_series_1_bare} for ${\Omega}^{1,2}$, the column vector with entries $\Omega^{1,2}(\tau)$, as
\begin{equation}
		{\Omega}^{1,2} = \sum_{n= 1}^{B+1} 2^{n-1}  (-\alpha k_3)^n (L \chi)^{n-1} LM
	\end{equation}

Note that for finite $\Delta t$ we in principle only have a finite number of terms in the bubble diagram series that contribute as shown because $L\chi$ is strictly lower triangular (i.e.\ has zeros on the diagonal), 
but for the same reason the sum is unchanged if we extend it to all $n\geq 1$. We can perform the resulting geometric summation to obtain
	\begin{equation}
		{\Omega}^{1,2} =  		
		(-\alpha k_3)  \left( \mathbb{I} - (-2\alpha k_3 L \chi )  \right)^{-1}  LM
	\end{equation}
In continuous time this gives \cref{Gamma_12_contTime_bare} of the main text
	\begin{equation}
		\Omega^{1,2}(\tau) = (-\alpha k_3) \int_0^\tau d\tau' \left( \delta(\tau-\tau') + 2\alpha k_3 R_0^2(\tau,\tau') \right)^{-1} \mu^2(\tau')
	\end{equation}
where the inverse is now in the operator sense. Self-consistent replacement of the bare propagator, $R_0$, by the dressed propagator, $R$, leads to \cref{Gamma_12_contTime} of the main text. If $k_3$ is a function of time $k_3(\tau)$, then this can be easily generalized to obtain
\begin{equation}{\label{omega_time_dependent}}
	\Omega^{1,2}(\tau) = -\alpha \int_0^\tau d\tau' k_3(\tau') \left( \delta(\tau-\tau') + 2\alpha k_3(\tau') R^2(\tau,\tau') \right)^{-1} \mu^2(\tau')
\end{equation}

In the following, we directly start with the dressed response, instead of $R_0$. To sum up the series of diagrams in \cref{sigma_series_1} one can proceed similarly. We define a new diagonal matrix $\zeta$ with elements $\zeta(\tau,\tau')$ as
	\begin{equation}
		\zeta (\tau,\tau') = \mu(\tau) \delta_{\tau,\tau'}/ (\Delta t) 
	\end{equation}
	Then we can express the self-energy $\Sigma^*$ with entries $\Sigma^*(\tau,\tau')$ as
 \begin{equation}		
		\Sigma^* = \sum_{n= 1}^{B+1} (-2\alpha k_3)^n  (L \chi)^{n-1} L \zeta
	\end{equation}
with $\chi$ now understood as the element-wise square of $R$ rather than $R_0$. This can be summed up to give
	\begin{equation}		
		\Sigma^* = \lim_{\Delta t \rightarrow 0} (-2\alpha k_3)  \left( \mathbb{I} - (-2\alpha k_3 L \chi )  \right)^{-1}  L \zeta
	\end{equation}		
	In continuous time this gives \cref{sigma_12_contTime} of the main text
	\begin{equation}
		\Sigma^*(\tau,\tau') =  (-2\alpha k_3) \left( \delta(\tau-\tau') + 2\alpha k_3 R^2(\tau,\tau') \right)^{-1} \mu(\tau')
	\end{equation}
which for time-dependent rates $k_3(\tau)$ would generalize to
	\begin{equation}{\label{sigma_time_dependent}}
		\Sigma^*(\tau,\tau') =  -2\alpha k_3(\tau) \left( \delta(\tau-\tau') + 2\alpha k_3(\tau') R^2(\tau,\tau') \right)^{-1} \mu(\tau')
	\end{equation}
In the simplest discrete-time implementation with $B={t}/{\Delta t}$ 
time steps, the above relations lead to a computational complexity for getting all copy number means and response functions of $O(B^3)$. This comes from an $O(B^2)$ effort for the inversion of the triangular matrices; since this needs to be done for every one of the $B$ time steps, we have $O(B^3)$ complexity overall. 
	
	\subsection{Poisson copy number distribution}{\label{poiss_dist_proof}}
	Following the same line of arguments as in \cref{response_correlation_relations_section} we will now show that the our path integral as defined in \cref{path_integral_main_text} or more explicitly in \cref{path_integral1} describes marginal Poisson copy number distributions for each species if the path integral is Gaussian, i.e.\ the action only has has linear and quadratic terms, and if the field correlation functions $C_i$ vanish. We start by taking the $k$ derivatives w.r.t $\tilde{\theta}$ and $\theta$ such that the operators are normal ordered, 
	\begin{equation}
		\begin{split}{\label{copy_number_dist_1}}
		\frac{1}{(\Delta t)^{2k}} \left( \frac{\partial }{\partial \tilde\theta_i(\tau_+)} + 1 \right)^k \frac{\partial^k }{\partial \theta_i^k(\tau)} \mathcal{Z} \Bigg|_{\theta,\tilde{\theta}=0} &= \langle \bm{1}| (\hat{a}_i^\dagger)^k (\hat{a}_i)^k| \bm{\phi}_{\tau} \rangle \\
		&= \langle n_i(\tau) (n_i(\tau)-1) \dots (n_i(\tau)-k+1) \rangle \\		
		&=\langle \lbrace n_i(\tau) \rbrace_k\rangle 
		\end{split}
	\end{equation}
where we acted on the state by the annihilation operator to lower it, then used the creation operator and finally exploited probability conservation as defined in \cref{prob_cons_proj}. $\langle \lbrace n_i \rbrace_k \rangle \equiv \langle n_i!/(n_i-k)! \rangle $ is the $k^{\text{th}}$ \textit{factorial moment}.
	
	Calculating the same derivatives from the path integral we have
		\begin{equation}
			\begin{split}
		\frac{1}{(\Delta t)^{2k}} \left( \frac{\partial }{\partial \tilde\theta_i(\tau_+)} + 1 \right)^k  \frac{\partial^k }{\partial \theta_i^k(\tau)} \mathcal{Z} \Bigg|_{\theta,\tilde{\theta}=0} &= \langle (\tilde\phi_i(\tau_+)+1)^k \phi_i^k(\tau) \rangle \\
		&= \langle \phi_i^k(\tau)\rangle \\
		& = \langle [\delta \phi_i(\tau) + \langle n_i(\tau)\rangle ]^k\rangle
		\end{split}
	\end{equation}
where we have used the causality property discussed in \cref{response_correlation_relations_section} in going from the first to the second line. 

The final average can now be expanded and evaluated using Wick's theorem if, as assumed, we have a Gaussian path integral. All terms involving $\delta\phi_i$ then reduce to powers of $C_i(\tau,\tau) = \langle \delta \phi_i(\tau) \delta \phi_i(\tau)\rangle$, so if the latter vanishes 
one has simply
			\begin{equation}
		\begin{split}
		\frac{1}{(\Delta t)^{2k}} \left( \frac{\partial }{\partial \tilde\theta_i(\tau_+)} + 1 \right)^k  \frac{\partial^k }{\partial \theta_i^k(\tau)} \mathcal{Z} \Bigg|_{\theta,\tilde{\theta}=0} = \langle n_i(\tau)\rangle^k
		\end{split}
	\end{equation}
	Comparing with \cref{copy_number_dist_1} we obtain that the $k^{\text{th}}$ factorial moment of the copy number distribution is equal to the mean to the power $k$, that is $\langle \lbrace n_i \rbrace_k\rangle = \langle n_i\rangle^k $ for all times, which uniquely characterizes a Poisson distribution~\cite{potts_note_1953} as claimed. Note that even though the marginal distributions are Poisson, there can still be correlations between species.
	 
	\subsection{Numerical solution of master equation}{\label{app_ME}}
	
In this appendix we explain how we construct numerical solutions of the master equation, using the single species system with the $A+A \xrightarrow{k_3} A$ reaction and the baseline reactions $\varnothing \xrightleftharpoons[k_2]{k_1} A$ as an example. For the particular case of single species and single step master equations -- where only one molecule can be created or destroyed in one time step -- a more efficient method~\cite{smith_general_2015} is available that relies on calculating only the eigenvalues (analytically or numerically) of the master operator. We stick to the significantly more general finite space projection method~\cite{munsky_finite_2006} here.

The master equation for the time evolution of the probabilities $P(n,\tau)$ of having $n$ number of molecules of $A$ in the reaction volume is given by 
	\begin{equation}
		\partial_\tau P(n,\tau) = k_{1}P(n-1,\tau) +[k_{2}(n+1) + k_3 (n+1)n] P(n+1,\tau)  - \left[ k_1 + n k_2 +k_3 n(n-1)\right]P(n,\tau)
	\end{equation}
	The state of the system is completely described by $n$, the number of molecules of $A$. To make the state space finite and thus allow a numerical approach to the problem, we impose an upper bound on the maximum copy number, calling this $n^{\text{max}}$. The state-space of the system is thus of size $\kappa = n^{\text{max}}+1$. This has to be chosen large enough to cover the bulk of the probability throughout the dynamics. The errors of this method are well controlled and its convergence to the true solution can be shown for many cases~\cite{munsky_finite_2006}. For the simulations for $A+A \rightarrow A$, for example, we use $\mathbf{n^{\text{max}} = 20}$. Calling the vector of state probabilities 
$| P(\tau) \rangle = | P(0,\tau), P(1,\tau), \ldots, P(n^{\text{max}},\tau) \rangle$, the master equation can then be written in matrix-vector form as
	\begin{equation}
		\partial_\tau |P(\tau) \rangle = M |P(\tau)\rangle
		\label{master_eqn_matrix_form}
	\end{equation}
with an appropriately defined master operator $M$, a matrix with elements $M_{nn'}$; the off-diagonal entries give the rates for transitions from $n'$ to $n$. To ensure probability conservation, any reactions that would violate the upper bound $n\leq n^{\text{max}}$ are removed; in the concrete example, this applies to the particle creation reaction $\emptyset \to A$ from $n=n^\text{max}$ to $n=n^\text{max}+1$.

The initial distribution over the state space is a Poisson distribution with mean $\bar{n}_0$ 
	\begin{equation}
		P(n,0) = \mathcal{N}^{-1} e^{- \bar{n}_0} \frac{ \bar{n}_0^{n} }{n!}
	\end{equation}

Due to the truncation of the state space an additional normalization factor $\mathcal{N} = \sum_{n=0}^{n^{\text{max}}} P(n,0)$ is in principle necessary here, though for sensibly chosen $n^\text{max}$ this will be very close to unity; 
 e.g.\ for the concrete system considered here one finds, with $\bar{n}_0=4/3$ as in the main text and $n^\text{max}=20$, that $1-\mathcal{N}$ is of the order of $10^{-18}$. (For the multispecies case, the initial is constructed analogously as a product of such truncated Poisson distributions.)
 
Given the initial condition, $|P(\tau)\rangle$ can be found by integrating the system \cref{master_eqn_matrix_form} of linear differential equations using any standard algorithm, in the simplest case Euler integration with fixed time step $\Delta t$. Alternatively, and this is the approach we use throughout, one can write the solution as  $ | P(\tau) \rangle =  e^{M \tau} | P(0) \rangle$ and evaluate the matrix exponential by diagonalizing $M$. To do this one decomposes
		\begin{equation}
			M = R \Lambda L^{\rm T}
		\end{equation}
where the columns of the matrices $R$ and $L$ are the right and left eigenvectors, respectively, normalized as $RL^{\rm T}=\mathbb{I}$, and $\Lambda$ is a diagonal matrix collecting the corresponding eigenvalues. We perform this diagonalization using standard numpy linear algebra packages for singular value decomposition. The matrix exponential is then 	$e^{M\tau} = R e^{\Lambda \tau} L^{\rm T} $ and accordingly 
		\begin{equation}
			|P(\tau)\rangle = R e^{\Lambda \tau} L^{\rm T}  | P(0)\rangle
		\end{equation}
			
Once $|P(\tau)\rangle$ is known, averages of one-time observables such as the mean copy number or its second moment can be evaluated straightforwardly as
		\begin{equation}
			\langle n(\tau) \rangle = \sum_{n} n P(n,\tau),  \qquad  \langle n^2(\tau) \rangle = \sum_{n} n^2 P(n,\tau)
		\end{equation}
To measure two-time quantities such as the connected number correlator, $\langle \delta n(\tau) \delta n(\tau')\rangle$ we use
		\begin{equation}
			\langle \delta n(\tau) \delta n(\tau')\rangle = \langle \bm{1} | \hat{n}\, e^{M(\tau-\tau')} \hat{n}\, e^{M \tau'} | P(0) \rangle - \langle n(\tau) \rangle \langle n(\tau') \rangle		
		\end{equation}
where $\hat{n}$ is the number operator, a diagonal matrix with the number of particles for each state ($n=0,1,\ldots n^\text{max}$) on the diagonal.

To find the response function $R(\tau,\tau')$, finally, we use the following procedure. Going back to the definition of the response as caused by temporary perturbations in the creation rate $k_1$, we perturb $k_1$ to $k_1'=k_1+\Delta k_1$ at time $\tau'$ for a time $\Delta t$ and call $M'=M+\Delta M$ the perturbed master operator. The response function for $\tau>\tau'$ is then
		\begin{equation}
			\Delta t \,\Delta k_1\, R(\tau,\tau') = \langle \bm{1} | \hat{n}\, e^{M(\tau-(\tau'+\Delta t))} e^{M' \Delta t} e^{M \tau'} | P(0) \rangle - \langle \bm{1} | \hat{n}\, e^{M(\tau-(\tau'+\Delta \tau))} e^{M \Delta \tau} e^{M \tau'} | P(0) \rangle
		\end{equation}
		In the limit $\Delta t \rightarrow 0$ this simplifies to
		\begin{equation}
				R(\tau,\tau') = \langle \bm{1} | \hat{n}\,e ^{M(\tau-\tau')} \frac{\Delta M}{\Delta k_1} e^{M \tau'} | P(0) \rangle
	\end{equation}
The result is independent of $\Delta k_1$ as it should be because $\Delta M$ is directly proportional to $\Delta k_1$.

	\end{fmffile}
\end{document}